\documentclass[
 reprint,
 superscriptaddress,
 nofootinbib,
 amsmath,amssymb,
 aps
]{revtex4-1}

\usepackage{nicematrix}
\usepackage{graphicx}
\usepackage{subcaption}   
\usepackage{xcolor}
\usepackage{dcolumn}
\usepackage{dsfont}
\usepackage{amsthm, amsfonts}
\usepackage{bbm}
\usepackage{comment}
\usepackage{eucal}
\usepackage{mathtools}
\usepackage{bm}

\usepackage{algorithm}
\usepackage{algpseudocode}

\newcommand{\grafe}[1]{\left\{ #1 \right\}}
\newcommand{\tonde}[1]{\left( #1 \right)}
\newcommand{\quadre}[1]{\left[ #1 \right]}

\makeatletter
\newcommand*\bigcdot{\mathpalette\bigcdot@{.5}}
\newcommand*\bigcdot@[2]{\mathbin{\vcenter{\hbox{\scalebox{#2}{$\m@th#1\bullet$}}}}}
\makeatother

\newlength{\myeqskip}  
\AtBeginDocument{%
    \setlength\abovedisplayskip{\myeqskip}%
    \setlength\belowdisplayskip{\myeqskip}%
    \setlength\abovedisplayshortskip{\myeqskip-\baselineskip}%
    \setlength\belowdisplayshortskip{\myeqskip}}

\usepackage[symbol]{footmisc}
\begin{document}

\preprint{APS/123-QED}
\title{Non-reciprocal interactions and high-dimensional chaos: \\ comparing dynamics and statistics of equilibria in a solvable class of models}

\author{Samantha J. Fournier$^*$}
\affiliation{Université Paris-Saclay, CNRS, CEA, Institut de physique théorique, 91191, Gif-sur-Yvette, France}

\author{Alessandro Pacco$^{*, \dagger}$}
\affiliation{Université Paris-Saclay, CNRS, LPTMS, 91405, Orsay, France}

\author{Valentina Ros}
\affiliation{Université Paris-Saclay, CNRS, LPTMS, 91405, Orsay, France}

\author{Pierfrancesco Urbani}
\affiliation{Université Paris-Saclay, CNRS, CEA, Institut de physique théorique, 91191, Gif-sur-Yvette, France}

\def\thefootnote{$*$}     \footnotetext{These authors contributed equally.}\def\thefootnote{\arabic{footnote}}
\def\thefootnote{$\dagger$}     \footnotetext{alessandro.pacco@universite-paris-saclay.fr}\def\thefootnote{\arabic{footnote}}

\date{\today}

\begin{abstract}
We investigate a model of high-dimensional dynamical variables with all-to-all interactions that are random and non-reciprocal. We characterize its phase diagram and show that the model can exhibit chaotic dynamics. We show that the equations describing the system’s dynamics exhibit a number of equilibria that is exponentially large in the dimensionality of the system, and these equilibria are all linearly unstable in the chaotic phase. Solving the effective equations governing the dynamics in the infinite-dimensional limit, we determine the typical properties (magnetization, overlap) of the configurations belonging to the attractor manifold. We show that these properties cannot be inferred from those of the equilibria, challenging the expectation that chaos can be understood purely in terms of the numerous unstable equilibria of the dynamical equations. We discuss the dependence of this scenario on the strength of non-reciprocity in the interactions. These results are obtained through a combination of analytical methods such as Dynamical Mean-Field Theory and the Kac-Rice formalism.
\end{abstract}

\maketitle
Systems with many units interacting through random and \emph{non-reciprocal} (i.e., asymmetric) couplings have been studied since the earlier works on neural networks~\cite{crisanti1987dynamics, hertz1986memory, ChaosSompo88, gardner1989phase}, which already highlighted how non-reciprocity leads to distinctive dynamical features such as chaos~\cite{ChaosSompo88}.
High-dimensional models of this sort have remained highly relevant in neural networks theory~\cite{rajan2010stimulus, crisanti2018path, marti2018correlations, MastrogiuseppeLink2018, AnnibaleDynamics2024, HeliasMemory18, aguirre2022satisfiability, pereira2023forgetting} and more recently also in machine-learning, in view of the fact that high-dimensional chaotic systems can be trained~\cite{sussillo2009generating, fournier2023statistical} and can serve as generative models~\cite{ fournier2024generative}.  Dynamical systems with non-reciprocal interactions are also at the center of a current trend of research in econophysics and ecology~\cite{ros2023generalized, garnier2024unlearnable, blumenthal2024phase, aguirre2024heterogeneous, poley2025interaction, castedo2024generalised, suweis2024generalized, arnoulx2024many, mahadevan2024continual}, where asymmetry of the interactions between the interacting agents or species is natural, as much as chaotic dynamics~\cite{rogers2022chaos}.

In high-dimensional random models, chaos emerges as a dynamical phase transition driven by the strength of random interactions: at weak randomness, the dynamics rapidly settles into a time-independent, stable fixed point. Certain properties of this asymptotic configuration concentrate in the high-dimensional limit and are fixed by the parameters describing the statistics of the interactions (rather than by their specific realization). As randomness increases beyond a critical level, this single equilibrium looses its dynamical stability and the system enters into a chaotic phase characterized by persistent endogenous fluctuations. This dynamical transition is often concomitant with the loss of uniqueness of the equilibrium and the emergence of \emph{glassiness}, intended here as the explosion in the number of equilibria of the equations describing the system's dynamics. The multiplicity of equilibria naturally raises the question of their classification in terms of their properties, such as their linear stability \cite{Fyodorov_2016, ben2021counting, lacroix2022counting}, and of whether this classification helps in elucidating features of the system's dynamics \cite{HuangSteadyState2024, HuangFranzParisiPot2024, HeliasFP2022}. This question is particularly compelling in presence of non-reciprocal interactions, since the system's dynamics is non-conservative and cannot be described in terms of the optimization of (gradient-based algorithms in) an underlying energy landscape. In systems with conservative, gradient-based dynamics the situation is clearer: the trajectories converge to equilibria, which correspond to the stationary points of the energy landscape.
When multiple stationary points are present, gradient-based algorithms tend to converge to those that are marginally stable; peculiar out-of-equilibrium behaviors of conservative dynamics, such as aging and avalanches, can be understood in terms of this marginal stability \cite{anderson1978houches, cugliandolo1993analytical, franz2021surfing, urbani2024statistical}. Several studies have focused on understanding how this behavior changes when non-conservative terms are introduced~\cite{crisanti1987dynamics, CugliandoloNonrelax97, berthier2000two, Fyodorov_2016, berlemont2022glassy, Fyodorov_resilient_2021}, and whether changes in the dynamics can be interpreted in terms of changes in the nature of the equilibria. More generally, when the optimization framework no longer holds, it remains an open question whether equilibria still play any role in shaping the system's dynamics \cite{HeliasFP2022, wainrib2013topological}.

Here, we tackle these questions in the context of a paradigmatic model with non-reciprocity, that is sufficiently simple to allow for an explicit analysis of both its dynamics and the distribution of its equilibria. We consider a model in which the chaotic drive is a non-linear force with Gaussian statistics and the amount of non-reciprocity is controlled by a parameter $\alpha$. For $\alpha=0$ the model has completely asymmetric interactions and it reduces to the one considered in \cite{fournier2023statistical} to study learning algorithms of recurrent neural networks; we derive correspondingly the dynamical phase diagram, by solving explicitly the Dynamical Mean-Field Theory (DMFT) equations in the large-time limit. We then present results on the topological complexity (the entropy of equilibria of the dynamical equations) obtained via a statistical physics approach relying on the replicated Kac-Rice (KR) formalism. Comparing these results allows us to elucidate that the dynamics in the chaotic phase converges to an attractor whose properties \emph{are not} those of the most numerous (or, typical) equilibria of the dynamical equations, even though connections between the dynamics and the distribution of the equilibria are present. We discuss the extension of our results to $\alpha>0$. In the SM \cite{SM_ours}, we provide a detailed derivation of our statements as well as a generalization to a broader family of models.

\begin{figure*}
    \centering
    \includegraphics[width=1\textwidth]{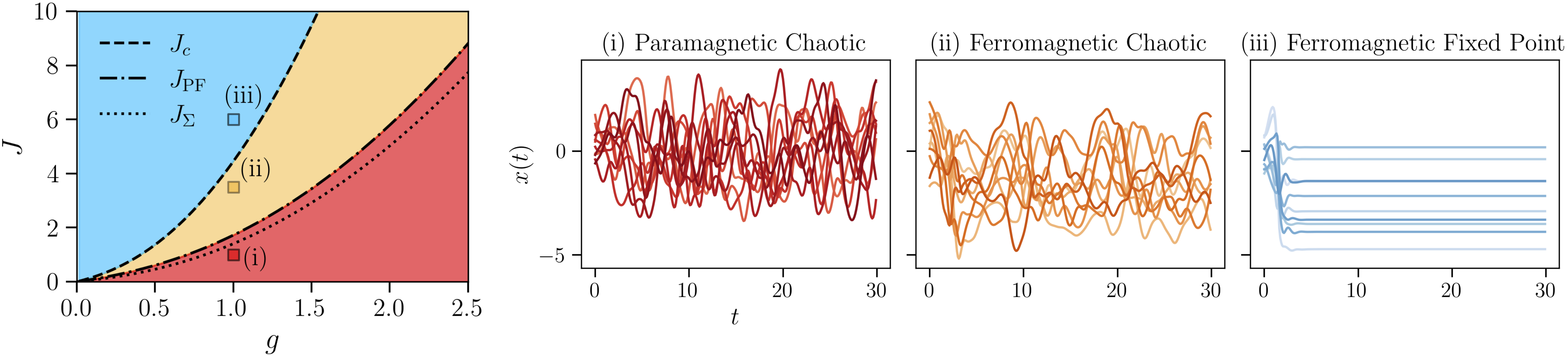}
    \caption{(\textit{Left}) Dynamical phase diagram of the confined model: at fixed $g$, increasing $J$ the system is in a (i) PC - Paramagnetic Chaotic  phase, (ii) FC - Ferromagnetic Chaotic phase, and (iii) FFP - Ferromagnetic Fixed Point phase.  
    (\textit{Right}) Traces of randomly selected $x_i(t)$ obtained from the numerical integration (Euler-discretized) of Eqs.~\eqref{eq:dynamical_system} with a time-step $dt=0.1$, $N=200$ neurons, $g=1$ and (i) $J=1$ (ii) $J=3.5$ (iii) $J=6$. In both figures, $\gamma=0.5$ and  $\alpha=0$.}
    \label{fig:dmft_phase}
\end{figure*}

\textbf{The model.} We consider $N \gg 1$ variables (neurons) $\{x_i\}_{i=1,..,N}$ evolving through the equations
\begin{align}
  \label{eq:dynamical_system}
  \partial_t x_i(t) = -\lambda({\bf x}) x_i(t) + f_i({\bf x}) := F_i({\bf x}),
\end{align}
where $f_i({\bf x})$ are random functions taking the form 
\begin{align}
  f_i({\bf x}) = \sqrt{1-\alpha}\, f^d_i({\bf x}) - \sqrt{\alpha} \, \frac{\partial H ({\bf x})}{\partial x_i}   +  \frac{J}{N}\sum_j x_j
\end{align}
with the parametrization $f^d_i({\bf x}) = g \,  N^{-1} \sum_{jk}^N J_{i}^{jk} x_j x_k,$ and $H({\bf x}) = - 2 g \, N^{-1} \sum_{i<j<k}^N S_{ijk} x_i x_j x_k$.
The couplings $J_{i}^{jk}$ 
and $S_{ijk}$ are random, centered and Gaussian. The first tensor satisfies $J_{i}^{jk}= J_{i}^{kj}$ and 
$\mathbb{E}[J_{i}^{jk}J_a^{bc}]=\delta_{ia}(\delta_{jb}\delta_{kc}+\delta_{jc}\delta_{kb})$. Due to the lack of symmetry in the lower index, the components $f^d_i({\bf x})$ can not be written as derivatives of a unique function: this term corresponds to the non-conservative part of the dynamics. The second tensor is instead symmetric, with independent entries with $\mathbb{E}[S_{ijk}^2]=1$. It corresponds to the conservative part of the force: a gradient term driving the system towards minima of the energy $H({\bf x})$. The parameters $\alpha, g $  control the strength of the conservative contribution and of the randomness~\cite{Fyodorov_2016, ben2021counting, lacroix2022counting}, while the term proportional to $J>0$ favors configurations with a non-zero magnetization. The first, deterministic term in~\eqref{eq:dynamical_system} provides a confinement to the dynamics.
We focus on the case $\lambda({\bf x})= N^{-1}{\bf x}^2 - \gamma$ with $\gamma>0$. This defines a generalization of the “confined model" considered in \cite{fournier2023statistical} as a simplification of standard recurrent neural network models \cite{ChaosSompo88}, obtained by replacing the neurons' firing rate function with a generic nonlinear function of the neuron's membrane potentials $x_i$. In the SM \cite{SM_ours}, we generalize our results to the alternative case in which $\lambda({\bf x})$ is a Lagrange multiplier chosen to enforce the spherical constraint ${\bf x}^2=N$, providing a strict confinement of the dynamics on the $(N-1)$-dimensional hypersphere, as well as to the case in which the random forces have general covariances.
The model \eqref{eq:dynamical_system} exhibits a rich dynamical phase diagram.
One can study its high-dimensional dynamics via DMFT when learning algorithms such as the one proposed in \cite{sussillo2009generating} update dynamically the couplings as a function of the activity of the neurons~\cite{fournier2023statistical}. Here we focus on the dynamics described by the equations \eqref{eq:dynamical_system} with couplings that are random, but fixed in time.

\textbf{Dynamics ($\alpha=0$).}  The numerical integration of Eqs.~\eqref{eq:dynamical_system} displays three qualitatively distinct regimes, shown in Fig.~\ref{fig:dmft_phase}~(\emph{Right}). At fixed $g, \gamma$, for large $J$ the variables $x_i(t)$ converge quickly to time-independent values that are different from zero, and temporal fluctuations are suppressed: the dynamics is attracted by a fixed point, with ferromagnetic properties. As $J$ decreases, the system transitions into a chaotic regime characterized by persistent oscillations of the $x_i(t)$, around values that are non-zero for intermediate $J$ (ferromagnetic chaos), and zero for smaller $J$ (paramagnetic chaos). When $N \to \infty$ these regimes correspond to three distinct dynamical phases separated by sharp transitions, see Fig.~\ref{fig:dmft_phase}~(\emph{Left}). A quantitative characterization of these phases is obtained by solving the DMFT equations describing the system's dynamics in the  large-$N$ limit, like in \cite{ChaosSompo88, crisanti2018path}. The DMFT equations are deterministic, coupled integro-differential equations describing the dynamics of the system's magnetization 
 $m(t)= N^{-1} \sum_i x_i(t)$,  correlation function $C(t,t')=N^{-1} \sum_i x_i(t) x_i(t')$, and response function $R(t,t') =N^{-1} \sum_i \delta x_i(t)/\delta h_i(t') \Big|_{h \to 0}$, where $h_i(t)$ are linear perturbations to the dynamical equations. When
$N \to \infty$, these stochastic functions concentrate around their typical values, which solve the DMFT equations. The latter are derived from a mean-field representation of Eqs.~\eqref{eq:dynamical_system} in terms of an effective single-site stochastic equation
\begin{align}
\begin{split}
\label{eq:effective_dynamical_system}
 & \partial_t{x}(t) =  -\lambda(t) x(t) + 4g^2\alpha\int_0^{t}ds\, C(t,s) R(t,s) x(s)\\
  & + J m(t) + \eta(t):= F(t), \quad \quad  \lambda(t)=C(t,t)-\gamma
\end{split}
\end{align}
where $\eta(t)$ is a zero-mean Gaussian process with variance $\langle \eta(t)\eta(t') \rangle = 2g^2C^2(t,t')$. The functions  $m(t) \equiv \langle x(t)\rangle$,  $C(t,t') \equiv \langle x(t)x(t')\rangle$ and  $R(t,t') \equiv \langle \delta x(t)/\delta \eta(t') \rangle$ are obtained self-consistently from \eqref{eq:effective_dynamical_system}, through averages $\langle \cdot \rangle$ with respect to $\eta(t)$, see Eqs.~\eqref{eq supp: equation for m} in the SM \cite{SM_ours}. The resulting equations can be integrated numerically for arbitrary $\alpha$; for $\alpha=0$, their asymptotic solutions can be determined explicitly. At large times, the dynamics reaches a stationary state in which one point functions become time-independent, $m(t) {\to} m_\infty$ and $\lambda(t) {\to} \lambda_\infty$, 
and two-point functions become Time Translational Invariant (TTI),  $C(t,t') \to C(|t-t'|)$ and $R(t,t') \to R(|t-t'|)$.  
When $\alpha=0$, we can follow the same approach as \cite{ChaosSompo88,crisanti2018path} and transform the DMFT equations into the form:
 \begin{equation}\label{eq:Pot}
     \partial^2_\tau C(\tau)= -  V'(C)
 \end{equation}
 with an effective potential\footnote{While in standard recurrent neural network models \cite{ChaosSompo88} the potential $V$ is defined only implicitly, here it is simple and explicit because we have chosen a Gaussian process as a forcing term.} $V(C) = \frac23 g^2 C^3 -\frac12 \lambda_{\infty}^2 C^2 + \left( J\, m_{\infty}\right)^2 C$ . The correlation $C(\tau)$ in the TTI regime evolves conserving the effective energy $E= (\partial_\tau C)^2/2 + V(C)$, from which it follows that $V(C_0)=V(C_\infty)$ where  $C_0= \lim_{\tau \to 0} C(\tau)$ and $C_\infty= \lim_{\tau \to \infty} C(\tau)$. This relation, in addition to the DMFT equations, allows one to get closed equations for the parameters $m_\infty, C_0$ and $C_\infty$ (with $\lambda_\infty=C_0-\gamma$ in the confined model)
 characterizing the system's behavior in the asymptotic, TTI regime. The explicit solutions to these equations for arbitrary $\gamma$, $g$ and $J$ are given in Sect.~\ref{sect: stationnary solution of the DMFT equations for alpha 0} of the SM \cite{SM_ours}. They confirm the existence of three distinct dynamical phases: (i) the Paramagnetic Chaotic (PC) phase with $m_\infty=0=C_\infty$; (ii) the Ferromagnetic Chaotic (FC) phase with $m_\infty, C_\infty>0$ and $C_0 >C_\infty$, and the (iii) Ferromagnetic Fixed Point (FFP) phase, with  $m_\infty>0$ and $C_\infty=C_0>0$. The fixed point phase is characterized by $C_0 =C(\tau)= C_\infty$, while in the chaotic phases $C_0 > C_\infty$, since the system decorrelates during the chaotic trajectories, either completely (paramagnet) or partially (ferromagnet). The transition (iii) $\to $ (ii) from the FFP to the FC phase occurs at
 \begin{equation}\label{eq:Jc}
     J_{\rm c} := 2g \left(g+\sqrt{\gamma+g^2}\right).
 \end{equation}
 Along this critical line, 
the fixed point solution $C_0=C_\infty$ develops a linear instability, corresponding to $V''(C) |_{C=C_0}=0$. 
The transition (ii) $\to $ (i) between the FC and PC phases occurs instead at 
\begin{equation}\label{eq:ChaosChaosT}
     J_{\rm PF} := \frac23g\left(g+\sqrt{3\gamma+g^2}\right),
 \end{equation}
as it follows from imposing the continuity of $m_\infty, C_0$ at the transition line, see Sec.~\ref{sect: stationnary solution of the DMFT equations for alpha 0} in the SM \cite{SM_ours}. From the DMFT, one can also derive the asymptotic value of the force term ${\bf F}({\bf x})$ in the RHS of Eq.~\eqref{eq:dynamical_system}, which reads 
\begin{equation}\label{eq:force}
    F^2_0 := \langle F(t)^2 \rangle= 2g^2 C_0^2 -\lambda_{\infty}^2 C_0 + \left( J m_{\infty}\right)^2.
\end{equation}
Given the explicit expressions of $C_0$ and $m_\infty$ in the various phases, one can check explicitly that $F_0^2=0$ in the FFP phase, as expected: the dynamics asymptotic converges to an equilibrium configuration ${\bf x}^*$, which satisfies ${\bf F}({\bf x}^*)=0$. However, \eqref{eq:force} is strictly positive in the chaotic phases, see Figs.~\ref{fig supp: stationnary order parameters and total force in the confined case} and \ref{fig supp: stationnary order parameters and total force in the spherical case} of the SM \cite{SM_ours}. This suggests a separation between the attractor manifold in the chaotic phase, and the equilibria of the dynamical equations, i.e. the configurations where the force vanishes.

\begin{figure*}
    \centering
    \includegraphics[width=1\textwidth, trim={4 4 4 4},clip]{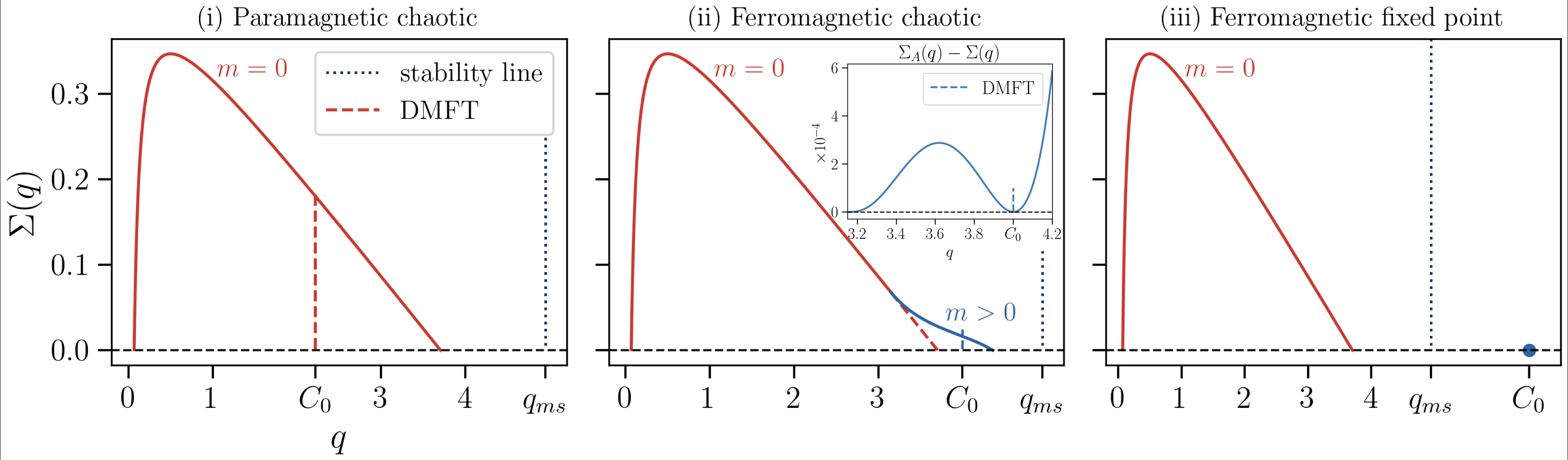}
    \caption{Complexity $\Sigma(q)=\Sigma(m_{\rm typ}(q),q)$, with $m_{\rm typ}=\arg\max_m \Sigma(m,q)$  for parameters corresponding to the points (i),(ii),(iii) in Fig.~\ref{fig:dmft_phase}. The red curves correspond to $m_{\rm typ}=0$ (paramagnetic equilibria), the blue curves to $m_{\rm typ} \neq 0$ (ferromagnetic equilibria). The blue dotted line is the marginal stability line $q_{\rm ms}$, separating unstable (left) to stable (right) equilibria. $q=C_0$ marks the value obtained from the DMFT solution. \textit{Inset}: Difference between quenched and annealed complexity in the FC phase, which vanishes for $q=C_0$.}
     \label{fig:kac_rice_pics_confined}
\end{figure*}

\textbf{Topological complexity ($\alpha=0$).} The topological complexity is the entropy of equilibria ${\bf x}^*$ satisfying $F_i({\bf x}^*)=0$ for $i=1, \cdots, N$. 
We denote with $\mathcal{N}(m,q)$ the number of equilibria with $m({\bf x}^*)=N^{-1}\sum_i x_i^*$ and $q({\bf x}^*)=N^{-1}\sum_i[x_i^*]^2$ at fixed values of $g,J,\alpha,\gamma$. The complexity is defined as
\begin{align}\label{eq:Comp}
    \Sigma(m,q):=\lim_{N\to \infty}\frac{\mathbb{E}[\log\mathcal{N}(m,q)]}{N}.
\end{align}
This quantity controls the asymptotics of the \emph{typical} value of $\mathcal{N}(m,q)$; in the 
language of disordered systems, it is a quenched (topological) complexity as opposed to the annealed one $ \Sigma_A(m,q):=\lim_{N\to \infty} N^{-1}\log\mathbb{E}\mathcal{N}(m,q)$, which controls the asymptotic of the average number. It holds $\Sigma_A(m,q)\geq \Sigma(m,q)$. The complexity \eqref{eq:Comp} can be computed via the replicated KR formalism \cite{ros2019complex, ros2025high}, which involves a combination of the 
replica trick $\log \mathcal{N}= \lim_{k \to 0} (\mathcal{N}^k-1)/k$ with the Kac-Rice formula for the moments $\mathbb{E}[\mathcal{N}^k]$. See the SM \cite{SM_ours} for details of this calculation, done within the Replica Symmetric (RS) assumption. \\
Fig.~\ref{fig:kac_rice_pics_confined} shows $\Sigma(q)=\Sigma(m_{\rm typ}(q),q)$, where $m_{\rm typ}(q)=\arg\max_m \Sigma(m,q)$ is the typical magnetization of equilibria in the shell parametrized by $q$. We find that Eqs.~\eqref{eq:dynamical_system} always admit an exponentially large number of equilibria with $m=0$ distributed in a range of $q$ values (red curves in Fig.~\ref{fig:kac_rice_pics_confined}). Their complexity $\Sigma(m=0, q)$ is independent of $J$, and coincides with the annealed one.
These exponentially numerous paramagnetic equilibria are all \emph{linearly unstable}. The linear stability of equilibria 
is given by the random matrix 
\begin{equation}\label{eq:StabMat}
   \frac{\partial F_i({\bf x})}{\partial x_j}
 \Big|_{{\bf x}={\bf x}^*}= \frac{\partial f_i({\bf x}^*)}{\partial x_j} - \lambda({\bf x}^*) \delta_{ij} -\frac{2}{N} x^*_j x^*_i,
 \end{equation}
 which arises when linearizing 
\eqref{eq:dynamical_system}.  
When evaluated at equilibria with $q({\bf x}^*)=q$, the fluctuating part of this matrix has the same statistics as random matrices extracted from a circular ensemble \cite{girko1986elliptic} with variance $\sigma^2= 4 g^2 q$: when $N \to \infty$, almost all\footnote{We focus here on the continuous part of the eigenvalue density of~\eqref{eq:StabMat}. Due to the presence of finite-rank perturbations, the matrix can exhibit isolated eigenvalues that do not belong to the support of the continuous part of the eigenvalue density. We neglect them in our discussion.} the eigenvalues of \eqref{eq:StabMat} are located within a circle in the complex plane \cite{sommers1988spectrum} 
centered at $q-\gamma$ and with radius $2 g \sqrt{q}$. The equilibrium ${\bf x}^*$ is linearly stable when the eigenvalues have negative real part, which corresponds to 
\begin{equation}\label{eq:stabi}
  q-\gamma - 2 g \sqrt{q}  \geq 0.
\end{equation}
Saturation of the inequality corresponds to marginal stability, occurring for $q_{\rm ms}=\gamma + 2 g  \left(g +\sqrt{\gamma + g^2}\right)$. One sees (Fig.~\ref{fig:kac_rice_pics_confined}) that the complexity of paramagnetic equilibria is 
positive for $q$s which violate \eqref{eq:stabi}, indicating that they are all unstable.\\
For small $J$, those at $m=0$ are the most abundant equilibria for all values of $q$, see Fig.~\ref{fig:kac_rice_pics_confined}.(i). As $J$ becomes large enough, $\Sigma(q)$ develops another branch corresponding to $m>0$ (blue curve in Fig.\ref{fig:kac_rice_pics_confined}.(ii)). Equating the expressions for the ferromagnetic and paramagnetic branches of the complexity, one finds that the two merge at $\overline{q}=(2 \gamma J + J^2)/[2 (J-g^2 )]$. The complexity vanishes at $\overline{q}$ when $J$ reaches $J_{\Sigma}=g \left(\sqrt{\ln 16} - \sqrt{\ln16-2}\right) \left(\sqrt{\gamma + g^2\ln2} + 
   g \sqrt{\ln2} \right)$, meaning that for $J \leq J_{\Sigma}$ the most numerous equilibria are paramagnetic for all $q$, while for $J>J_{\Sigma}$,
   for sufficiently large $q$
   the ferromagnetic equilibria are the most abundant, as
   $\Sigma(m,q)$ is maximal at values of $m>0$. 
   The complexity of ferromagnetic equilibria (optimized over $m$) is supported in the region where \eqref{eq:stabi} is violated: these equilibria are also linearly unstable.
  Further increasing $J$, the branch of ferromagnetic equilibria splits into two components with disconnected support, and eventually collapses to a single point of vanishing complexity: there are single values of $q=q_{\rm ffp}=\gamma + J$,  $m=m_{\rm ffp}=
{\sqrt{(\gamma+ J) [J \left(J - 2 g^2 \right)-2 \gamma g^2]}}/{J}$ at which the complexity is exactly zero, indicating the existence of a sub-exponential number of equilibria at those values of parameters. At $J=J_c$, it holds $q_{\rm ffp}=q_{\rm ms}$, and for larger $J$, $q_{\rm ffp}$ satisfies \eqref{eq:stabi}, see Fig.~\ref{fig:kac_rice_pics_confined}.(iii). The FFP phase is thus captured exactly by the complexity: the unique, stable ferromagnetic fixed point that attracts the dynamics corresponds to a local maximum of the complexity, where the latter is exactly zero.

\begin{figure*}
\centering
\includegraphics[width=0.53\textwidth]{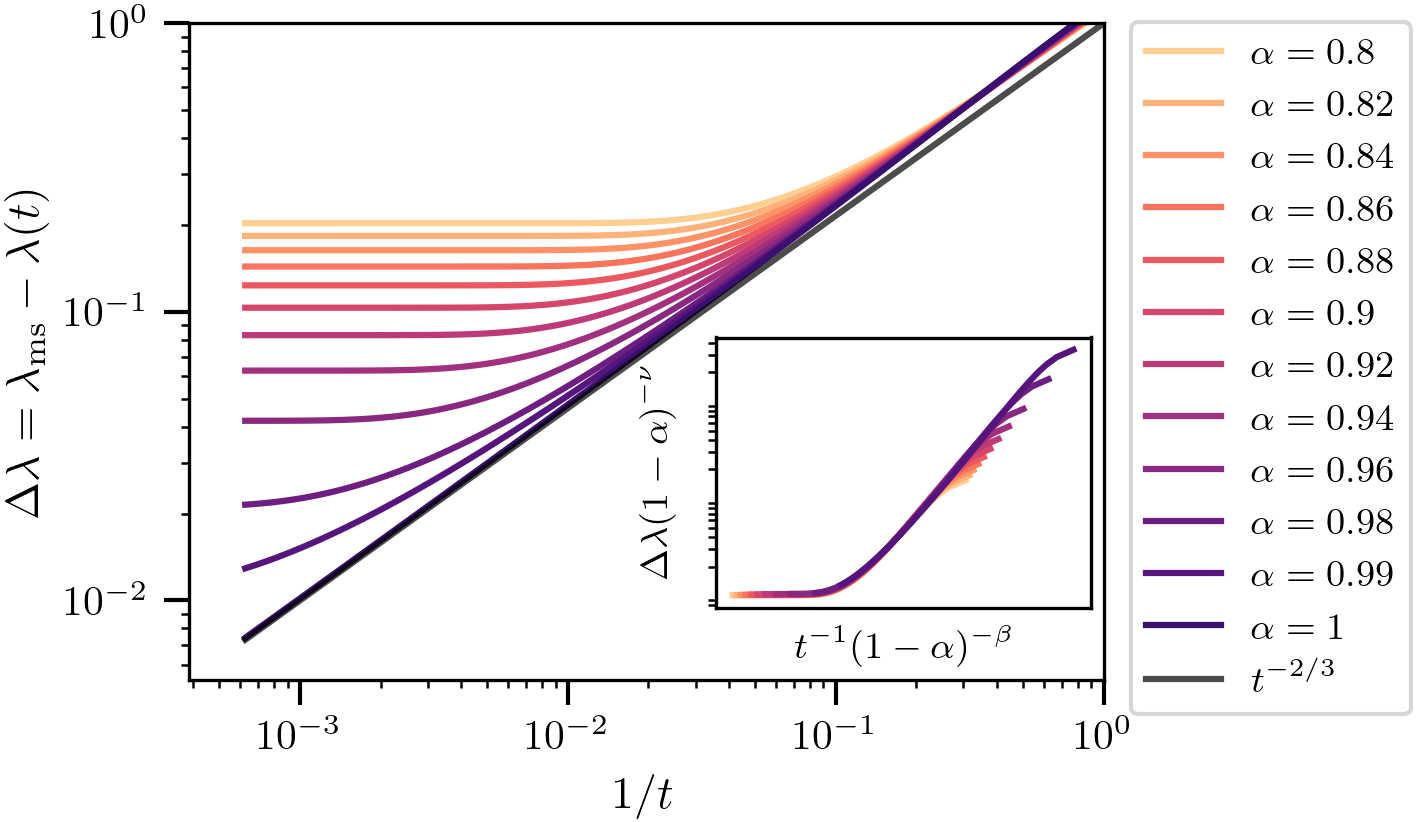}
\includegraphics[width=0.45\textwidth]{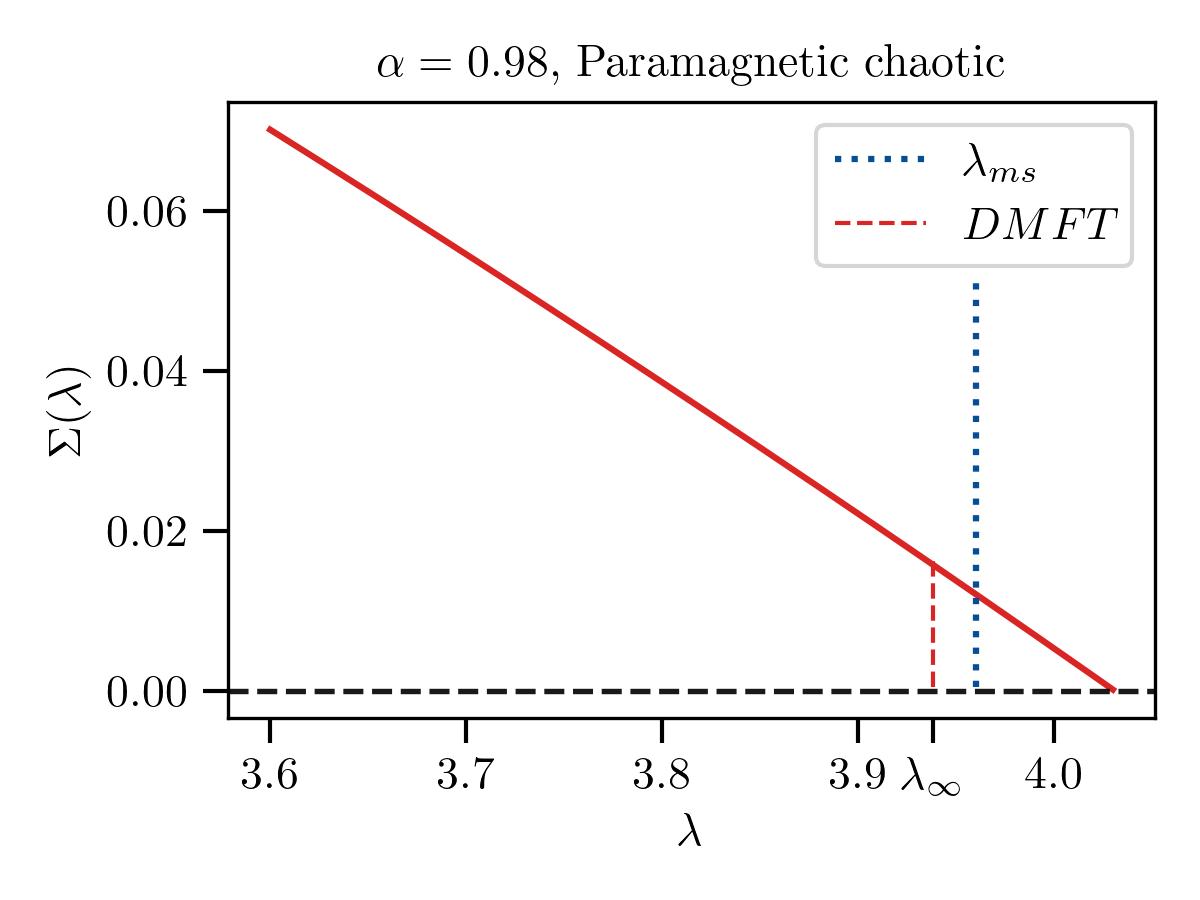}
\caption{ \textit{Left.} $\Delta \lambda(t)= \lambda_{\rm ms}-\lambda(t)$ obtained integrating numerically the DMFT equations for the spherical model with $g=1$, $J=0$ and various $\alpha$. The inset shows collapse of the curves when rescaling  $t \to t(1-\alpha)^\beta$ and  $\Delta \lambda \to \Delta \lambda(1-\alpha)^{-\nu}$, where $\beta=3/2$ and $\nu=1$. \textit{Right.} Complexity of equilibria $\Sigma(\lambda):= \Sigma(m_{\rm typ})$ of the spherical model with $\lambda$ fixed, $m$ optimized ($q=1$ and $m_{\rm typ}=0$ in this case) and $\alpha=0.98$. 
The marginal line (blue dotted line) separates unstable (left) from stable (right) equilibria. $\lambda_\infty$ denotes the asymptotic value of $\lambda$ selected by the dynamics.}
\label{fig:diff_alphas}
\end{figure*}

 \textbf{Comparing dynamics and complexity ($\alpha=0$). } In the TTI regime, the dynamics visits configurations with average magnetization $m \to m_\infty$ and squared norm $q \to C_0$. One can compare the explicit value of these quantities provided by the DMFT  with the information on equilibria at fixed $m,q$ encoded in the complexity. The comparison is consistent in the FFP phase (see Fig.~\ref{fig:kac_rice_pics_confined}.(iii)): the complexity has a branch of paramagnetic unstable equilibria, but also a local maximum (where $\Sigma(m_{\rm ffp}, {q_{\rm ffp}})=0$) at the value of parameters that correspond to the unique stable equilibrium attracting the dynamics: $m_{\rm ffp}=m_\infty$ and $q_{\rm ffp}=C_0=C_\infty$. The dynamical transition $J_c$ to the chaotic phase is signaled by the fact that \eqref{eq:stabi} is saturated by $q_{\rm ffp}$: the equilibrium ceases to be stable. The transition \eqref{eq:ChaosChaosT} is instead not captured by $\Sigma$: one would expect it to be signaled by the disappearance of the complexity branch corresponding to equilibria with $m>0$, occurring at $J_\Sigma$; however, this 
 is not the case, as shown in Fig.~\ref{fig:dmft_phase}(\textit{Left}). Instead, $J_\Sigma< J_{PF}$: an exponentially large number of ferromagnetic equilibria appears at values of $J$ for which dynamically the system is  in a paramagnetic chaotic phase. In the chaotic phases, the values $q=C_0, m=m_\infty$ obtained from the DMFT belong to the range where the complexity of equilibria is positive: there are exponentially many equilibria with the parameters selected by the dynamics. However,  
in the PC phase, $q=C_0$ does not coincide with the maximum nor the minimum of $\Sigma(m=0,q)$, see Fig.~\ref{fig:kac_rice_pics_confined}.(i). Analogously, in the FC phase, at $q=C_0$ the quenched complexity is optimized at a $m_{\rm typ}(q=C_0) \neq m_\infty$, meaning that the dynamics visits configurations whose magnetization is different with respect to the typical magnetization of equilibria in that shell.
This reinforces the idea that the dynamical trajectories do not dwell near equilibria, as suggested by the asymptotic value of the force~\eqref{eq:force} being non-zero in the chaotic phase. 
Some intriguing connections emerge nevertheless from the comparison: in the FC phase, exactly at $q=C_0$ the quenched complexity $ \Sigma(q)= \Sigma(m_{\rm typ}(q),q) $ coincides with the annealed one, see the inset of Fig.~\ref{fig:kac_rice_pics_confined}.(ii). Due to this coincidence, it is natural to expect that the RS assumption made to compute the quenched complexity is exact at $q=C_0$. We also find that in the PC phase, the dependence of the DMFT solution $C_0$ on $\gamma, g$ is such that $\partial_g\Sigma(0,q=C_0)=\partial_\gamma\Sigma(0,q=C_0)=0$, see Sec. \ref{app:dmft_vs_kac_confined} in the SM \cite{SM_ours}: the complexity of equilibria in the shell selected by the dynamics remains constant in $\gamma, g$. 

In Ref.~\cite{wainrib2013topological} it was conjectured that at the chaotic transitions the growth of the complexity of equilibria mirrors the increase of the maximal Lyapunov exponent characterizing chaos. In the SM we analyze two models within the Gaussian family, for which we can compute the maximal Lyapunov exponent explicitly and show that they furnish two distinct counterexamples: in one case, the critical exponent controlling the growth of the complexity close to the chaotic transition and the one controlling the growth of the Lyapunov are different; in the case, at the chaotic transition the complexity of critical points does not even vanish, but it is strictly positive.

\textbf{Dynamics and marginality ($\alpha>0$). } The spherical version of model \eqref{eq:dynamical_system} has been studied extensively in the conservative limit $\alpha=1$~\cite{crisanti1992spherical, cugliandolo1993analytical, cavagna1998stationary}: when $J=0$, it exhibits an exponentially large number of marginally stable equilibria, that dominate the large-time limit of low-temperature Langevin dynamics giving rise to aging~\cite{bouchaud1998out}. By contrast, for $\alpha=0$ and $J<J_c$, all equilibria are unstable and the dynamics is chaotic. 
What happens for intermediate  \( \alpha \in (0,1) \)? Do (marginally) stable equilibria appear at a critical asymmetry \( \alpha_c \), and is the dynamics attracted by them?
 The replicated KR calculation for generic $\alpha$ given in the SM \cite{SM_ours} shows that \emph{stable} paramagnetic equilibria have a positive complexity 
 for $\alpha >\alpha_c=(1 - \ln 2)/(2 \ln 2-1)\approx 0.8$, for both the confined and spherical model \cite{fyodorov2016nonlinear, Kivimae2024} see Fig.~\ref{fig:diff_alphas} (\textit{Right}). 
In the spherical model, marginally stable equilibria are characterized by the specific value  $\lambda_{\rm ms}=2g(1+\alpha)$.   Fig.~\ref{fig:diff_alphas} shows, for the region $\alpha \in (\alpha_c, 1)$ in which
marginal equilibria are present, the behavior of $\lambda(t)$ obtained from 
numerical integration of the  DMFT equations (paramagnetic $J=0$). The dynamics converges to configurations with $\lambda_\infty \neq \lambda_{\rm ms}$: marginal stability does not appear to dominate the dynamics away from the reciprocal limit. According to the common belief that aging is linked to marginality, this aligns with the claim that even a minimal degree of non-reciprocity disrupts aging in these types of models \cite{crisanti1987dynamics, CugliandoloNonrelax97, berthier2000two}. 
The Inset in Fig.~\ref{fig:diff_alphas}.({\it Left}) shows that $\lambda(t)$ for $\alpha\to 1$ collapse on a master curve, indicating that the following limit exists:
\begin{equation}
\tilde \lambda(t):=\lim_{\alpha\to 1} (1-\alpha)^{\nu}(\lambda_{\textrm{ms}}-\lambda(t(1-\alpha)^{\beta}))
\end{equation}
with $\nu\simeq 1$ and $\beta\simeq 3/2$. This implies that there exists a timescale $\hat \tau$ diverging for $\alpha\to 1$ as $\hat \tau\sim (1-\alpha)^{-\beta}$, in agreement with the analysis for the relaxation time in~\cite{berthier2000two}.

{\bf Conclusions.}
For a high-dimensional system with chaotic dynamics, we have shown explicitly that the properties of the attractor manifold can not be understood solely based on the distribution of typical equilibrium configurations of the dynamical equations. We argued that this mismatch does not rely on the RS approximation used to compute the distribution of equilibria, and it suggests that chaotic dynamics cannot be interpreted straightforwardly as trajectories lingering near unstable equilibria. 
This work opens several perspectives:
first, whether a static characterization of the attractor manifold could be achieved by studying measures over configurations \cite{franz1995recipes} that do not select equilibria but states with stationary magnetization. The characterization of the scaling function governing the dynamics of $\lambda(t)$ close to reciprocity is also an interesting open problem. Finally, it is worth investigating connections between the phenomenology of the ferromagnetic chaotic phases and that of ecological models~\cite{mallmin2024chaotic, mallmin2025fluctuating}.

\paragraph*{Acknowledgments --} All the authors thank Alessia Annibale for very insightful discussions on these issues in the context of a different model.
VR and PU acknowledge funding by the French government under the France 2030 program (PhOM - Graduate School of Physics) with reference ANR-11-IDEX-0003.

\bibliography{refs.bib}
\widetext
\clearpage

\setcounter{equation}{0}
\setcounter{figure}{0}
\setcounter{section}{0}
\setcounter{table}{0}
\renewcommand{\theequation}{S-\arabic{equation}}
\renewcommand{\thefigure}{S-\arabic{figure}}
\renewcommand{\thetable}{S-\arabic{table}}

\begin{center}
  {\LARGE Supplemental Material for “Non-reciprocal interactions and high-dimensional chaos: comparing dynamics and statistics of equilibria in
a solvable class of models"\par}
\end{center}

\centerline{Samantha Fournier$^{*,1}$, Alessandro Pacco$^{*,2}$, Valentina Ros$^{2}$, Pierfrancesco Urbani$^{1}$}
\vspace{3mm}

\centerline{$^1$\textit{Université Paris-Saclay, CNRS, CEA, Institut de physique théorique, 91191, Gif-sur-Yvette, France}}
\vspace{3mm}

\centerline{$^2$\textit{Université Paris-Saclay, CNRS, LPTMS, 91405, Orsay, France}}
\vspace{5mm}

\def\thefootnote{$*$}     \footnotetext{These authors contributed equally.}\def\thefootnote{\arabic{footnote}}

\tableofcontents

This Supplemental Material is structured as follows: in Sec.~\ref{app:general_model} we introduce a broader family of models, that includes the ones discussed in the main text as special cases. In Sec.~\ref{sect app: DMFT} we derive the Dynamical Mean Field Theory equations for this general family of models, and we present the solution of these equation in the long-time limit for the models discussed in the main text, in the purely non-conservative case ($\alpha=0$). This section presents a detailed discussion of the dynamical phase diagrams and of the properties of the attractor manifold of the chaotic phases, for both the Confined Model (CM) and the Spherical Model (SpM). In Sec.~\ref{app:kac_rice_computation} we provide the expressions of the annealed and quenched topological complexity for the broad family of models. Specific results on the models discussed in the main text are analyzed in Sec.~\ref{app:COMP}, where the comparison with the asymptotic solutions of the DMFT is also discussed. Finally, 
Sec.~\ref{eq:Conto} presents the detailed  derivation of the quenched complexity. Finally, in Sec.~\ref{sec:touboul} we discuss the relation between the complexity and the Lyapunov exponent at the chaotic transition(s). 

\section{A broader family of models}
\label{app:general_model}
\noindent In this section of the Supplemental Material we introduce a broader family of models that generalize those discussed in the main text. This family of models is almost analogous to that considered in \cite{Fyodorov_2016, ben2021counting, lacroix2022counting}. The dynamical equations take the general form:
\begin{align}\label{eqapp:mod}
    \partial_tx_i(t)=-\lambda({\bf x}) \, x_i(t)+f_i({\bf x})
\end{align}
where now $f_i({\bf x})$ is a Gaussian vector field of mean and covariance given by 
\begin{equation}
\label{app:eq:def_Covariance}
\mathbb{E}\left[f_i({\bf x})\right] = \frac{J }{N}\sum_{j=1}^N{x_j}  = J\, m({\bf x}), \quad \quad\text{Cov}\left[f_i({\bf x}), f_j({\bf y}) \right] =\delta_{ij} \Phi_1\tonde{\frac{{\bf x} \cdot {\bf y}}{N}}+ \frac{x_j \, y_i}{N} \Phi_2\tonde{\frac{{\bf x} \cdot {\bf y}}{N}}, 
\end{equation}
with $J \geq 0$ and $\Phi_1$ and $\Phi_2$ suitably chosen functions. In the following, we derive several results (in particular on the topological complexity) keeping these functions generic; we  implicitly assume that the functions $\Phi_1$ and $\Phi_2$ satisfy all the conditions required for these expressions to be well-defined. This is of course true for the specific choices discussed in the main text.  In particular, we always assume that $\Phi_1,{\Phi}'_1$ are positive. We consider two variations of this model:

\begin{itemize}
    \item {\bf Confined Model (CM). } In this case, ${\bf x} \in \mathbb{R}^N$ and $\lambda({\bf x})$ in \eqref{eqapp:mod} is a confining term of the form 
    \begin{equation}
\lambda({\bf x})= \frac{||{\bf x}||^2}{N}- \gamma, \quad \quad \gamma \geq 0.
    \end{equation}
    \item {\bf Spherical Model (SpM). } In this case, ${\bf x} \in \mathcal{S}_N(\sqrt N)= \left\{ {\bf x}: ||{\bf x}||^2=N \right\}$. To enforce the spherical constraint, $\lambda({\bf x})$ in \eqref{eqapp:mod} is chosen as a Lagrange multiplier, satisfying at each time:
       \begin{equation}
\lambda({\bf x})= \frac{{\bf f} ({\bf x}) \cdot {\bf x}}{N}.
    \end{equation}
\end{itemize}
The dynamics described by this model is non-conservative for generic $\Phi_1(u), \Phi_2(u)$: the conservative limit corresponds to choosing $\Phi_2(u) =\Phi_1'(u)$. To control deviations from this limit, following \cite{Fyodorov_2016} we introduce the ratio $\alpha(u):={\Phi_2(u)}/{\Phi'_1(u)}$.
When the dynamics is conservative, meaning that the (random) force field is the derivative of a (random) energy field $H({\bf x})$ with isotropic covariances, then $\alpha(u)=1$. Indeed, assume that  
\begin{align}
    f_k({\bf x})=-\frac{\partial H({\bf x})}{\partial x_k}, \quad \quad  \text{Cov}\left[H({\bf x}),H({\bf y})\right]=Nh\tonde{\frac{{\bf x}\cdot{\bf y}}{N}}. 
\end{align}
 By differentiating with respect to $x_k$ and $y_l$, one gets
\begin{align}
    \text{Cov}\left[f_k({\bf x}),f_l({\bf y})\right]=\delta_{kl}h'\left(\frac{{\bf x}\cdot{\bf y}}{N}\right) + h''\left(\frac{{\bf x}\cdot{\bf y}}{N}\right)\frac{y_kx_l}{N}
\end{align}
from which we can immediately identify $\Phi_1(u)=h'(u)$ and $\Phi_2(u)=h''(u)=\Phi'_1(u)$ which clearly implies $\alpha(u)=1$. When $\alpha(u)$ is not constant and equal to unity, the dynamics is, in general, non-gradient. We focus in particular on the sub-family of models defined by: 
\begin{equation}\label{eq:App:alp}
   \Phi_2(u)=\alpha \Phi'_1(u) \quad \quad \alpha \in [0,1]. 
\end{equation}
 In this case, the force ${\bf f}({\bf x})$ can be decomposed in a conservative contribution (the gradient of an energy landscape) and a non-conservative contribution,
 \begin{align}\label{eqapp:decoF}
f_i({\bf x})=\sqrt{1-\alpha}f^d_i({\bf x})-\sqrt{\alpha}\frac{\partial H({\bf x})}{\partial x_i}+\frac{J}{N}\sum_j x_j,
\end{align}
where both $f_i^d({\bf x})$ and $H({\bf x})$ are independent Gaussian fields with zero average and covariances satisfying:
\begin{align}\label{eq:CovRel}
 \text{Cov}[f_i^d({\bf x})f_j^d({\bf y})]=\Phi_1\left(\frac{{\bf x}\cdot{\bf y}}{N}\right)\delta_{ij}, \quad \quad 
\text{Cov}[H({\bf x}),H({\bf y})]&= Nh\left(\frac{{\bf x}\cdot{\bf y}}{N}\right), \quad \quad  \text{with} \quad \quad h'(u)=\Phi_1(u).
\end{align}
 It is simple to check that in this case both \eqref{app:eq:def_Covariance} and \eqref{eq:App:alp}  hold true. The parameter $\alpha$ measures the strength of the conservative part of the dynamics: for $\alpha=0$ we have pure non-conservative (non-gradient) dynamics, while for $\alpha=1$ we have conservative (gradient) dynamics. The results presented in the main text correspond to the choice $\Phi_1(u)= 2 g^2 u^2$ and $h(u)= 2 g^2 u^3/3$. Within this choice, the random terms appearing in \eqref{eqapp:decoF} can be parametrized as $f^d_i({\bf x}) = g \,  N^{-1} \sum_{jk}^N J_{i}^{jk} x_j x_k$ and 
$H({\bf x}) = - 2 g \, N^{-1} \sum_{i<j<k}^N S_{ijk} x_i x_j x_k$, where $J_{i}^{jk}$ are components of an asymmetric tensor in the upper index, while $S_{ijk}$ are components of a symmetric tensor. Up to symmetry, the entries of the tensors are independent and Gaussian, with zero average and unit variance.

\section{Dynamical Mean-Field Theory}
\label{sect app: DMFT}
In this section of the Supplemental Material, we derive the Dynamical Mean Field Theory (DMFT) equations and discuss their asymptotic solution. 

\subsection{General equations}
We consider the setup presented in Sec.~\ref{app:general_model} with $\Phi_2(u)=\alpha\Phi'_1(u)$. We recall that in this case $h'(u)=\Phi_1(u)$ and so $h''(u)=\Phi'_1(u)$. 
The DMFT equations can be derived following standard treatments~\cite{mezard1987spin, cugliandolo2002dynamics}. By means of a Path Integral approach, one finds the effective dynamics of a single (typical) neuronal degree of freedom $x(t)$, obeying the following stochastic integro-differential equation
\begin{equation}
\label{eq supp: effective single-site equation}
  \partial_t x(t) = -\lambda(t) x(t) + \alpha \int_0^{t}ds\, \Phi_1'(C(t,s)) R(t,s) x(s)  + \eta(t) + J m(t).
\end{equation}
We assume that the initial condition $ x(0)$ for this equation is a Gaussian variable with zero average and unit variance. 
This assumption is sufficient in the spherical model to describe an initial condition for the vector $\bm{x}(0)$ extracted uniformly on the hypersphere. In the context of the confined model, it is just a choice. Different choices could have been considered but they do not change the stationary state reached by the dynamics.
In Eq.~\eqref{eq supp: effective single-site equation}, the noise $\eta(t)$ is Gaussian with zero mean and variance self-consistently defined as $\langle \eta(t)\eta(t') \rangle = \Phi_1\left(C(t,t')\right) $. Eq.~\eqref{eq supp: effective single-site equation}, can be used to obtain a set of closed equations for the high-dimensional limit of correlation functions.
We therefore define the following dynamical observables
\begin{align}
    C_N(t,t')&=\frac{1}{N}\sum_{i=1}^N x_i(t)x_i(t'),\\
    R_N(t,t')&=\frac{1}{N}\sum_{i=1}^N \frac{\delta x_i(t)}{\delta h_i(t')}\label{empirical_resp},\\
    m_N(t)&=\frac{1}{N}\sum_{i=1}^Nx_i(t),    
\end{align}
where in Eq.~\eqref{empirical_resp}, $h_i(t)$ is a linear perturbation added to the corresponding equation of motion Eq.~\eqref{eq supp: effective single-site equation} to compute the response function.
In the high-dimensional limit, $C_N$, $R_N$ and $m_N$ concentrate around their limiting values
\begin{equation}
    C(t,t') := \lim_{N\to \infty}C_N(t,t')\:,\ \ \ R(t,t') := \lim_{N\to \infty}R_N(t,t')\:,\ \ \ m(t) := \lim_{N\to \infty}m_N(t)\:.
\end{equation}
Moreover we have that
\begin{align}
    C(t,t') =\langle x(t)x(t')\rangle\:, \ \ \ 
    R(t,t') =\left\langle \frac{\delta x(t)}{\delta \eta(t')}\right\rangle\:,\ \ \ m(t)=\langle x(t)\rangle
\end{align}
where the brackets denote the average over $\eta$.
Using Eq.~\eqref{eq supp: effective single-site equation}, one can obtain equations for these dynamical observables which are given by
\begin{equation}\label{eq supp: equation for m}
\begin{split}
  &\partial_t m(t) = -\lambda(t) m(t) + \alpha\int_0^{t}ds\,  \Phi_1'(C(t,s)) R(t,s) m(s) + J m(t),\\
  &\partial_t C(t,t') = -\lambda(t) C(t,t') + \alpha\int_0^{t}ds\,  \Phi_1'(C(t,s)) R(t,s) C(s,t')  + \int_0^{t'}ds\,  \Phi_1\left(C(t,s)\right) R(t',s) + J m(t) m(t'),\\
  &\partial_t R(t,t') = -\lambda(t) R(t,t') +  \alpha\int_{t'}^{t}ds\,  \Phi_1'(C(t,s)) R(t,s) R(s,t') + \delta(t-t').
  \end{split}
\end{equation}
These equations depend on $\lambda(t)$. The form of this function depends on whether we consider the confined or spherical model and it is given by 
\begin{equation}
  \begin{cases}
    &\text{CM}: \quad \quad \lambda(t) = C(t,t)-\gamma ,\\
     &\text{SpM}: \quad \quad\lambda(t) = \alpha\int_0^{t}ds\,  \Phi_1'(C(t,s)) R(t,s) C(s,t)  + \int_0^{t}ds\,  \Phi_1\left(C(t,s)\right) R(t,s) + J m^2(t).
  \end{cases}
  \label{contraints_DMFT}
\end{equation}
In particular, the equation for the spherical case is obtained by evaluating the second of Eqs.~\eqref{eq supp: equation for m} at $t'=t$ and by setting $C(t,t)=1$.\\
In the main text, we study the particular case where $\Phi_1(u)=2g^2u^2$ which is the minimal non-linear model. We stick to this case in the following but we stress that what we derive can be easily extended to more generic covariance structure for the forcing term $\bm{f}^d$ of the dynamical system Eq.~\eqref{eqapp:mod}.

Eqs.~\eqref{eq supp: equation for m}-\eqref{contraints_DMFT} have some remarkable properties:
\begin{itemize}
    \item They have a causal structure which is inherited from the causal structure of the original system of ODEs.
    \item They are closed.
\end{itemize}
These two properties allow to integrate such equations numerically in an efficient and simple way by discretizing the time-derivatives and integrals via an Euler discretization scheme.
This simplicity in the structure of the DMFT equations is not shared by generic models: in particular, for the model considered in \cite{ChaosSompo88}, this does not happen and the numerical integration of the corresponding DMFT equations has to be done by integrating a self-consistent stochastic process. We refer the readers to Ref.~\cite{kamali2023dynamical} for a discussion on the computational complexity classes of DMFT equations.

\subsection{Stationary solution of the DMFT equations for $\alpha=0$}
\label{sect: stationnary solution of the DMFT equations for alpha 0}
\begin{figure*}
  \centering
  \includegraphics[width=0.85\textwidth]{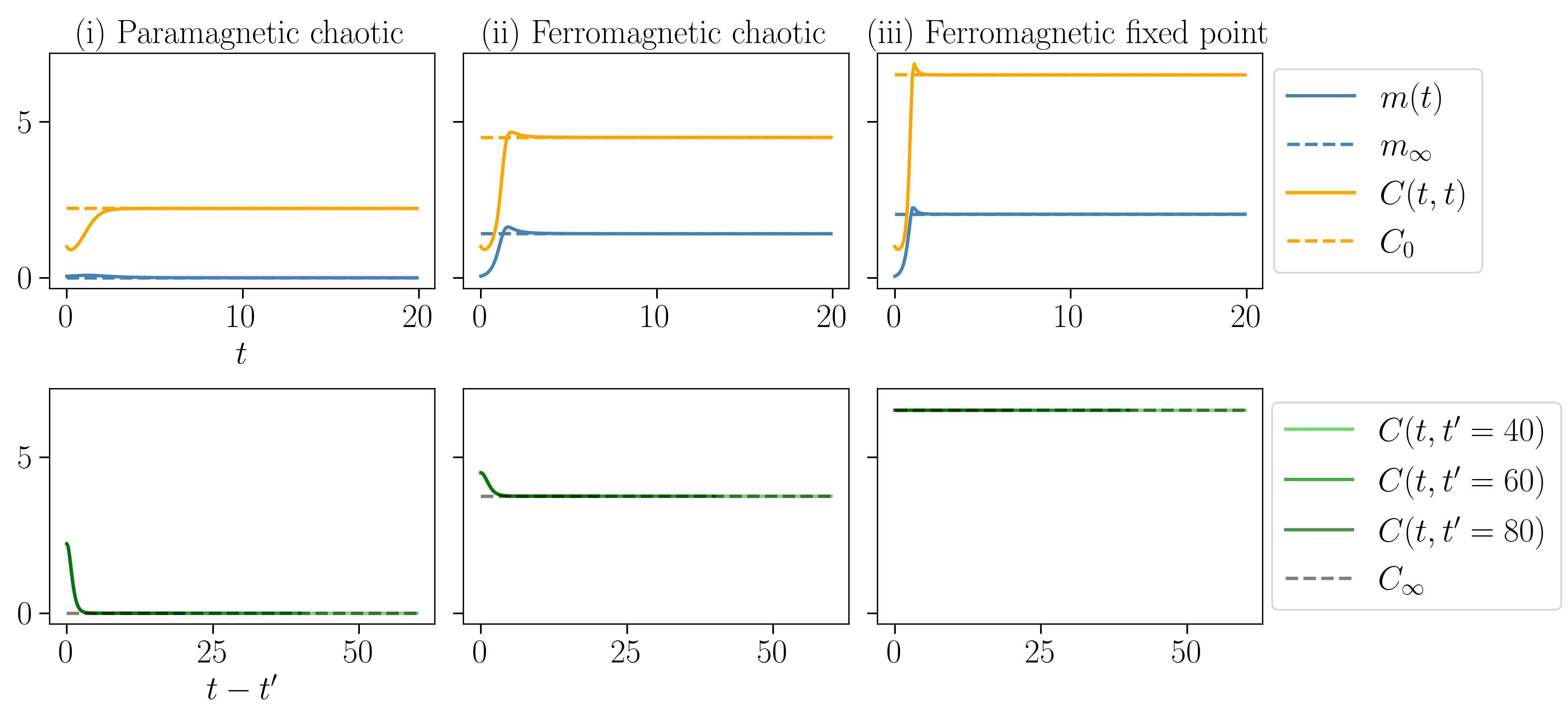}
  \caption{Numerical integration of the DMFT Eqs.~\eqref{eq: DMFT when alpha is 0}-\eqref{eq: DMFT of lambda} in the confined model. We show the time evolution of the dynamical order parameters when $\gamma=0.5$, $\alpha=0$ and (i) $g=1$, $J=1$ (ii) $g=1$, $J=4$ (iii) $g=1$, $J=6$. The DMFT equations  have been integrated numerically with and Euler discretization and a timestep $dt=0.1$.}
  \label{fig supp: dmft of confined model}
\end{figure*}

\begin{figure*}
  \centering
  \includegraphics[width=0.85\textwidth]{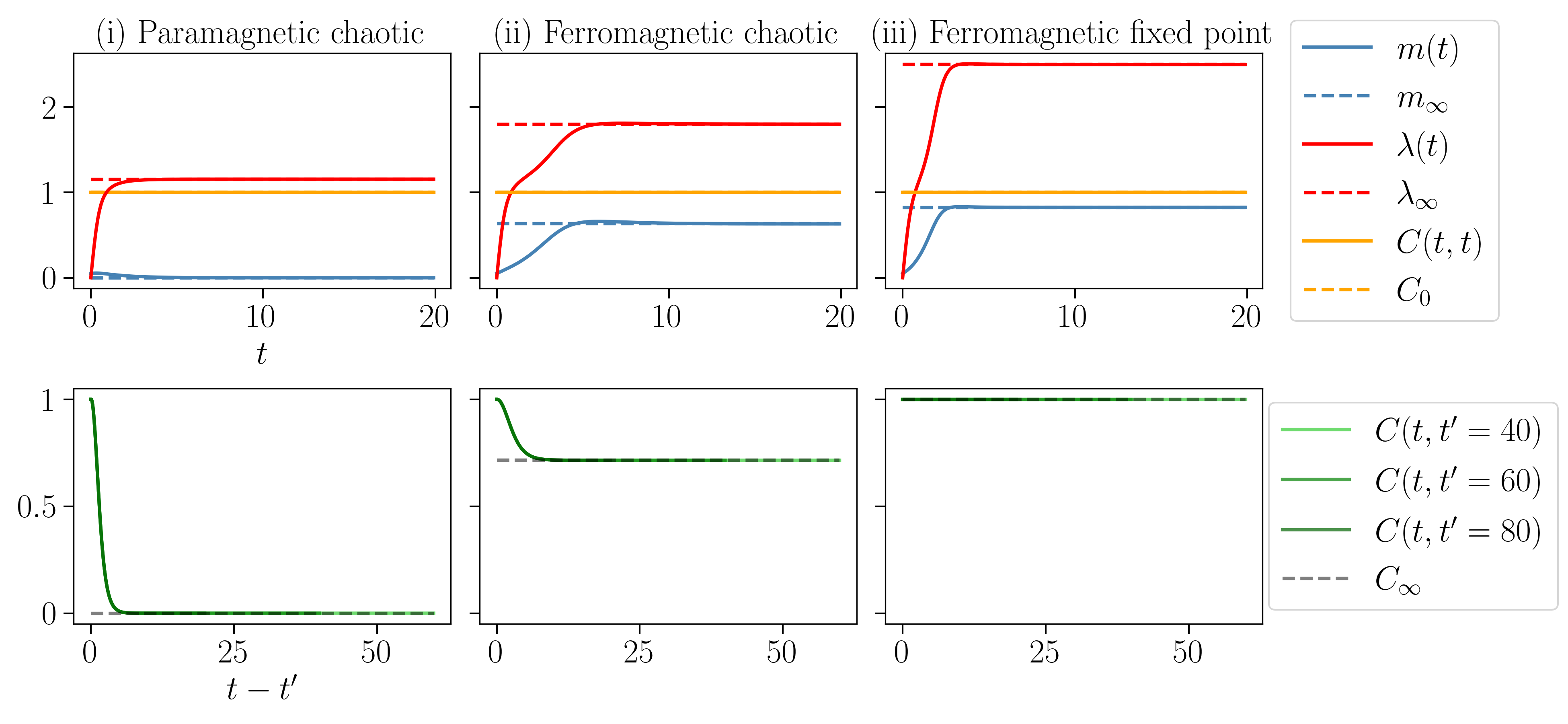}
  \caption{Numerical integration of the DMFT Eqs.~\eqref{eq: DMFT when alpha is 0}-\eqref{eq: DMFT of lambda} in the spherical case. We show the dynamical order parameters with $\alpha=0$ and (i) $g=1$, $J=0.5$ (ii) $g=1$, $J=1.8$ (iii) $g=1$, $J=2.5$. The DMFT equations have been integrated numerically with an Euler discretization and a timestep $dt=0.1$.}
  \label{fig supp: dmft of spherical model}
\end{figure*}

The DMFT equations drastically simplify when $\alpha=0$. In this case the memory term in Eq.~\eqref{eq supp: effective single-site equation} disappears and we are left with
\begin{equation}
  \label{eq: effective dynamical system for alpha 0}
  \partial_t x(t) = -\lambda(t) x(t) + \eta(t) + Jm(t) := F(t)\:.
\end{equation}
This implies that the equations for the dynamical observables introduced above become
\begin{align}
\label{eq: DMFT when alpha is 0}
\begin{split}
&\partial_t m(t) = -\lambda(t) m(t) + J\, m(t),\\
&\partial_t C(t,t') = -\lambda(t) C(t,t') + 2g^2 \int_0^{t'}ds\, C^2(t,s) R(t',s) + J\, m(t) m(t'),\\
&\partial_t R(t,t') = -\lambda(t) R(t,t') + \delta(t-t'),
\end{split}
\end{align}
with 
\begin{equation}
\label{eq: DMFT of lambda}
\begin{cases}
  &\text{CM}: \quad \quad\lambda(t) = C(t,t) - \gamma \\
  &\text{SpM}: \quad \quad\lambda(t) = 2g^2\int_0^{t}ds\, C^2(t,s) R(t,s) + J m^2(t).
\end{cases}
\end{equation}
The disappearance of the memory term in Eq.~\eqref{eq: effective dynamical system for alpha 0} has the important effect that the dynamics for the effective process of $x(t)$ does not slow down due to conservative forces. Instead, it is entirely driven by the non-conservative terms in the original dynamical system. The result is that the dynamics quickly evolves to a stationary regime where the dynamical order parameters become Time-Translational Invariant (TTI), described by
\begin{align*}
  \lambda_\infty:=\lim_{t\to\infty} \lambda(t), \quad \quad 
  m_\infty:=\lim_{t\to\infty} m(t), \quad \quad 
  C(|t-t'|):=\lim_{t,t'\to\infty} C(t,t'), \quad \quad 
  R(|t-t'|):=\lim_{t,t'\to\infty} R(t,t').
\end{align*}
We note that we are abusing with the notation above indicating with $C$ and $R$ the limiting TTI form of the correlation and response functions, since we believe that the resulting equations are not ambiguous.
The fact that the dynamics reaches a TTI stationary regime can be seen in Figs.~\ref{fig supp: dmft of confined model} and \ref{fig supp: dmft of spherical model} where the DMFT equations \eqref{eq: DMFT when alpha is 0} have been integrated numerically both for the confined model and for the spherical one.


\begin{figure}
  \centering
  \includegraphics[width=0.75\textwidth]{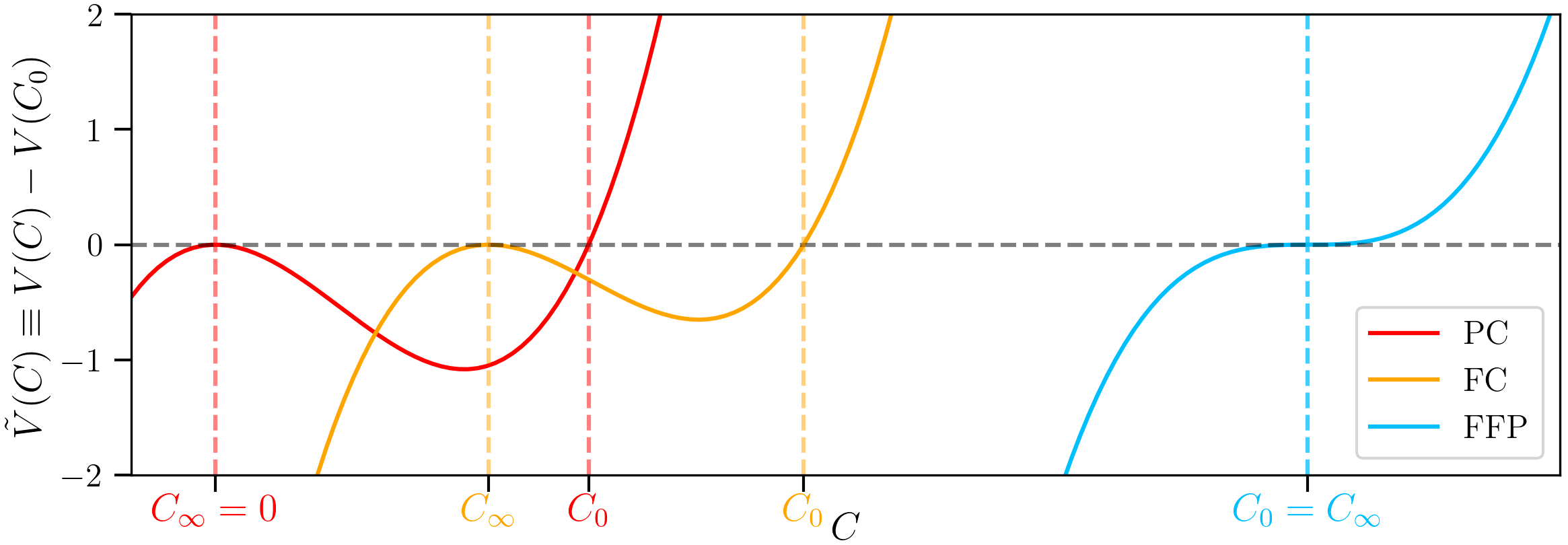}
  \caption{Form of the potential $\tilde{V}(C) \equiv V(C)-V(C_0) = \frac23 g^2 (C-C_0)(C-C_\infty)^2$ in the different phases. \textit{Red:} Paramagnetic Chaotic (PC) phase. \textit{Orange:} Ferromagnetic Chaotic (FC) phase with $J=3$. \textit{Blue:} Ferromagnetic Fixed Point (FFP) phase with $J=6$. In all three cases $\gamma=0.5$ and $g=1$.}
  \label{fig supp: potential}
\end{figure}

The equations governing the asymptotic TTI form of the dynamical observables can be obtained by taking $t,t'\to\infty$ with $\tau=t-t'$ fixed. In the steady state, the DMFT equations become
\begin{align}
  \label{eq: equation for m infinity}
  &0 = (-\lambda_{\infty} + J) m_{\infty},\\
  \label{eq: equation for dC of tau}
  &\partial_{\tau} C(\tau) = -\lambda_{\infty} C(\tau) + 2g^2 \int_0^{\infty}ds\, C^2(\tau+s) e^{-\lambda_{\infty}s} + J m_{\infty}^2.
\end{align}
In writing  Eq.~\eqref{eq: equation for dC of tau} we solved for $R(\tau)$, which obeys a simple ordinary differential equation describing an exponential relaxation, and we injected the corresponding solution in the equation for $C(\tau)$. It is convenient now to introduce two additional order parameters 
\begin{align*}
  C_0 := \lim_{\tau\to0} C(\tau) \quad \quad  \mathrm{and} \quad \quad  C_\infty := \lim_{\tau\to\infty} C(\tau).
\end{align*}
Because $C(\tau)$ is TTI, we know that $C(\tau)=C(-\tau)$ which implies $\partial_{\tau} C(\tau)|_{\tau=0}=0$. 
Apart from the full function $C(\tau)$, we are interested in obtaining the unknowns $\lambda_\infty$, $C_\infty$, $m_\infty$ and $C_0$. We note that the effective number of unknowns is three: indeed in the spherical model, $C_0=1$ while in the confined model, we have trivially $\lambda_\infty=C_0-\gamma$.
Therefore we need to obtain three equations that relate these four unknowns. The first of them is Eq.~\eqref{eq: equation for m infinity}. The second one can be obtained by considering the long time limit of Eq.~\eqref{eq: equation for dC of tau}. Indeed given that $\lim_{\tau\to \infty}\partial_\tau C(\tau)=0$, we get
\begin{equation}
  \label{eq: equation for C infinity}
  0 = -\lambda_{\infty} C_{\infty} + 2g^2 \frac{C^2_{\infty}}{\lambda_{\infty}} + J m_{\infty}^2\:.
\end{equation}
The last equation can be obtained following the same steps as in \cite{crisanti2018path}.
Multiplying Eq.~\eqref{eq: effective dynamical system for alpha 0} by itself and averaging over $\eta$ we get
\begin{align}
    (\partial_t+\lambda(t))(\partial_{t'}+\lambda(t'))C(t,s)=2g^2C^2(t,t')+J^2 m(t)m(t').
\end{align}
So taking the TTI limit $t,t'\to\infty$ with $\tau=t-t'$, we obtain
\begin{equation}
  (\lambda_{\infty}^2 - \partial^2_{\tau}) C(\tau) = 2g^2 C^2(\tau) + (J m_{\infty})^2\:.
\end{equation}
This equation can be conveniently re-written in the form of a particle moving under the influence of a potential $V(C)$, as
\begin{equation}\label{Newton}
  \partial^2_{\tau} C(\tau) = - \frac{\partial V}{\partial C},
\end{equation}
where we defined
\begin{equation}
  V(C) = \frac23 g^2 C^3 -\frac12 \lambda_{\infty}^2 C^2 + \left( J m_{\infty}\right)^2 C.
\end{equation}
All acceptable solutions $C(\tau)$ must be bounded $|C(\tau)| \leq C_0$ and must conserve a fictitious energy defined as
\begin{equation}
\mathcal E = \frac12 \left( \partial_{\tau} C(\tau) \right)^2 + V(C).
\end{equation}
In particular, since $\partial_{\tau} C(\tau)|_{\tau\to\infty}=0$ and $\partial_\tau C(\tau)|_{\tau=0}=0$, the conservation of $\mathcal E$ implies that
\begin{equation}
  V(C_0) = V(C_\infty)\:.
\end{equation}
This is the last equation that we need.
It can be rewritten as
\begin{equation}
  \label{eq: equation for C0 or lambda infinity}
  (C_\infty-C_0)^2\left[ \frac43g^2C_\infty - \frac12\lambda_\infty^2 + \frac23g^2C_0 \right] = 0,
\end{equation}
which shows that it has a possible solution $C_0=C_\infty$. This solution describes a stationary state where the dynamics reaches a fixed point. However this solution only appears if $V'(C_0)=0$ which is not always the case (see Fig.~\ref{fig supp: potential}, orange and red curves).

\begin{figure}
    \centering
    \includegraphics[width=0.35\textwidth]{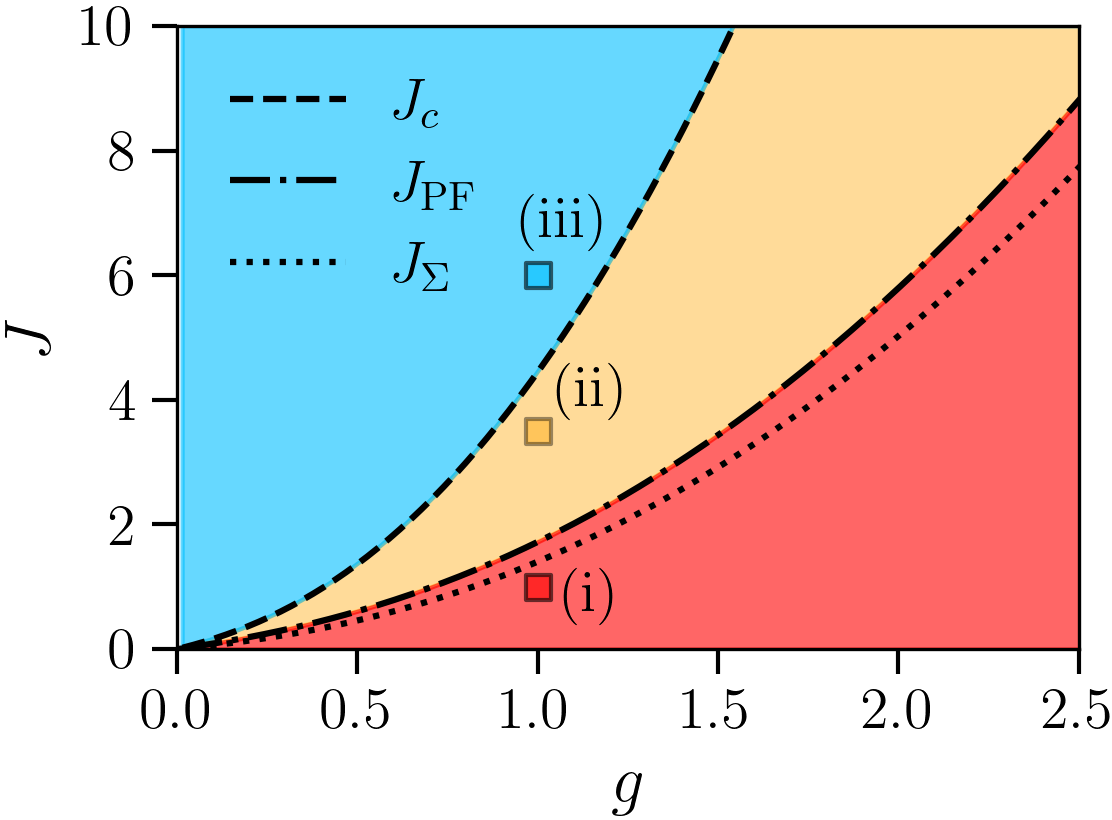}
    \includegraphics[width=0.35\textwidth]{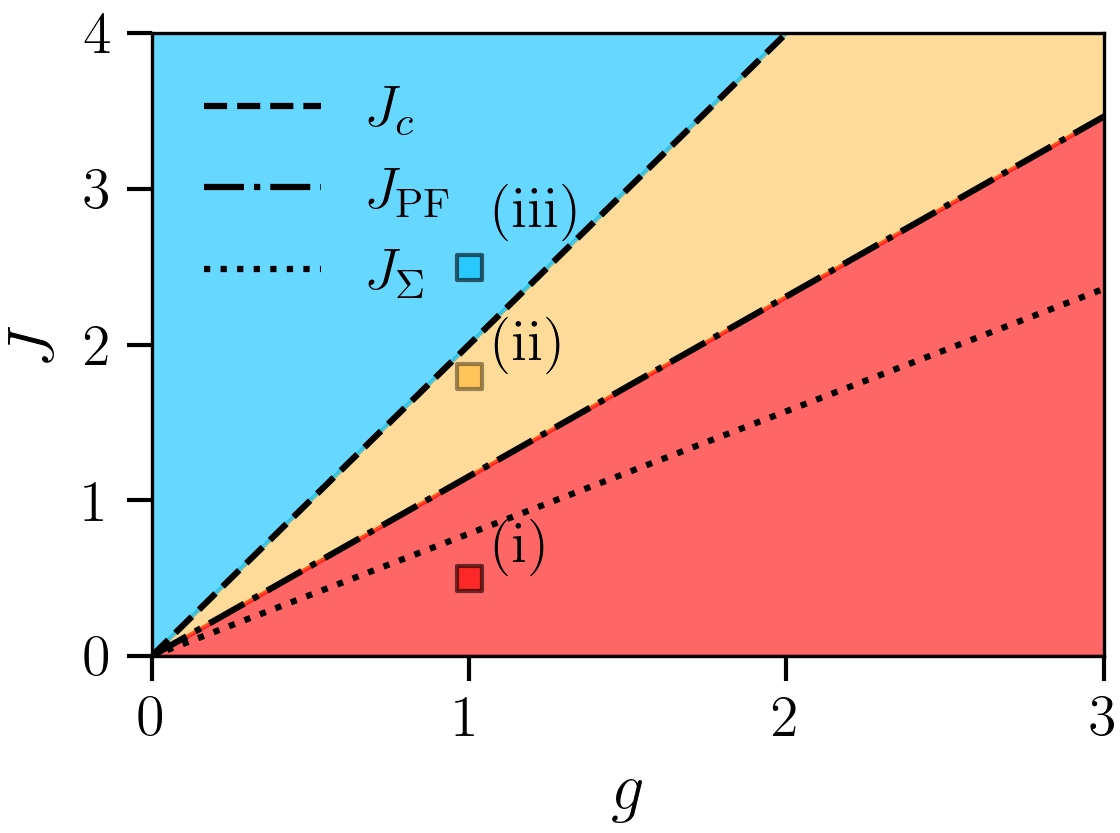}
    \caption{(\textit{Left.}) Dynamical phase diagram of the confined model with $\gamma=0.5$. The squares correspond to (i) $g=1$, $J=1$ (ii) $g=1$, $J=4$ (iii) $g=1$, $J=6$. (\textit{Right.}) Dynamical phase diagram of the spherical model. Squares correspond to (i) $g=1$, $J=0.5$ (ii) $g=1$, $J=1.8$ (iii) $g=1$, $J=2.5$. The red region corresponds to the PC phase, the orange region to the FC phase and the light-blue region to the FFP phase.}
    \label{fig: confined and spherical phase diagrams appendix}
\end{figure}

In summary the equations determining $\lambda_\infty$, $C_\infty$, $m_\infty$ and $C_0$ are
\begin{equation}
    \begin{split}\label{final_eq}
        &0 = (-\lambda_{\infty} + J) m_{\infty},\\
        &0 = -\lambda_{\infty} C_{\infty} + 2g^2 \frac{C^2_{\infty}}{\lambda_{\infty}} + J m_{\infty}^2,\\
        &0=(C_\infty-C_0)^2\left[ \frac43g^2C_\infty - \frac12\lambda_\infty^2 + \frac23g^2C_0 \right],\\
        &\begin{cases}
        &\text{CM:} \quad \lambda_{\infty}=C_0-\gamma \\
        &\text{SpM:} \quad  C_0=1.
         \end{cases}
    \end{split}
\end{equation}
Solving these equations gives access to the dynamical phase diagram of the models (see Fig.~\ref{fig: confined and spherical phase diagrams appendix}). Correspondingly, one can get the shape of the potential $V(C)$. Some examples of its form can be found in Fig.\ref{fig supp: potential}.\\
In the steady-state $t,t'\to\infty$ with $\tau=t-t'$, we can study the dynamical correlation function of the driving forces. 
We can define 
\begin{equation}
    \Gamma(t,t') := \lim_{N\to \infty}\frac 1N \sum_{i=1}^N f_i({ \bf x}(t))f_i({ \bf x}(t'))
\end{equation}
which within the DMFT formalism is given by
\begin{equation}
    \Gamma(t,t') = \langle F(t) F(t')\rangle\:.
\end{equation}
Using Eq.~\eqref{eq: effective dynamical system for alpha 0} multiplied by itself and taking the long time limit we get 
\begin{equation}
  \hat \Gamma(\tau):=\lim_{t\to \infty}\Gamma(t+\tau,t)  = -\partial^2_\tau C(\tau) = \frac{\partial V}{\partial C} = 2g^2 C^2(\tau) -\lambda_{\infty}^2 C(\tau) + \left( J m_{\infty}\right)^2.
\end{equation}
Now taking $\tau\to0$, we get the average force at equal times reported in Eq.~\eqref{eq:force} of the main text
\begin{equation}
\label{eq supp: expression for the force}
 F^2_0:=\hat \Gamma(0) = 2g^2 C_0^2 -\lambda_{\infty}^2 C_0 + \left( J m_{\infty}\right)^2.
\end{equation}
We also note that
\begin{equation}
    \lim_{\tau\to \infty}\Gamma(\tau)=0
\end{equation}
which implies that in the long time limit there is no correlation between the driving forces evaluated at asymptotically large time separation.

\begin{table}
    \centering
    \begin{tabular}{ |c|c|c|c|c|  }
        \hline
         $C_0 =0$ & $C_\infty=0$ & $m_\infty=0$ & $F_0=0$ & Paramagnetic fixed point phase (PFP)\\
        \hline
         $C_0>0$ & $C_\infty=C_0$ & $m_\infty>0$ & $F_0=0$ &  Ferromagnetic fixed point phase (FFP)\\
        \hline
         $C_0>0$ & $C_\infty=0$ & $m_\infty=0$ &  $F_0\neq 0$ & Paramagnetic chaotic phase (PC)\\
        \hline
         $C_0>0$ & $C_\infty>0$ & $m_\infty>0$ & $F_0\neq 0$ & Ferromagnetic chaotic phase (FC)\\
        \hline
    \end{tabular}
    \caption{Summary of the possible phases arising in the models that we consider, with the corresponding properties of the asymptotic solution of the DMFT equations.}
    \label{tab:summary_phases}
\end{table}

\subsection{Dynamical phase diagrams}
Eqs.~\eqref{final_eq} can be solved analytically to get the dynamical phase diagram of the models.
According to the solution, we can distinguish a set of dynamical phases which are summarized in the Table \ref{tab:summary_phases}.
The corresponding phase diagrams can be found in Fig.\ref{fig: confined and spherical phase diagrams appendix}.
We now describe in detail what are the phases, the order parameters, the transition lines between the different phases, as well as how to get them.

\subsubsection{Phase diagram of the confined model}
\begin{figure*}
  \centering
  \includegraphics[width=0.9\textwidth]{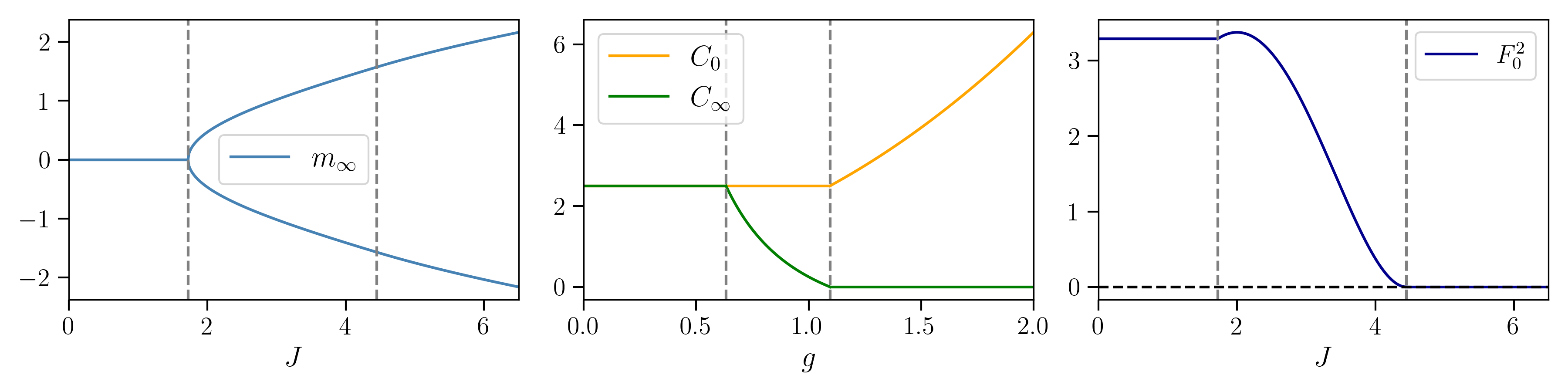}
  \caption{Stationary order parameters and total equal times force in the confined case with $\gamma=0.5$, $\alpha=0$ and (\emph{Left}) $g=1$, (\emph{Middle}) $J=2$, (\emph{Right}) $g=1$. The vertical gray dashed lines represent a change in the phase of the dynamics which are: (for left and right panels) PC, FC and FFP in order from left to right; (middle panel) FFP, FC and PC in order from left to right.}
  \label{fig supp: stationnary order parameters and total force in the confined case}
\end{figure*}

The solution to Eqs.~\eqref{final_eq} gives rise to four qualitatively different dynamical phases, each characterized by different values of the stationary order parameters. In the following, we detail the properties of the different phases.
\paragraph*{\textbf{Paramagnetic fixed point phase (PFP) --}} In this case the dynamics relaxes to the equilibrium $\bm{x}^*=0$. Therefore
  \begin{equation}
    m_{\infty}=0,\quad \quad C_0 = C_{\infty} = 0.
  \end{equation}
  However this solution is always only marginally stable: only if the dynamics is initialized exactly at $\bm{x}(0)=0$, the system does not evolve. A little perturbation destabilizes this fixed point. Therefore this phase is not seen unless the special initialization $\bm{x}(0)=0$ is chosen.
  We note that the equations in this regime admit also a solution $C_0=C_\infty=\gamma + g^2\left(1+\sqrt{1+\frac{2\gamma}{g^2}}\right)$, which is always unstable.
\paragraph*{\textbf{Paramagnetic chaotic phase (PC) --}} In this case the system is found in an endogenously fluctuating stationary state with zero magnetization. The corresponding order parameters are
  \begin{equation}\label{ParaChaPhaseConfPar}
    m_{\infty}=0,\quad \quad C_{\infty} = 0,\quad \quad C_0 = \gamma + \frac23g^2 \left(1+\sqrt{1+\frac{3\gamma}{g^2}}\right).
  \end{equation} 
\paragraph*{\textbf{Ferromagnetic fixed point phase (FFP) --}}
  In this case the dynamics lands on a fixed point that has non-vanishing magnetization. Given that the statistics of the chaotic drive of the dynamics that we are focusing on is invariant under $\mathbb{Z}_2$ (because we are considering $\langle \eta(t)^2\rangle\propto C^2(t,t)$) the magnetization can take two opposite values
  \begin{equation}\label{DMFT_fixFFCM}
    C_{\infty}= C_0 = J+\gamma,\quad \quad m_{\infty} = \pm \sqrt{J+\gamma - \frac{2g^2}{J^2}(J+\gamma)^2}.
  \end{equation}
\paragraph*{\textbf{Ferromagnetic chaotic phase (FC) --}} In this case the dynamics reaches a stationary state with fluctuations that are persistent in time, in a region of phase space having a fixed magnetization. The corresponding order parameters are
  \begin{equation}\label{FChaPhaseConfPar}
    C_0 = J_0+\gamma, \quad \quad C_\infty=\frac12\left( \frac{3 J^2}{4 g^2} - J -\gamma \right), \quad \quad m_\infty = \pm \frac{1}{2}\sqrt{\frac12\frac{3J^2}{8 g^2} + J+\gamma - \frac{2g^2}{J^2}(J+\gamma)^2}.
  \end{equation}
The values of the stationary order parameters as a function of the control parameters $g$, $J$ are presented in Fig.~\ref{fig supp: stationnary order parameters and total force in the confined case}. The force $F_0$ is given by Eq.~\eqref{eq supp: expression for the force} and is represented in Fig.~\ref{fig supp: stationnary order parameters and total force in the confined case}-right as a function of $J$. 
In Fig.~\ref{fig supp: dmft of confined model} instead we plot the numerical integration of the DMFT equations and we compare the asymptotic behavior of the solution with the order parameters computed by solving Eqs.~\eqref{final_eq}. We find an excellent agreement.
The phase diagram of the confined model is shown in Fig.~\ref{fig supp: dmft of confined model}-left.

We now turn to the description of the phase transition lines between the different phases. Given that the PFP phase is not found, we have just three lines to determine.
\paragraph*{\textbf{Phase transition from FFP to FC. --}} 
This line can be found by looking at the stability of the fixed point solution $C(\tau)=C_0=J_0+\gamma$
\begin{equation}\label{eq:sphTR}
  \left. \frac{\partial^2 V}{\partial C^2}\right\vert_{C=C_0} = 0  \implies J_c = 2g \left(g+\sqrt{\gamma+g^2}\right).
  \end{equation}
This condition corresponds to the situation in which the local maximum in $V(C)$ disappears and becomes a saddle (see the \textit{blue} curve in Fig.~\ref{fig supp: potential}).
\paragraph*{\textbf{Phase transition from PC to FC --}} 
This transition line can be found coming from the FC phase and imposing that $m_\infty\to 0$. This gives the critical value of $J$ as a function of $g$ given by
\begin{equation}
  J_{\text{PF}} = \frac23g\left(g+\sqrt{3\gamma+g^2}\right).
\end{equation}

\begin{figure*}
  \centering
  \includegraphics[width=0.8\textwidth]{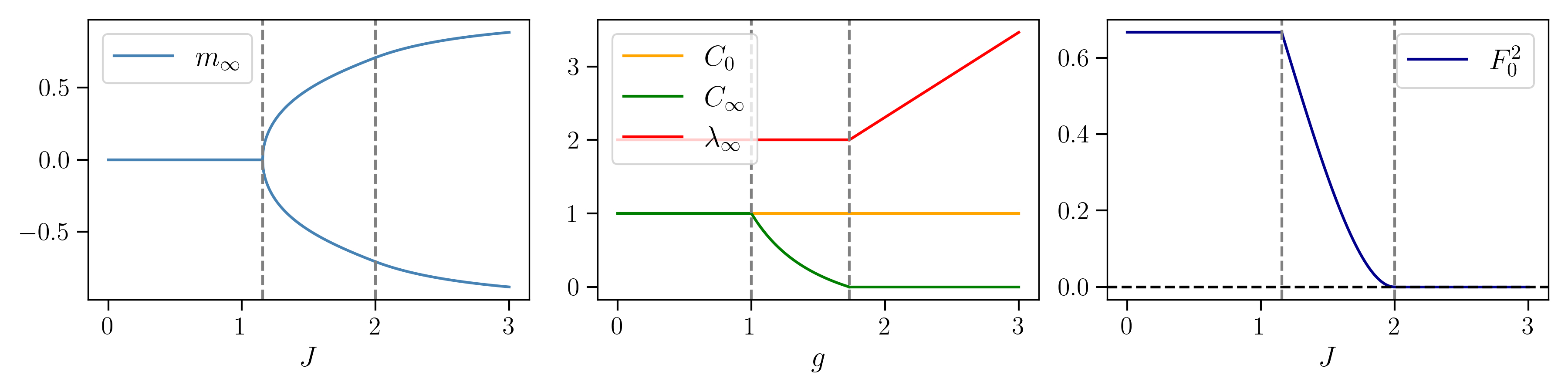}
  \caption{Stationary order parameters and total equal times force in the spherical case with (\emph{Left}) $g=1$, (\emph{Middle}) $J=2$, (\emph{Right}) $g=1$. The vertical gray dashed lines represent a change in the phase of the dynamics which are: for (left and right) in order from left to right PC, FC and FFP; and (middle) in order from left to right FFP, FC and PC.}
  \label{fig supp: stationnary order parameters and total force in the spherical case}
\end{figure*}

\subsubsection{Phase diagram of the spherical model}
In this case, $\lambda(t)$ is chosen to be a Lagrange multiplier that enforces spherical constraint, i.e. $C(t,t)=1$. Therefore $C_0=1$ always, and we need to determine $\lambda_{\infty}$. The solution is completely determined by the three equations Eqs.~\eqref{final_eq}.
Given that $C_0=1$, one cannot have a PFP phase.
The phase diagram contains three phases.
\paragraph*{\textbf{Paramagnetic chaotic phase (PC)} --}
In this case, the dynamics reaches a dynamical steady state at zero magnetization. The only order parameter that we need to determine is $C_\infty$. We have
\begin{equation}
\label{eq supp: stationary order parameters in PC phase for spherical model}
  m_{\infty}=0,\quad \quad C_0 = 1,\quad \quad C_{\infty} = 0,\quad \quad \lambda_\infty = \frac{2}{\sqrt{3}}g.
\end{equation} 

\paragraph*{\textbf{Ferromagnetic fixed point phase (FFP)}--} Here the dynamics reaches a magnetized fixed point. The corresponding order parameters are given by
\begin{equation}\label{DMFT_fixFFSM}
  \lambda_\infty = J,\quad \quad C_{\infty}= C_0 = 1,\quad \quad m_{\infty} = \pm \sqrt{1 - \frac{2g^2}{J^2}}.
\end{equation}

\paragraph*{\textbf{Ferromagnetic chaotic phase (FC) --}} In this case, the dynamics reaches a steady state with persistent fluctuations and non-vanishing magnetization. The corresponding order parameters are:
\begin{equation}\label{dmftsphfc}
  \lambda_\infty = J, \quad \quad C_0=1,\quad \quad C_\infty= \frac{3J^2}{8g^2} - \frac12,\quad \quad m_\infty = \pm \sqrt{\frac14 + \frac{3J^2}{32g^2} - \frac{g^2}{2J^2}}.
\end{equation}

The values of the stationary order parameters as a function of the control parameters $g$, $J$ are presented in Fig.~\ref{fig supp: stationnary order parameters and total force in the spherical case}. The force $F_0$ is given by Eq.~\eqref{eq supp: expression for the force} and is represented in Fig.~\ref{fig supp: stationnary order parameters and total force in the spherical case}-right as a function of $J$. The phase diagram is shown Fig.~\ref{fig supp: dmft of spherical model}-right.\\

We now detail the equations giving the phase transition lines in the phase diagram.

\paragraph*{\textbf{Phase transition from FFP to FC. --}} This line can be found by looking at the stability of the fixed point solution $C(\tau)=C_0=1$
\begin{equation}\label{eq:traSM}
\left. \frac{\partial^2 V}{\partial C^2}\right\vert_{C=1} = 4g^2 - J^2 = 0 \implies J_c = 2g.
\end{equation}

\paragraph*{\textbf{Phase transition from PC to FC. --}} This transition can be found again coming from the FC phase and taking the limit $m_\infty\to 0$ which gives
\begin{equation}
J_{\text{PF}} = \frac{2}{\sqrt{3}}g\:.
\end{equation}

\section{Quenched and annealed complexity: expressions}
\label{app:kac_rice_computation}
In this section we discuss the complexity of equilibria of the dynamical equations \eqref{eqapp:mod} for both the confined (CM) and spherical model (SpM).
\subsection{Definition of the complexity}\label{Sec:IntroKR}
 In order to treat the two models on the same footing, we introduce the random variable $\mathcal{N}(m, \lambda, q)$ counting the number of solutions of the equations
 \begin{equation}\label{eq:General2}
    f_i({\bf x})-\lambda \, x_i=0 \quad \quad  i=1, \cdots, N  \quad \quad {\bf x} \in \mathbb{R}^N
    \end{equation}
for arbitrary $\lambda \in \mathbb R$, such that  
\begin{equation}\label{eq:momments}
    m({\bf x})= \frac{1}{N} \sum_{i=1}^N x_i, \quad \quad q({\bf x})= \frac{1}{N} \sum_{i=1}^N x_i^2. 
\end{equation}
From this number, we can obtain the number of equilibria of the CM and SpM as follows:

\begin{itemize}
    \item For the confined model, equilibria are solutions of the equations
    \begin{equation}
    f_i({\bf x})-\lambda({\bf x}) \, x_i=0 \quad \quad  i=1, \cdots, N  \quad \quad {\bf x} \in \mathbb{R}^N
    \end{equation}
    such that $ m({\bf x})=m$ and $q({\bf x})=q$, with $\lambda({\bf x})=q({\bf x})-\gamma$.  Given the variable $\mathcal{N}(m, \lambda, q)$, the number of equilibria at fixed $m, q$ in the CM is obtained as $\mathcal{N}(m, \lambda=q-\gamma, q)$.
    \item For the spherical model, equilibria are solutions of the equations
    \begin{equation}
    f_i({\bf x})-\lambda \, x_i=0 \quad \quad   i=1, \cdots, N  \quad \quad \sum_{i=1}^N x_i^2=N
    \end{equation}
    such that $m({\bf x})=m$, and where the Lagrange multiplier equals to $\lambda=N^{-1} \, {\bf f}({\bf x}) \cdot {\bf x}$, i.e., it is the projection of the vector field ${\bf f}({\bf x})$ along the radial direction ${\bf x}$. This choice indeed guarantees that $\partial_t ||{\bf x}||^2= 2 {\bf x} \cdot \partial_t {\bf x} =0$. Since $ {\bf f}({\bf x}) \cdot {\bf x}$ is a random quantity, we count the number of equilibria of the SpM at fixed value of $\lambda$, which is given by $\mathcal{N}(m, \lambda, q=1)$. In this calculation, $\lambda$ is a parameter that, when tuned, determines the stability properties of the counted equilibria, as we discuss more extensively below. The total number of equilibria at fixed $m$ is obtained integrating $\mathcal{N}(m, \lambda, q=1)$ over the parameter $\lambda$.
\end{itemize}
In glassy phases, the typical number of solutions of \eqref{eq:General2} scales exponentially with $N$. The associated quenched complexity is defined by
\begin{align}\label{app.SQ}
\Sigma(m,\lambda,q):=\lim_{N\to\infty}\frac{\mathbb{E}\log\mathcal{N}(m,\lambda,q)}{N}=\lim_{N\to\infty,\,n \to 0}\frac{\mathbb{E}[\mathcal{N}(m,\lambda,q)^n]-1}{Nn}=\lim_{n\to0,N\to\infty}\frac{\log\mathbb{E}\mathcal{N}(m,\lambda,q)^n}{Nn}
\end{align}
where the identities imply the use of the replica trick, namely that one considers integer $n$ copies (replicas) of the system to determine $\mathcal{N}^n$, and then sends $n\to 0$ at the end of the calculation. The quenched complexity is in general different from the annealed one, defined by
\begin{align}\label{app.SA}
\Sigma_A(m,\lambda,q):=\lim_{N\to\infty}\frac{\log\mathbb{E}\mathcal{N}(m,\lambda,q)}{N}. 
\end{align}
As usual in the replica formalism, the calculation of the quenched complexity is mapped into a variational problem for an $n \times n$ overlap matrix $\hat Q$, which encodes for the distribution of scalar products ${\bf x}^a \cdot {\bf x}^b$ extracted with uniform measure among the family of equilibria with given $m, \lambda$ and $q$. We solve this variational problem within the so called Replica Symmetric (RS) assumption, which corresponds to assuming that the overlaps ${\bf x}^a \cdot {\bf x}^b$ between different equilibria $a \neq b$ have a unique typical value independently of the choice of $a,b$, denoted with $\tilde Q$ below, implying that the matrix $\hat Q$ is symmetric. In the following section, we report the resulting explicit expressions for the quenched (and annealed) complexity.

\subsection{The complexity for the general family of models}
\label{app:sec:general_family}
\noindent \paragraph*{\underline{{\bf Quenched complexity.}}} Within the Replica Symmetric (RS) assumption, the quenched complexity \eqref{app.SQ} is expressed as a variational problem over a parameter $ \tilde Q \in [-1,q]$, 
\begin{align}\label{eq.compvar}
\begin{split}
&\Sigma(m,\lambda,q)=\text{extr}_{\tilde{Q}}\tilde{\Sigma}(m,\lambda,q,\tilde{Q}).
\end{split}
\end{align}
The explicit expression of $\tilde{\Sigma}(m,\lambda,q,\tilde{Q})$ depends on the functions  $\Phi_1(u), \Phi_2(u)$ in  \eqref{app:eq:def_Covariance}, evaluated at either $q$ or $\tilde{Q}$. We introduce the following compact notation: 
\begin{equation}\label{eq:Notation}
    \alpha_q:= \frac{\Phi_2(q)}{\Phi_1'(q)}; \quad \quad  \Phi_i^q:=\Phi_i(q), \quad  \tilde \Phi_i:=\Phi_i(\tilde Q), \quad  \dot{\Phi}^q_i:= \Phi_i'(q)\quad   \text{for} \quad  i=1,2; \quad \quad   \kappa(\lambda, q)=\frac{\lambda}{\sqrt{\Phi_1'(q)}}.
\end{equation}
Then 
\begin{align}\label{eq.CompQgen}
\begin{split}
&\tilde{\Sigma}(m,\lambda,q,\tilde{Q})=\mathcal{V}(m, q, \tilde Q)+\mathcal{P}(m,\lambda,q,\tilde{Q})+\Theta (\kappa(\lambda, q), q)
\end{split}
\end{align}
where 
\begin{align}
\label{app:det_eqn}
    \Theta(\kappa, q)=\begin{cases}
    \frac{\log\dot{\Phi}_1^q}{2}+\frac{1}{2}\left(\frac{\kappa^2}{1+\alpha_q}-1\right)\quad\quad\text{if } |\kappa|\leq 1+\alpha_q\\
     \frac{\log\dot{\Phi}_1^q}{2}+\frac{1}{8\alpha_q}\left(\kappa-\text{sign}(\kappa)\sqrt{\kappa^2-4\alpha_q}\right)^2+\log\left|\frac{\kappa+\text{sign}(\kappa)\sqrt{\kappa^2-4\alpha_q}}{2}\right|\quad\quad\text{if }|\kappa|>1+\alpha_q,
    \end{cases}
\end{align} 
\begin{align}\label{eq.vol}
&\mathcal{V}(m, q, \tilde Q)= \frac{q-m^2 + (q - \tilde{Q}) \log(2 \pi) + (q - \tilde{Q}) \log\left(q - \tilde{Q}\right)}{2 (q - \tilde{Q})},
\end{align} 
and
\begin{align}\label{eq.Proba}
&\mathcal{P}(m,\lambda,q,\tilde{Q})=-\frac{1}{2}\left\{\log(2\pi)+\log(\Phi_1^q-\tilde{\Phi}_1)+\frac{\tilde{\Phi}_1}{\Phi_1^q-\tilde{\Phi}_1}+\lambda^2U_{11}-\lambda J mU_{12}+J^2m^2U_{22}\right\}
\end{align}
with
\begin{equation}
\begin{split}
&U_{11}=\frac{q \Phi_1^q - 2 q \tilde{\Phi}_1 + 
\tilde{Q} \tilde{\Phi}_1 + (q - \tilde{Q})^2 \tilde{\Phi}_2}{\mathcal{A}}\\
&U_{22}=\frac{1}{\Phi_1^q - \tilde{\Phi}_1}-\frac{m^2(\alpha_q\dot{\Phi}^q_1-\tilde{\Phi}_2)}{\mathcal{A}}\\
&U_{12}=2m\frac{\tilde{\Phi}_2(q-\tilde{Q})+\Phi_1^q-\tilde{\Phi}_1}{\mathcal{A}}\\
&\mathcal{A}=(\Phi_1^q)^2 + (\tilde{\Phi}_1)^2 - 2 q \tilde{\Phi}_1 \alpha_q\dot{\Phi}^q_1 + (q - \tilde{Q})^2 \alpha_q\dot{\Phi}^q_1 \tilde{\Phi}_2 + \tilde{Q} \tilde{\Phi}_1 (\alpha_q\dot{\Phi}^q_1 + \tilde{\Phi}_2) + \Phi_1^q (-2 \tilde{\Phi}_1 - 2 \tilde{Q} \tilde{\Phi}_2 + q (\alpha_q\dot{\Phi}^q_1 + \tilde{\Phi}_2)).\\
\end{split}
\end{equation}

\noindent \paragraph*{\underline{{\bf Annealed complexity. }}} 
The annealed complexity \eqref{app.SA} reads instead:
\begin{align}\label{eq: AnnCompGen}
\Sigma_A(m,\lambda,q)=\mathcal{P}_A(m,\lambda,q)+\mathcal{V}_A(m,q)+\Theta (\kappa(\lambda, q), q),
\end{align}
where now
\begin{equation}\label{eq.pA}
\begin{split}
&\mathcal{V}_A(m,q)=\frac{1}{2}+\frac{1}{2}\log(2\pi (q-m^2)),\\
&\mathcal{P}_A(m,\lambda,q)=-\frac{1}{2}\left[\log(2\pi)+\log(\Phi_1^q)+\lambda^2U_{11}^A-\lambda J mU_{12}^A+J^2m^2U_{22}^A \right]
\end{split}
\end{equation}
and 
\begin{equation}
\begin{split}
U_{11}^{A}=\frac{q}{\Phi_1^q + q \alpha_q\dot{\Phi}^q_1}, \quad \quad
U_{12}^A =\frac{2m}{\Phi_1^q + q \alpha_q\dot{\Phi}^q_1}, \quad \quad 
U_{22}^A=\frac{\Phi_1^q + (q-m^2) \alpha_q\dot{\Phi}^q_1}{\Phi_1^q (\Phi_1^q + 
q \alpha_q\dot{\Phi}^q_1)}.
\end{split}
\end{equation}
For $m=0$ and $q=1$, this reduces to 
\begin{align}\label{eq: AnnCompGenParaGFyod}
\Sigma_A(m=0,\lambda,q=1)=\frac{1}{2}-\frac{1}{2}\log(\Phi_1^1)-\frac{\dot{\Phi}^1_1 \kappa^2}{2[\Phi_1^1 + \alpha_1\dot{\Phi}^1_1]}+\Theta (\kappa, 1), \quad \quad \kappa= \frac{\lambda}{\sqrt{\dot{\Phi}^1_1}},
\end{align}
which is consistent with the results of \cite{Fyodorov_2016} (see Eq. (3.3) of \cite{Fyodorov_2016}).\\

\noindent \paragraph*{\underline{ {\bf Linear stability of the equilibria.}} }  
The linear stability of the equilibria counted by the complexity \eqref{eq.CompQgen} and its annealed counterpart \eqref{eq: AnnCompGen} is controlled by $\lambda$. As we show in detail in section \ref{app:determinant_calc}, the matrix controlling the linear stability of equilibria  (obtained linearizing the dynamical equations around each equilibrium configuration) is an asymmetric random matrix with Gaussian entries; the  eigenvalues of this matrix are uniformly distributed in a region of compact support on the complex plane. This support has the shape of an ellipse, centered on a point that depends on $\lambda$.  An equilibrium is linearly stable if all the eigenvalues of this matrix have positive real part, and unstable otherwise. In the SpM, $\lambda$ can be kept as a free parameter, which can be tuned to study the complexity of linearly unstable, marginal and stable equilibria. We find that  
\begin{equation}
\begin{cases}
       &\lambda>\sqrt{\dot{\Phi}_1^q}(1+\alpha_q) \quad \longrightarrow \quad   \text{ linearly stable equilibria SpM}\\
    &\lambda<\sqrt{\dot{\Phi}_1^q}(1+\alpha_q) \quad \longrightarrow \quad   \text{ linearly unstable equilibria SpM}.
    \end{cases}
    \end{equation}
 Marginally stable equilibria correspond to saturation of the inequality, which occurs at 
 \begin{equation}
     \lambda_{\rm ms}:=\sqrt{\dot{\Phi}^q_1}(1+\alpha_q).
 \end{equation}
  In the CM, $\lambda = q-\gamma$, and thus the stability of the equilibria is controlled by the parameter $q$. It holds: 
 \begin{equation}
\begin{cases}
       &q>\gamma+\sqrt{\dot{\Phi}_1^q}(1+\alpha_q) \quad \longrightarrow \quad   \text{ linearly stable equilibria CM}\\
    &q<\gamma +\sqrt{\dot{\Phi}_1^q}(1+\alpha_q) \quad \longrightarrow \quad   \text{ linearly unstable equilibria CM}.
    \end{cases}
    \end{equation}
  Marginally stable equilibria correspond to saturation of the inequality, which occurs at 
 \begin{equation}
    q_{\rm ms}:=\gamma +\sqrt{\dot{\Phi}_1^q}(1+\alpha_q).
 \end{equation}

\subsection{The complexity for the models discussed in the main text}\label{Sec:ResultsSpecific}
We now specify the general results of the above section to the models that are discussed in the main text. 
We set
\begin{equation}\label{app:spher_rnn_phi}
    \alpha(u)=\alpha\in[-1,1], \quad \quad \Phi_1(u)=2g^2u^2, \quad \quad \Phi_2(x)=\alpha\Phi_1'(u)=4g^2\alpha u,
\end{equation}
which fixes the covariances of the force to
\begin{align}
    \text{Cov}\left[ f_i({\bf x}),f_j({\bf y})\right]_c=2g^2\delta_{ij}\left(\frac{{\bf x}\cdot{\bf y}}{N}\right)^2+4g^2\alpha\left(\frac{{\bf x}\cdot{\bf y}}{N}\right)\frac{y_ix_j}{N}.
\end{align}
We consider separately the Spherical Model and the Confined Models.\\

\paragraph*{The Spherical Model. } The quenched complexity for the SpM is obtained from \eqref{eq.CompQgen} setting $q \to 1$. This is easily done in the terms \eqref{app:det_eqn} and \eqref{eq.vol}. The term \eqref{eq.Proba} for this particular choice of the model takes the form:
\begin{align}
\begin{split}
    \mathcal{P} \to& -\frac{1}{2}\Bigg(\log \quadre{4\pi g^2(1-\tilde{Q}^2)}+\frac{\tilde{Q}^2}{1-\tilde{Q}^2}\\
    &-\frac{1}{{(2 g^2 (1 - 
   \tilde{Q}^2) (1 + 2 \alpha) ((1 + \tilde{Q})^2 + 2 \tilde{Q} \alpha)}}\Bigg[2 J \lambda m^2 (1 + \tilde{Q}) (1 + \tilde{Q} + 2 \tilde{Q} \alpha)\\
   &+ J^2 m^2 (-(1 + \tilde{Q})^2 + 
    2 (-1 + m^2 (1 + \tilde{Q}) - \tilde{Q} (3 + \tilde{Q})) \alpha - 4 \tilde{Q} \alpha^2) + 
 \lambda^2 (-1 - 2 \tilde{Q} (1 + \alpha) + \tilde{Q}^3 (1 + 2 \alpha))\Bigg],
 \end{split}
\end{align}
where we remind that $\tilde{Q}$ is a variational parameter to be  optimized over, see \eqref{eq.compvar}. In the annealed case, from \eqref{eq.pA} we obtain instead:
\begin{align}
\begin{split}
\mathcal{P}_A \to -\frac{1}{2}\left(\log(4\pi g^2) +\frac{ \lambda^2 -2 J\lambda m^2 + J^2 m^2 (1 - 2 m^2 \alpha + 2  \alpha)}{2 g^2  (1 + 2 \alpha)}\right).
\end{split}
\end{align}
A plot of the resulting complexity as a function of $\lambda$ for this particular model for $\alpha=0$ is given in Fig.~\ref{fig:diff_alphas}(\emph{Right}) in the main text. The comparison of the complexity with the dynamics for this model is discussed in Sec.\ref{app:kac_dyn_spher}. \\

\paragraph*{ The confined model. } The quenched complexity for the CM is instead obtained from \eqref{eq.CompQgen} setting $\lambda \to q-\gamma$ for some $\gamma>0$. The term \eqref{eq.vol} is not affected by this substitution. The term \eqref{app:det_eqn} becomes instead
\begin{align}
    \Theta \to \begin{cases}
        \frac{1}{2} \left(-1 + \frac{(\gamma - q)^2}{4 g^2 q (1 + \alpha)} + \log(4 g^2 q)\right)\quad\quad\text{if }\quad|q - \gamma| < 2 g (1 + \alpha) \sqrt{q}\\
        \frac{\left(\gamma - q + \sqrt{\gamma^2 - 2 \gamma q + q^2 - 16 g^2 q \alpha}\right)^2}{32 g^2 q \alpha} 
+ \log\left(\frac{1}{2} \left(q-\gamma + \sqrt{(\gamma - q)^2 - 16 g^2 q \alpha}\right)\right)\quad\quad\text{otherwise}.
    \end{cases}
\end{align}
In the quenched case, from \eqref{eq.Proba} we obtain:
\begin{align}
\begin{split}
\mathcal{P}&\to-\frac{1}{4 g^2 (q - \tilde{Q}) (q + \tilde{Q}) (1 + 2 \alpha) ((q + \tilde{Q})^2 + 2 q \tilde{Q} \alpha)} 
\Bigg( 
(q + \tilde{Q}) \big(q^4 + q^3 \tilde{Q} - q (-2 g^2 + q) \tilde{Q}^2 + 2 g^2 \tilde{Q}^3\big)\\& 
+ 2 \tilde{Q} \big(q^4 + q^2 (2 g^2 - \tilde{Q}) \tilde{Q} + 6 g^2 q \tilde{Q}^2 + 2 g^2 \tilde{Q}^3\big) \alpha 
+ 8 g^2 q \tilde{Q}^3 \alpha^2 
- 2 J m^2 q (q + \tilde{Q}) (q + \tilde{Q} + 2 \tilde{Q} \alpha)\\& 
+ J^2 m^2 \big((q + \tilde{Q})^2 + 2 (q^2 + 3 q \tilde{Q} + \tilde{Q}^2 - m^2 (q + \tilde{Q})) \alpha 
+ 4 q \tilde{Q} \alpha^2\big) \\&
+ \gamma^2 (q + \tilde{Q}) \big(q^2 - \tilde{Q}^2 (1 + 2 \alpha) + q (\tilde{Q} + 2 \tilde{Q} \alpha)\big) 
+ 2 \gamma (q + \tilde{Q}) \big(J m^2 (q + \tilde{Q} + 2 \tilde{Q} \alpha)\\& 
- q (q^2 - \tilde{Q}^2 (1 + 2 \alpha) + q (\tilde{Q} + 2 \tilde{Q} \alpha))\big) 
+ 2 g^2 (q - \tilde{Q}) (q + \tilde{Q}) (1 + 2 \alpha) \big((q + \tilde{Q})^2 + 2 q \tilde{Q} \alpha\big) 
\log\big(4 g^2 \pi (q - \tilde{Q}) (q + \tilde{Q})\big)
\Bigg),
\end{split}
\end{align}
while in the annealed case,  \eqref{eq.pA} reduces to
\begin{align}
\begin{split}
\mathcal{P}_A&\to-\frac{q \left(J (2 \gamma + J) m^2 + (\gamma^2 - 2 J m^2) q - 2 \gamma q^2 + q^3\right) 
+ 2 J^2 m^2 (-m^2 + q) \alpha 
+ 2 g^2 q^3 (1 + 2 \alpha) \log\left(4 g^2 \pi q^2\right)}
{4 g^2 q^3 (1 + 2 \alpha)}.
\end{split}
\end{align}

\section{Analysis of the complexity, and comparison with the dynamics}\label{app:COMP}

\subsection{Analysis of the annealed complexity for the spherical model}\label{sse:Ann_SM}
In this section we discuss in detail the behavior of the annealed complexity \eqref{eq: AnnCompGen} for the spherical model. In this case, $\lambda$ is a free parameter controlling the stability of the counted equilibria. The study of the annealed complexity is motivated by several reasons: (1) the annealed complexity gives an upper bound to the quenched complexity; if the study of the annealed complexity reveals regimes in which no single stable equilibrium exists, one knows that this statement remains true also when the quenched calculation is performed. (2) The annealed complexity fully describes those phases in which stable fixed points exist for the dynamics, such as the Ferromagnetic Fixed Point (FFP) phase of the models we are considering. (3) The annealed calculation often describes correctly  the complexity of “paramagnetic" equilibria with $m=0$, which are the relevant equilibria when $J=0$. (4) From the analysis of the annealed complexity at arbitrary values of the  parameter $\alpha$, we can derive a lower bound on the critical value $\alpha_c$ at which a family of (marginally) stable equilibria with positive complexity exists. For $J=0$, given that the dominant equilibria are paramagnetic and their annealed complexity is exact, the bound is actually saturated. This information is relevant when discussing the dynamics when approaching the conservative limit $\alpha \to 1$, as reported in the final part of the main text. \\
In the spherical model, $q$ is fixed ($q=1$), and the annealed complexity is a function of two parameters, $m$ and $\lambda$. Below, we give the expressions for arbitrary value of $q$, assuming it to be fixed but possibly different from one. We also assume $\lambda \geq 0$.\\
Typical equilibria (i.e., the most numerous) are obtained optimizing the complexity over $m$ and $\lambda$. The corresponding equations are:
\begin{align}
\label{app:eq:ann_der_m}
\frac{\partial\Sigma_A(m,\lambda,q)}{\partial m}=m \left(\frac{1}{m^2 - q} - J\frac{
      J [(q-2 m^2) \alpha_q \dot{\Phi}_1^q + \Phi_1^q]-2 \lambda \Phi_1^q}{\Phi_1^q (q \alpha_q \dot{\Phi}_1^q + \Phi_1^q)}\right),
\end{align}
and 
\begin{align}
\label{app:eq:ann_der_lambda}
    \frac{\partial\Sigma_A(m,\lambda,q)}{\partial\lambda}=\begin{cases}
        \frac{2}{\lambda + \sqrt{\lambda^2 - 4 \alpha_q \dot{\Phi}_1^q}} + \frac{
 J m^2 - q \lambda}{q \alpha_q \dot{\Phi}_1^q + \Phi_1^q}\quad &\text{ if }\quad \lambda>\sqrt{\dot{\Phi}_1^q}(1+\alpha_q)\\
    \frac{\lambda}{\dot{\Phi}_1^q + \alpha_q \dot{\Phi}_1^q} + \frac{
 J m^2 - q \lambda}{q \alpha_q \dot{\Phi}_1^q + \Phi_1^q}\quad &\text{ if }\quad0 \leq \lambda<\sqrt{\dot{\Phi}_1^q}(1+\alpha_q).
    \end{cases}
\end{align}
From \eqref{app:eq:ann_der_m} it appears that, at fixed $\lambda$, the annealed complexity is always stationary at $m=0$ (paramagnetic equilibria). For $J=0$, this is the only solution of the stationarity equation. For $J>0$, two other solutions appear at values of $m$ that are non-zero and opposite in sign. The optimization over $\lambda$ has to be performed taking into account the different domains. We recall that these domains define also the linear stability of the equilibria counted by the complexity: equilibria are linearly unstable for $\lambda<\sqrt{\dot{\Phi}_1^q}(1+\alpha_q)$, and linearly stable otherwise. In the following subsections, we analyze the annealed complexity of linearly unstable, marginal and stable equilibria, respectively. 

\subsubsection{Annealed complexity of unstable equilibria}
For values of $\lambda$ that correspond to linearly unstable equilibria, the complexity reads: 
\begin{align}
\label{app:eq:unstable_compl}
\begin{split}
\Sigma_A(m,\lambda,q)&=\frac{1}{2} \Bigg\{
\frac{J^2 m^2 (1 + \alpha_q) \dot{\Phi}_1^q \left((m^2 - q) \alpha_q \dot{\Phi}_1^q - \Phi_{1}^{q}\right) + 
2 J m^2 (1 + \alpha_q) \lambda \dot{\Phi}_1^q \Phi_{1}^{q} + 
\lambda^2 \Phi_{1}^{q} \left( \Phi_{1}^{q}-q \dot{\Phi}_1^q\right)}{
(1 + \alpha_q) \dot{\Phi}_1^q \Phi_{1}^{q} \left(q \alpha_q \dot{\Phi}_1^q + \Phi_{1}^{q}\right)}\\
&+ 
\log{\left[\frac{(q-m^2) \dot{\Phi}_1^q}{\Phi_{1}^{q}}\right]}
\Bigg\}, \quad \quad \quad 0 \leq \lambda \leq \sqrt{\dot{\Phi}_1^q}(1+\alpha_q).
\end{split}
\end{align}
This function is symmetric in $\pm m$. As a function of $\lambda$, $\Sigma_A$ is a parabola with negative curvature, which attains its maximum at 
\begin{align}
\label{app:eq:max_lamb_ann_unst}
    \lambda_{\text{typ}}^{\rm{unst}}(m):=\frac{J m^2 (\dot{\Phi}_1^q + \alpha_q \dot{\Phi}_1^q)}{q \dot{\Phi}_1^q - \Phi_1^q} \, \theta \tonde{\sqrt{\dot{\Phi}_1^q}(1+\alpha_q)-     \lambda_{\text{typ}}^{\rm{unst}}},
\end{align}
which solves the stationarity equation \eqref{app:eq:ann_der_lambda}. \\

\paragraph*{Paramagnetic equilibria.}
The maximum of the complexity for $m=0$ is given by $\lambda_{\text{typ}}^{\rm{unst}}(0)=0$, and the corresponding value of the complexity is
\begin{align}
    \Sigma_A(m=0,\lambda=0,q)=\frac{1}{2}\log\left(\frac{q\,\dot{\Phi}_1^q}{\Phi_1^q} \right),
\end{align}
which does not depend on $\alpha_q$. This means that the annealed complexity can not be tuned to zero by changing the relative strength of the conservative part of the dynamics. The complexity of paramagnetic equilibria vanishes at
\begin{align}
\label{app:eq:lambda_roots_m=0}
    \lambda_{0,\pm}:=\pm
\sqrt{\frac{\log{\left(\frac{\Phi_{1}^{q}}{q \dot{\Phi}_1^q}\right)}(1 + \alpha_q) \dot{\Phi}_1^q \left(q \alpha_q \dot{\Phi}_1^q + \Phi_{1}^{q}\right)}{
  \Phi_{1}^{q}}-q \dot{\Phi}_1^q},
\end{align}
and we are assuming that these quantities satisfy $ 0 \leq \lambda_{0,\pm} <\sqrt{\dot{\Phi}_1^q}(1+\alpha_q)$ in order to belong to the region of linear instability.

\paragraph*{Ferromagnetic equilibria. }
At the value \eqref{app:eq:max_lamb_ann_unst} the  complexity reads:
\begin{align}
\Sigma_A(m,\lambda_{\text{typ}}^{\rm{unst}}(m),q)=
\frac{J^2 m^2 \left(m^2 \dot{\Phi}_1^q - q \dot{\Phi}_1^q + \Phi_{1}^{q}\right) - \Phi_{1}^{q} \left( \Phi_{1}^{q}-q \dot{\Phi}_1^q\right) \left(\log{[(q-m^2) \dot{\Phi}_1^q]} - \log{\Phi_{1}^{q}}\right)}{
2 \left(q \dot{\Phi}_1^q - \Phi_{1}^{q}\right) \Phi_{1}^{q}}.
\end{align}
One can then study the maximum of this function, as well as the values at which it vanishes, as a function of $m$; however, the specific results are model dependent and we leave the corresponding analysis to the interested reader. 

\subsubsection{Annealed complexity of marginally stable equilibria, and the critical value $\alpha_c$ }

Here, we aim at determining the minimal value of $\alpha_q$ for which marginally stable equilibria appear. We focus separately on paramagnetic and ferromagnetic equilibria.

\paragraph*{Paramagnetic equilibria.} For $m=0$, the complexity is symmetric with respect to $\lambda$. At the threshold value $\lambda_\text{ms}=\sqrt{\dot{\Phi}_1^q}(1+\alpha_q)$
the complexity of marginal equilibria reads:
\begin{align}
\label{app:eq:sigma_marginal_m=0}
\Sigma_A\left(m=0,\lambda_\text{ms},q\right)=
\frac{1}{2} \left(
\alpha_q - \frac{q (1 + \alpha_q)^2 \dot{\Phi}_1^q}{q \alpha_q \dot{\Phi}_1^q + \Phi_{1}^{q}} + 
\log{\left(\frac{e q \dot{\Phi}_1^q}{\Phi_{1}^{q}}\right)}
\right).
\end{align}
This function has a simple dependence on $\alpha_q$. The derivative of Eq.~\eqref{app:eq:sigma_marginal_m=0} with respect to $\alpha_q$ is:
\begin{align}
\partial_{\alpha_q}\Sigma_A=\frac{(-q \dot{\Phi}_1^q + \Phi_{1}^{q})^2}{2 \left(q \alpha_q \dot{\Phi}_1^q + \Phi_{1}^{q}\right)^2} \geq 0,
\end{align}
thus the annealed complexity of marginal equilibria increases as $\alpha_q$ increases. Hence, there exists (depending on the specific choice of the model) a critical alpha
$\alpha_c^q(m=0)$ such that for $1>\alpha_q>\alpha^q_c(0)$, there is an exponentially large (in $N$) number of marginal equilibria with complexity~\eqref{app:eq:sigma_marginal_m=0}. Such critical value is easily found imposing \eqref{app:eq:sigma_marginal_m=0} to be equal to zero, and it reads:
\begin{align}
\alpha_c^q(0)=
\frac{q \dot{\Phi}_1^q - \Phi_{1}^{q} \log{\left(\frac{e \,q \,\dot{\Phi}_1^q}{\Phi_{1}^{q}}\right)}}{ \Phi_{1}^{q}-q \dot{\Phi}_1^q + q \dot{\Phi}_1^q \log{\left(\frac{q \dot{\Phi}_1^q}{\Phi_{1}^{q}}\right)}}.
\end{align}

\paragraph*{Ferromagnetic equilibria.}
One can ask the same question as above, but for a generic value $m\neq 0$. For each value of $m$, we determine the minimal value of $\alpha_c^q$ for which exponentially many marginally stable equilibria appear, according to the annealed complexity. Also for general $m$ the complexity has positive derivative with respect to $\alpha_q$ at the threshold value $\lambda_{\rm s}$:
\begin{align}
    \partial_{\alpha_q}\Sigma_A\left(m, \lambda_\text{ms},q\right)=
\frac{\left(J m^2 \sqrt{\dot{\Phi}_1^q} +(- q \dot{\Phi}_1^q + \Phi_{1}^{q})\right)^2}{2 \left(q \alpha \dot{\Phi}_1^q + \Phi_{1}^{q}\right)^2} \geq 0.
\end{align}
The critical value (whenever it exists) of $\alpha_c^q$ for which the annealed complexity of marginal equilibria is zero (and no longer negative) reads:
\begin{align}
\alpha_c^q(m)=
\frac{\Phi_{1}^{q} \left(J^2 m^2 - 2 J m^2 \sqrt{\dot{\Phi}_1^q} + q \dot{\Phi}_1^q - \Phi_{1}^{q} - \Phi_{1}^{q} \log{[(q-m^2) \dot{\Phi}_1^q]} + \Phi_{1}^{q} \log{\Phi_{1}^{q}}\right)}{
J^2 m^2 (m^2 - q) \dot{\Phi}_1^q + 2 J m^2 \sqrt{\dot{\Phi}_1^q} \Phi_{1}^{q} + \Phi_{1}^{q} \left( \Phi_{1}^{q}-q \dot{\Phi}\right) + q \dot{\Phi}_1^q \Phi_{1}^{q} \log\left(\frac{(q-m^2) \dot{\Phi}_1^q}{\Phi_{1}^{q}}\right)}.
\end{align}

\subsubsection{Annealed complexity of stable equilibria and isolated fixed points}
As shown above, stable equilibria of magnetization $m$ have a positive annealed complexity as soon as $\alpha_q>\alpha^q_c(m)$. In the {stable} region defined by $\lambda>\lambda_\text{ms}=\sqrt{\dot{\Phi}_1^q}(1+\alpha_q)$, the annealed complexity reads:
\begin{align}
\begin{split}
\Sigma_A(m,\lambda,q)&=\frac{1}{8} \Bigg( 
\frac{(\lambda - \sqrt{\lambda^2 - 4 \alpha_q \dot{\Phi}_1^q})^2}{\alpha_q \dot{\Phi}_1^q} + 
\frac{8 J m^2 \lambda}{q \alpha_q \dot{\Phi}_1^q + \Phi_{1}^{q}} - 
\frac{4 q \lambda^2}{q \alpha_q \dot{\Phi}_1^q + \Phi_{1}^{q}} - 
\frac{4 J^2 m^2 \left((-m^2 + q) \alpha_q \dot{\Phi}_1^q + \Phi_{1}^{q}\right)}{\Phi_{1}^{q} \left(q \alpha_q \dot{\Phi}_1^q + \Phi_{1}^{q}\right)}\\
&+ 
8 \log{\left(\frac{1}{2} \left(\lambda + \sqrt{\lambda^2 - 4 \alpha_q \dot{\Phi}_1^q}\right)\right)} + 
4 \log{\left(\frac{e (-m^2 + q)}{\Phi_{1}^{q}}\right)}
\Bigg) \quad \quad \quad \lambda>\sqrt{\dot{\Phi}_1^q}(1+\alpha_q).
\end{split}
\end{align}
Now, a positive annealed complexity of stable equilibria can arise in two different scenarios: either (i) the complexity curve of unstable equilibria crosses $\lambda_{\rm ms}$ before vanishing, and thus a branch of that curve corresponds to stable equilibria; or (ii) there is a region entirely supported in $\lambda>\lambda_{\rm ms}$ where the annealed complexity is positive. We are interested in identifying such scenarios, and in particular the maximum of the complexity in the scenario (ii). By imposing that equations Eqs.~\eqref{app:eq:ann_der_lambda}, \eqref{app:eq:ann_der_m} are zero in the stable region, we can calculate such maxima. When the corresponding complexity is exactly zero, the corresponding point identifies an isolated equilibrium. This situation occurs, in particular, in the FFP phase of the dynamics.\\

\paragraph*{Paramagnetic isolated fixed point.}
In this paramagnetic case we find that the complexity is stationary at  
\begin{align}
    \lambda_{\text{pfp}}= \frac{q \alpha_q \dot{\Phi}_1^q + \Phi_{1}^{q}}{ \sqrt{q\Phi_{1}^{q}}}, 
\end{align}
which is in the correct region of $\lambda$ (thus acceptable) only if 
\begin{align}
    \alpha_q>\sqrt{\frac{\Phi_1^q}{q\dot{\Phi}_1^q}}.
\end{align}
One can additionally verify that for $\alpha_q>\frac{\Phi_1^q}{q\dot{\Phi}_1^q}$, the annealed complexity at $\lambda_{\text{pfp}}$ is zero, meaning that there is a paramagnetic isolated equilibrium, which is linearly stable. We notice that this condition is not satisfied for $\alpha_q=0$. In fact, in the dynamics of the model studied in the main text for completely non-conservative forces, there is no trace of a paramagnetic fixed point attracting the dynamical trajectories. \\


\paragraph*{Ferromagnetic isolated fixed point.} We assume $J>0$. A way to obtain the solution of $\partial_m\Sigma_A=0, \,\,\partial_\lambda\Sigma_A=0$ is to solve for $\lambda$ in both Eqs.~\eqref{app:eq:ann_der_m},\eqref{app:eq:ann_der_lambda}, and then to match the two results in order to solve for $m$. Solving for $\lambda$ the two equations gives:
\begin{align}
\begin{split}
&\lambda=\frac{J m^2 (\Phi_{1}^q-q \alpha_q \dot{\Phi}_{1}^q) \pm (q \alpha_q \dot{\Phi}_{1}^q + \Phi_{1}^q) \sqrt{J^2 m^4 + 4 q \Phi_{1}^q}}{2 q \Phi_{1}^q}\\
&\lambda=
-\frac{q \alpha_q \dot{\Phi}_1^q + \Phi_{1}^{q}}{2 J m^2 - 2 J q}+ 
\frac{J \left(-2 m^2 \alpha_q \dot{\Phi}_1^q + q \alpha_q \dot{\Phi}_1^q + \Phi_{1}^{q}\right)}{2 \Phi_{1}^{q}}
\end{split}
\end{align}
now we equate them and solve for $m$, to get
\begin{align}
\label{app:eq:fp_general}
\begin{split}
 \lambda_{\text{ffp}}=J+\frac{\alpha_q\dot{\Phi}_1^q}{J}, \quad \quad \quad    m_{\text{ffp}}=\pm\frac{\sqrt{J^2-q\Phi_1^q}}{J}.
\end{split}
\end{align}
We see that the bound $J>\sqrt{\Phi_1^qq}$ must be satisfied for the existence of this solution. One can also check that the complexity becomes zero at these values of the order parameters, meaning that they correspond to an isolated stable fixed point. We can also verify that, in general, this point is stable for $J>\sqrt{\dot{\Phi}_1^q}$.


\subsection{Comparing complexity and dynamics: the spherical model}\label{app:kac_dyn_spher}
We now consider the complexity for the specific model discussed in the main text, characterized by the choices \eqref{app:spher_rnn_phi}. The corresponding complexity functions are given explicitly in Sec.~\ref{Sec:ResultsSpecific}. We begin with the spherical case, for which $\lambda$ is a tunable parameter and $q=1$. 

\subsubsection{Purely non-conservative case: $\alpha=0$. }

\paragraph*{Isolated equilibrium and the FFP phase. }From the analysis of the annealed complexity of Sec. \ref{sse:Ann_SM} it follows that for $\alpha=0$, the only stable equilibrium, if it exists, is the isolated ferromagnetic fixed point. As it follows from Eq.\eqref{app:eq:fp_general}, this isolated equilibrium has parameters
\begin{equation}\label{eq:DMFT_fix_SM}
    \lambda_{\text{ffp}}=J, \quad \quad \quad m_\text{ffp}=\pm\sqrt{1-\frac{2 g^2}{J^2}}.
\end{equation}
This isolated fixed point is linearly stable whenever $J>J_c=2g$: this is precisely the regime of parameters which corresponds to the FFP phase identified by the DMFT solution, see \eqref{eq:traSM}. In particular, the parameters \eqref{eq:DMFT_fix_SM} coincide with the values predicted by the asymptotic solution of the DMFT for the SpM, see Eq.~\eqref{DMFT_fixFFSM}. In the region $J>2g$ unstable equilibria also exist and have a positive complexity; however, only the stable fixed point matters for the large time dynamics.\\

\begin{figure*}
    \centering
    \includegraphics[width=1\textwidth, trim={4 4 4 4},clip]{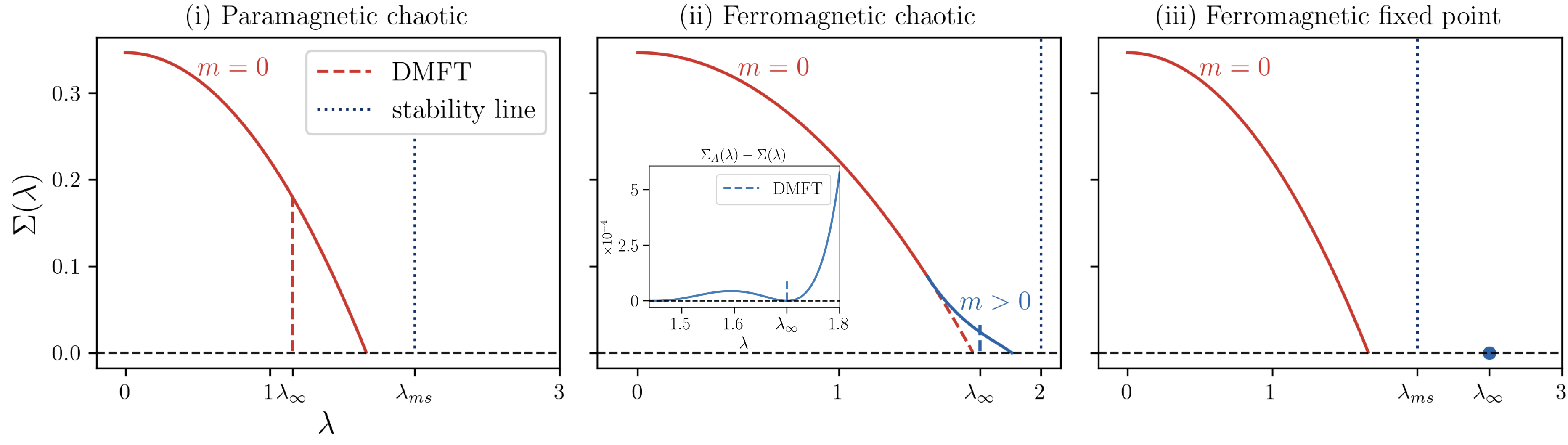}
    \caption{Complexity curves of the spherical model in the three different dynamical phases, (i) paramagnetic chaotic phase, (ii) ferromagnetic chaotic phase, and (iii) ferromagnetic fixed point phase. The red curve, identical in all plots, is the complexity of (unstable) equilibria with $m=0$, the blue curve with $m\neq 0$. The inset in the second panel shows $D(\lambda)$, the difference between the quenched and annealed complexities of ferromagnetic fixed points, which vanishes at $\lambda_\infty$. In the third panel, the is an isolated point in the stable region,where the complexity is exactly zero.}
    \label{app:fig:kac_spherical}
\end{figure*}

\paragraph*{Unstable equilibria and the chaotic phases. } As one lowers $J$ below $J_c=2g$, the system enters the chaotic phase (dynamically). Consistently, the complexity calculation shows that the dynamical equations admit exponentially many equilibria, but they are \emph{all} linearly unstable: the complexity is non-negative only for $\lambda<2g$. We consider fixed $\lambda \geq 0$, and optimize the {quenched} complexity over $\tilde{Q}$ and $m$. We denote with $\tilde{Q}_{\text{typ}}$ and $m_{\text{typ}}$ the values where the optimum is attained. The optimization gives, in general, two solutions: the first is paramagnetic, with $m_{\text{typ},P}=0,\tilde{Q}_{\text{typ},P}=0$ (here “P" stands for paramagnetic). The corresponding quenched complexity coincides with the annealed one,
\begin{align}
    \Sigma(m=0,\lambda)=\frac{1}{8}\left(-\frac{\lambda^2}{g^2} + \log(16)\right).
\end{align}
This function has a maximum in $\lambda=0$ and vanishes at $\lambda =g\sqrt{\log(16)}$. The second solution is ferromagnetic (here “F" stands for ferromagnetic), and reads:
\begin{align}
\begin{split}
&\tilde{Q}_{\text{typ},F}=\frac{-2 g^2 - J^2 + 2 J \lambda}{2 g^2}\\
&m_{\text{typ},F}=\sqrt{\frac{\tilde{Q}_\text{typ,F} (2 g^2 (1 + \tilde{Q}_\text{typ,F} (3 + \tilde{Q}_\text{typ,F} - \tilde{Q}_\text{typ,F}^2)) + (-3 + \tilde{Q}_\text{typ,F}) \tilde{Q}_\text{typ,F} \lambda^2)}{2(1 + \tilde{Q}_\text{typ,F}) (g^2 (1 + \tilde{Q}_\text{typ,F})^2 + 
J \tilde{Q}_\text{typ,F} (J - 2 \lambda))}}.
\end{split}
\end{align}
In this case the complexity takes the form
\begin{align}
\begin{split}
\label{QCSF}
\Sigma(m_{\text{typ},F},\lambda)&=\frac{
16 g^6 (J- \lambda)^2 
+ 8 g^2 J^2 (J - 2 \lambda)^2 \lambda^2 
+ J^3 (J - 2 \lambda)^3 \lambda^2 
+ 8 g^4 J (J - 2 \lambda) (J^2 - 2 J \lambda + 3 \lambda^2)
}{
8 g^2 J^2 (4 g^2 + J (J - 2 \lambda)) (J - 2 \lambda)^2
} \\
&+ \frac{1}{2} \log\left(-\frac{4 g^2}{J^2 - 2 J \lambda}\right).
\end{split}
\end{align}
For $\lambda=(2 g^2 + J^2)/(2 J)$ it holds $m_{\text{typ, F}}=0$, meaning that at this value of $\lambda$ the ferromagnetic and paramagnetic solutions of the optimization problem coincide: the optimization of the complexity over $m$ gives only the paramagnetic solution. Ferromagnetic equilibria are present at smaller values of $\lambda$, but the paramagnetic ones outnumber them. \\
At variance with the paramagnetic case, the ferromagnetic quenched complexity \eqref{QCSF} does not coincide with the annealed one, optimized over $m$ at the corresponding value of $\lambda$. Indeed, the annealed complexity is optimized at $m_{\text{typ, F}}^A=(2 g^2 + J^2 - 2 J \lambda)/(J^2 - 2 J \lambda)$, and the difference between the annealed and quenched complexities at the corresponding optimal values of parameters reads:
\begin{align}
D(\lambda):=\Sigma_A(m_{\text{typ, F}}^A,\lambda)-\Sigma(m_{\text{typ, F}},\lambda)=-\frac{
(2 g^2 + J (J - 2 \lambda))^3 (J - \lambda)^2
}{
4 g^2 J^2 (4 g^2 + J (J - 2 \lambda)) (J - 2 \lambda)^2
}.
\end{align}
In Fig.~\ref{app:fig:kac_spherical}.(ii) (inset) we plot $D(\lambda)$ for values of $J, g$ which correspond to the Ferromagnetic Chaotic phase: the difference vanishes at $\lambda=(2 g^2 + J^2)/(2 J)$ or $\lambda=J$. Intriguingly, the latter is the asymptotic value selected by the DMFT in the ferromagnetic chaotic phase, see \eqref{dmftsphfc}. However, the magnetization and the overlap between the most numerous equilibria at this value of $\lambda$ are 
\begin{align}
\begin{split}
    m_{\text{typ,F}}(\lambda=\lambda_\infty=J)=\sqrt{1 - 2 g^2/J^2}, \quad \quad \quad 
    \tilde{Q}_{\text{typ,F}}(\lambda=\lambda_\infty=J)=-1 + J^2/(2 g^2),
\end{split}
\end{align}
and do not coincide with the values of $m_\infty$ and $C_\infty$  characterizing the attractor manifold. One can further try to impose that $\tilde{Q}(\lambda=\lambda_\infty=J)=C_\infty$ and solve for the $m_{\text{typ}}$ such that $\partial_{\tilde{Q}}\tilde{\Sigma}(m_{\text{typ}},\lambda=\lambda_\infty=J,\tilde{Q})|_{\tilde{Q}=C_\infty}=0$; this results in 
\begin{align}
m_{\text{typ}}(\lambda=\lambda_\infty,\tilde{Q}=C_\infty)=\sqrt{\frac{81}{50} - \frac{3 J^2}{40 g^2} + 
 g^2 \left(-\frac{16}{4 g^2 + 3 J^2} + \frac{288}{25 (4 g^2 + 15 J^2)}\right)},
\end{align}\\
which again differs from the ferromagnetic chaotic DMFT value of $m_\infty$, see Sec.~\ref{sect: stationnary solution of the DMFT equations for alpha 0}.\\
In the paramagnetic chaotic phase, we make a further interesting observation: when computing the complexity at the values of $m, \lambda$ selected asymptotically by the dynamics we find 
\begin{align}
    \Sigma\left(m_\infty=0,\lambda_\infty=\frac{2}{\sqrt{3}}g\right)=\frac{1}{6} (3\log2-1),
\end{align}
which is independent of $g$. For all values of $g$, the dynamics selects values of parameters such that the corresponding complexity takes the same value. This suggests that one can use the complexity to guess the dependence  $\lambda(g)$ of the DMFT solution imposing $d\Sigma(0,\lambda(g))/dg=0$. This condition imposes a linear relation $\lambda(g)=cg$, with $c$ a constant that needs to be fixed with a boundary condition (to fix it, one must use the DMFT solution $\lambda_\infty|_{g=1}=\frac{4}{3}$).\\
 
\paragraph*{Ferromagnetic to Paramagnetic transition.} We now determine for which values of $J$ the maximal complexity corresponds to paramagnetic equilibria for \emph{all} values of $\lambda$. The ferromagnetic branch of the complexity collapses into the paramagnetic one when 
\begin{equation}
\frac{2 g^2 + J^2}{2 J}=g\sqrt{\log(16)}\Rightarrow J_\pm=g \left(2\sqrt{\log 2}\pm\sqrt{4\log 2-2}\right),
\end{equation}
which implies that values of $\lambda$ for which the majority of the equilibria are ferromagnetic exist only for $J_-<J<J_+$. To be consistent with the main text, we call $J_\Sigma:=J_-$. The curve $J_\Sigma$ would be a natural guess for the dynamical transition between the ferromagnetic chaotic phase and the paramagnetic chaotic phase, that the DMFT solution locates at $J_\text{PF}$. 
The curve of $J_\Sigma$ is plotted in Fig.~\ref{fig: confined and spherical phase diagrams appendix} (\textit{right}), black dashed line. As one sees\footnote{If we were to plot the curve of $J_-$, we would see that it belongs to the ferromagnetic fixed point region, meaning that in that region there are already unstable ferromagnetic fixed points.}, it lies strictly below $J_\text{PF}$: when the system is in the paramagnetic chaotic phase,  ferromagnetic unstable fixed points still exist and are still the most numerous ones for certain $\lambda$s. Notice also that $J_\text{PF}$ does not coincide with the value of $J$ at which the most numerous equilibria at $\lambda_\infty=J$ become paramagnetic, which is given by  
  $(2g^2+J^2)(2J)=J\Rightarrow J=\sqrt{2}g$.\\

A comparison between the DMFT solution and the complexity for the Spherical Model is given in Fig.~\ref{app:fig:kac_spherical}. In the figure we show the complexity in the three different regions of the dynamical phase diagram, as a function of $\lambda$. We highlight with $\lambda_\infty$ the value corresponding to the DMFT solution. The red curves correspond to paramagnetic equilibria, the blue curves to ferromagnetic. We see that: (i) in the Paramagnetic Chaotic phase $\lambda_\infty$ does not correspond to either the maximum or minimum of the red curve, but it lies in the middle of the complexity curve; (ii) in the ferromagnetic chaotic phase, at $\lambda_\infty$ quenched and annealed complexity (maximized over all parameters but $\lambda$) coincide, $D(\lambda_\infty)=0$. However  $\lambda_\infty$ does not coincide with the point where the blue curve vanishes, that is, with the parameter corresponding to the “less unstable" equilibria; (iii) in the ferromagnetic stable phase, where one isolated fixed point of zero complexity exists and is linearly stable, and we see that it coincides with the asymptotic solution found by the DMFT. In this region there are also exponentially many unstable equilibria (see the red curve). \\
We see that at the $\lambda= \lambda_\infty$ selected by the dynamics the complexity has some trivialization properties: in the ferromagnetic chaotic phase quenched and annealed complexities match;  in the paramagnetic chaotic phase the complexity is constant in $g$. Whether these simplifications are accidental or have a deeper meaning remains unclear and further analysis of the phase space is needed.  Irrespective of this, as seen in Sec.~\ref{sect: stationnary solution of the DMFT equations for alpha 0} in Fig.~\ref{fig supp: stationnary order parameters and total force in the spherical case} the asymptotic value of the force reached in the dynamics in the paramagnetic chaotic region reads: $F_0^2=2 g^2/3$, which is non-zero, suggesting that at large times the dynamical trajectories are not dominated by the equilibria counted by the complexity. \\

\subsubsection{Partially conservative case: $\alpha\neq 0$.}
Also in this case, the complexity of paramagnetic fixed points equals to the annealed complexity, with $\tilde{Q}_\text{typ,P}=0$, and reads:
\begin{align}
   \Sigma(m=0,\lambda)= \frac{1}{8} \left(-\frac{\lambda^2}{g^2 (1 + 3 \alpha + 2 \alpha^2)} + 4 \log 2\right).
\end{align}
Marginally stable equilibria have
$\lambda_{\text{ms}}=2g(1+\alpha)$. The paramagnetic complexity curve touches zero for 
\begin{align}
    \lambda=2g \sqrt{1 + 3 \alpha + 2 \alpha^2} \sqrt{\log 2},
\end{align}
which means that marginal states appear for $\alpha_c=(1 - \log 2)/(\log4-1 ) \approx 0.8$. This is the value used in the main text.\\

An isolated stable fixed point of ferromagnetic nature exists for
\begin{align}
    \lambda_{\text{ffp}}=J+\frac{4g^2\alpha}{J}, \quad \quad \quad m_{\text{ffp}}=\sqrt{1-\frac{2 g^2}{J^2}}
\end{align}
This fixed point becomes unstable when $J=2g$. In the region of $\lambda$ that corresponds to unstable equilibria, the quenched and annealed ferromagnetic complexities are equal at $\lambda=\lambda_{\text{ffp}}$, with quenched order parameters $m_{\text{typ,F}}=m_{\text{ffp}}$ and $\tilde{Q}_\text{typ,F}=(-2 g^2 + J^2)/(2 g^2)$, just like in the $\alpha=0$ case. \\

\begin{figure*}
  \centering
  \includegraphics[width=0.4\textwidth]{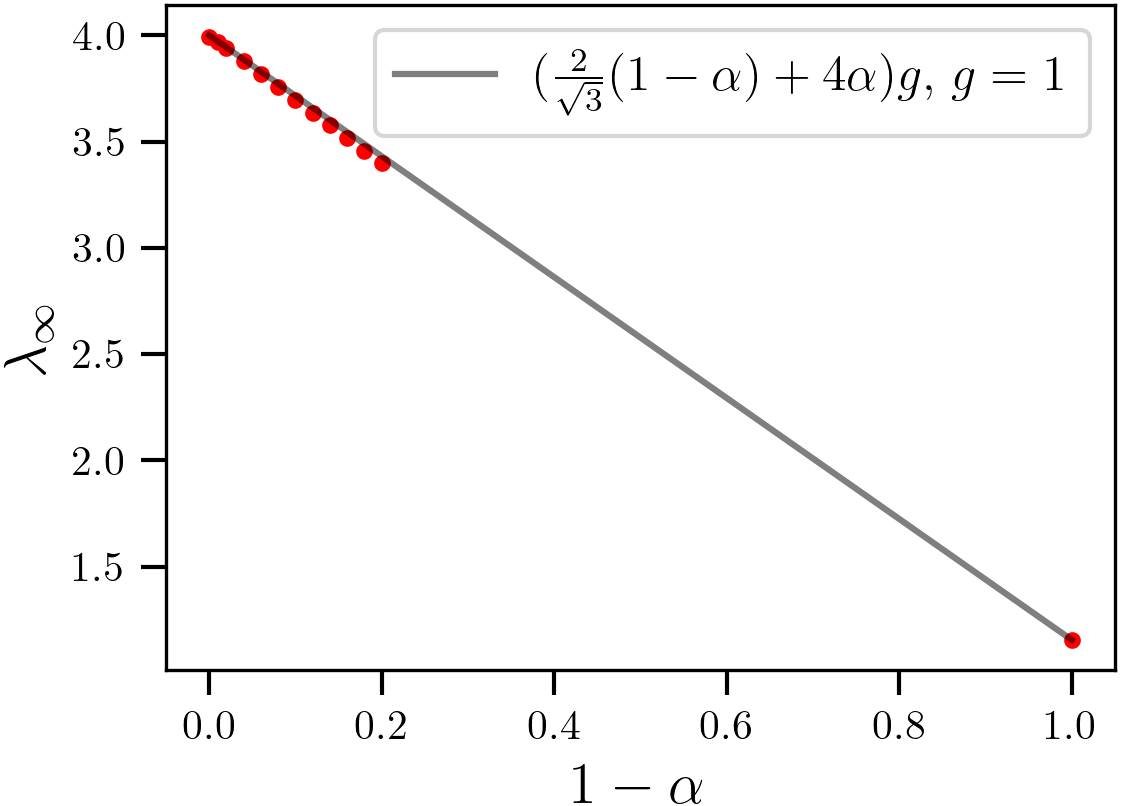}
  \caption{Behavior of $\lambda_\infty(\alpha,g)$ in the spherical model, for $g=1$. The red dots for $1-\alpha \in (0,0.2)$ correspond to the asymptotics of $\lambda(t)$ shown in Fig.~\ref{fig:diff_alphas} (\emph{Left}) in the main text, obtained by a numerical integration of the DMFT equations. The value of $\lambda_\infty$ at $1-\alpha=1$ is exact and given by Eq.~\eqref{eq supp: stationary order parameters in PC phase for spherical model}. The solid line is the conjectured linear behavior $ \lambda_\infty(\alpha,g=1)=2(1-\alpha)/\sqrt{3}+4\alpha$. }
  \label{fig supp: linear lambda infinity}
\end{figure*}

For the spherical model, a comparison between the dynamics and the complexity has been presented in the main text (see Fig.~\ref{fig:diff_alphas}) for $J=0$, where the system is in the paramagnetic chaotic phase. The complexity at the DMFT value $\lambda_\infty$ is $g$ independent, but not $\alpha$ independent. This suggests a conjecture for the asymptotic value $\lambda_\infty$, of the form $\lambda_\infty(
\alpha,g)=\eta(\alpha)g$. If we assume a linear behavior for $\eta$, we get $ \lambda_\infty(\alpha,g)=\left[\frac{2}{\sqrt{3}}(1-\alpha)+4\alpha\right]g$
where, to make the interpolation, we used knowledge of $\lambda_\infty$ for $\alpha=0, 1$. A numerical check of this conjectured linear behavior is given in Fig. \ref{fig supp: linear lambda infinity}. It remains an open problem to verify analytically whether this conjecture is true.

\subsection{Comparing complexity and dynamics: the confined model}
\label{app:dmft_vs_kac_confined}

We consider again the choices \eqref{app:spher_rnn_phi}, but for the confined case. We remind that in this case $q$ is a tunable parameter and $\lambda=q-\gamma$. The results of this analysis are discussed in the main text. In this appendix we present the essential steps of the derivation, in analogy to the spherical case discussed above.

\subsubsection{Purely non-conservative case: $\alpha=0$. }

\paragraph*{Isolated stable equilibrium and the FFP phase. }
As for the spherical model, also in this case for $\alpha=0$ the only stable equilibrium, if it exists, is the isolated ferromagnetic equilibrium. This isolated equilibrium has parameters
\begin{equation}\label{eq:DMFT_fix_CM}
  q_{\text{ffp}}=J+\gamma, \quad \quad \quad m_{\text{ffp}}=\pm \sqrt{\gamma + J} \frac{\sqrt{-2 \gamma g^2 + (J-2 g^2) J}}{J}, 
\end{equation}
that is linearly stable whenever $J>J_c=2 g (g + \sqrt{\gamma + g^2})$: this is precisely the regime of parameters which corresponds to the FFP phase identified by the DMFT solution, see \eqref{eq:sphTR}. Once more, the parameters \eqref{eq:DMFT_fix_CM} coincides with the values predicted by the asymptotic solution of the DMFT for the CM, see Eq.~\eqref{DMFT_fixFFCM}. Again, in the region $J>2 g (g + \sqrt{\gamma + g^2})$ unstable equilibria also exist and have a positive complexity; however, only the stable isolated equilibrium matters for the large time dynamics.\\

\paragraph*{Unstable equilibria and the chaotic phases. } As one lowers $J$ below $J_c$ and the system enters the chaotic phase (dynamically), the complexity calculation shows that the dynamical equations admit exponentially many equilibria, but they are \emph{all} linearly unstable: the complexity is non-negative only for $q<q_{\rm ms}=\gamma+2 g (g + \sqrt{\gamma + g^2})$. In this case, we consider fixed $q$, and optimize the {quenched} complexity over $\tilde{Q}$ and $m$, denoting again  with $\tilde{Q}_{\text{typ}}$ and $m_{\text{typ}}$ the values where the optimum is attained. \\
The optimization gives again a paramagnetic solution, $m_{\text{typ},P}=0,\tilde{Q}_{\text{typ},P}=0$, and as in the spherical case the corresponding quenched complexity coincides with the annealed one,
\begin{align}
 \Sigma(m=0,q)=\frac{1}{8} \left(-\frac{(\gamma - q)^2}{g^2 q} + 4\log2\right).
\end{align}
This curve has a maximum in $q=\gamma$ and goes to zero at $q=\gamma + g \left(2 \sqrt{\log2} \sqrt{\gamma + g^2 \log2} + g \log4\right)$. The second solution is ferromagnetic, and reads:
\begin{align}
&\tilde{Q}_\text{typ,F}=-\frac{J (2 \gamma + J - 2 q)}{2 g^2} - q,\\
&m_\text{typ,F}
=\pm\sqrt{\frac{\tilde{Q}_\text{typ,F} \left[-(\gamma - q)^2 (3q - \tilde{Q}_\text{typ,F}) \tilde{Q}_\text{typ,F} + 2g^2 \left(q^3 + 3q^2 \tilde{Q}_\text{typ,F} + q (\tilde{Q}_\text{typ,F})^2 - (\tilde{Q}_\text{typ,F})^3\right)\right]}
{2(q + \tilde{Q}_\text{typ,F}) \left[J (2\gamma + J - 2q) \tilde{Q}_\text{typ,F} + g^2 (q + \tilde{Q}_\text{typ,F})^2\right]}}.
\end{align}
The formula for the complexity is a bit cumbersome and so we avoid to report it explicitly.
For $q=(2 \gamma J + J^2)/[2 (J-g^2)]$ it holds $m_{\text{typ, F}}=0$, meaning that at this value of $q$ the optimization of the complexity over $m$ gives only the paramagnetic solution: ferromagnetic equilibria are present at smaller values of $q$, but the paramagnetic ones outnumber them. \\
As in the spherical case, the ferromagnetic quenched complexity  does not coincide with the annealed one, optimized over $m$ at the corresponding value of $q$. Indeed, the annealed complexity is optimized at $m_{\text{typ, F}}^A=\pm\sqrt{\frac{q (
 2 \gamma J + J^2 + 2 g^2 q - 2 J q)}{J (2 \gamma + J - 2 q)}}$, and the difference between the annealed and quenched complexities at the corresponding optimal values of parameters reads:
\begin{align}
D(q)\equiv \Sigma_A(m_\text{typ,F}^A,q)-\Sigma(m_\text{typ,F},q)=-\frac{(\gamma + J - q)^2 (J (2 \gamma + J - 2 q) + 2 g^2 q)^3)}{
 4 g^2 J^2 (2 \gamma + J - 2 q)^2 q (J (2 \gamma + J - 2 q) + 4 g^2 q)}.
\end{align}
 A plot of $D(q)$ is shown in the main text, see Fig.~\ref{fig:kac_rice_pics_confined} (\emph{Middle}). The difference $D(q)$ vanishes at $q=J+\gamma$ and $q=(2 \gamma J + J^2)/[2 (J-g^2)]$. Once more, the first value corresponds to the asymptotic value of $C_0$ selected by the dynamics in the ferromagnetic chaotic phase, see \eqref{DMFT_fixFFCM}. 
 However, one can verify that the magnetization and the overlap $\tilde{Q}$  between the most numerous equilibria at the shell $q=C_0=J+\gamma$ are
\begin{align}
m_\text{typ,F}(q=J+\gamma)=\pm\sqrt{(\gamma + J) (-2 \gamma g^2 - 2 g^2 J + J^2)}/J, \quad \quad 
\tilde{Q}_\text{typ,F}(q=J+\gamma)=-\gamma + \frac{1}{2} J (-2 + J/g^2),
\end{align}
at clear difference with the values of $m_\infty$ and $C_\infty$ characterizing the attractor manifold. 
 One can further try to impose that $\tilde{Q}(q=C_0)=C_\infty$ and then solve for the $m_\text{typ}$ such that $\partial_{\tilde{Q}}\tilde{\Sigma}_Q(m_\text{typ},q=C_0,\tilde{Q})|_{\tilde{Q}=C_\infty}=0$: the $m_\text{typ}$ found in this way does not coincide with the DMFT value $m_\infty$.\\
 As in the spherical case, in the paramagnetic chaotic region the complexity evaluated at the DMFT value for $q=C_0$ reads 
\begin{align}
    \Sigma\left(m=0,q=4g^2/3\right)=\frac{1}{6} (3\log2-1),
\end{align}
which is again independent of $g$.\\

\paragraph*{Ferromagnetic to Paramagnetic transition.} 
We can determine for which values of $J$ the maximal complexity corresponds to paramagnetic equilibria for \emph{all} values of $q$, i.e., when the ferromagnetic branch of the complexity collapses into the paramagnetic one. This value of $J$ can be found imposing that the point where the paramagnetic and ferromagnetic complexity branches cross is exactly the point where the paramagnetic complexity vanishes. Hence:
\begin{align*}
&\frac{2 \gamma J + J^2}{2 (J-g^2)}=\gamma + g \left(2 \sqrt{\log2} \sqrt{\gamma + g^2 \log2} + g \log4\right)\\
&\Rightarrow  J_{\pm}:=g \left(\sqrt{\ln 16} \pm \sqrt{\ln16-2}\right) \left(\sqrt{\gamma + g^2\ln2} + g \sqrt{\ln2} \right).
\end{align*}
This implies that a region of (large) $q$ exists in which ferromagnetic unstable equilibria are most abundant with respect to paramagnetic ones only  whenever $J_-<J<J_+$. In particular, $J_\Sigma=J_-$ is the black dotted line in Fig.~\ref{fig:dmft_phase} of the main text. This line lies below the critical line $J_\text{PF}$ of the dynamics transitions from a ferromagnetic chaotic phase, to a paramagnetic one. This means that in the paramagnetic chaotic phase (sufficiently close to the transition line), there are still values of $q$ for which the dominant equilibria are ferromagnetic unstable ones. The region where these ferromagnetic unstable equilibria exist extends up to $J_+$, which lies above $J_c$, meaning that in the ferromagnetic fixed point phase, there are also exponentially many ferromagnetic unstable equilibria.\\
In the shell selected asymptotically by the dynamics, $q=C_0=J+\gamma$, the equilibria transition from ferromagnetic to paramagnetic at 
\begin{align}
    \frac{2\gamma J+J^2}{2(J-g^2)}=J+\gamma\Rightarrow J=g \left(g + \sqrt{2 \gamma + g^2}\right),
\end{align}
which again does not correspond to the dynamical ferromagnetic-to-paramagnetic chaotic transition line $J_\text{PF}$.\\

\begin{figure*}
    \centering
    \includegraphics[width=1\textwidth, trim={4 4 4 4},clip]{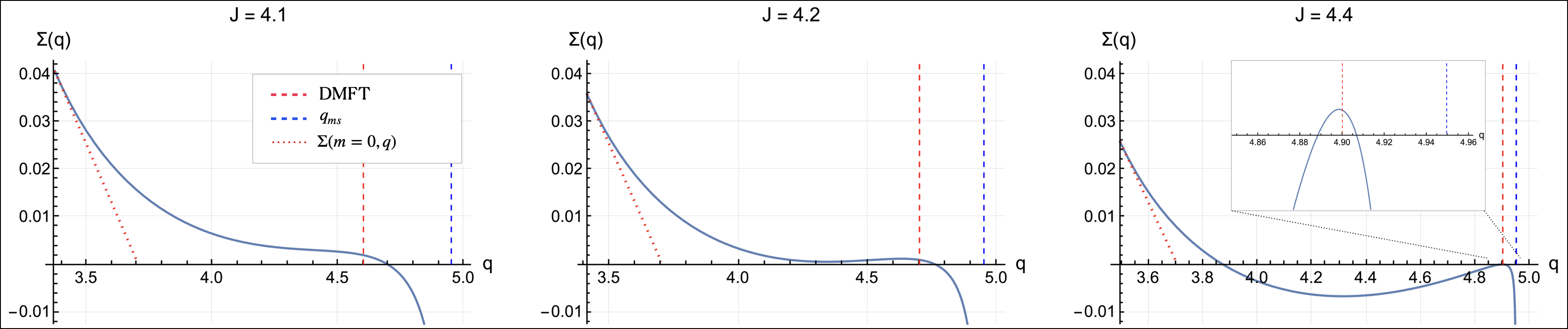}
    \caption{Complexity curves $\Sigma(q)=\Sigma(m_\text{typ}(q),q)$ of the confined model for $g=1$ and $\gamma=0.5$, in the ferromagnetic chaotic phase. Blue lines correspond to ferromagnetic equilibria, red dotted lines to paramagnetic ones. The blue dashed marks the marginal stability line $q_{\text{ms}}$, red dashed lines indicate the position of the dynamical order parameter $C_0$. From left to right, as $J$  increases the ferromagnetic complexity develops a local maximum. As $J\to J_c$, the local maximum converges to $q_\text{ms}$ and the complexity vanishes.}
    \label{app:fig:ferro_steps}
\end{figure*}

\begin{figure*}
    \centering
    \includegraphics[width=0.8\textwidth, trim={4 4 4 15},clip]{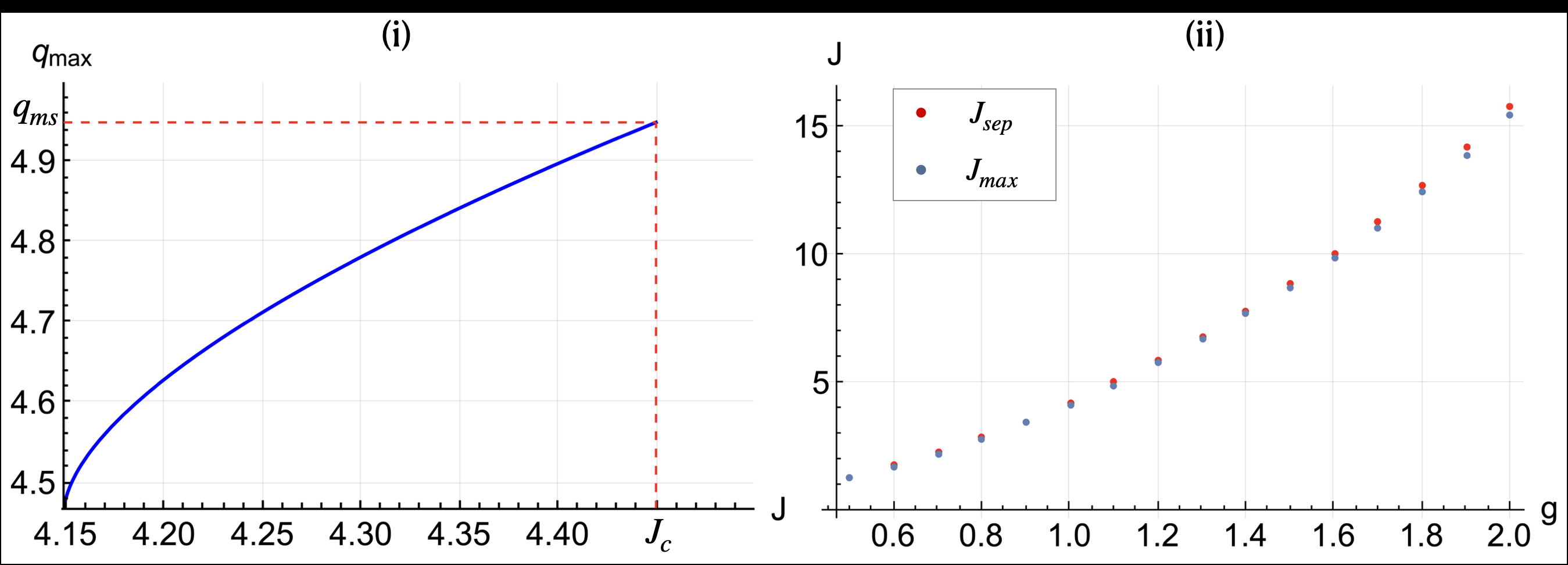}
    \caption{(i). Local maximum of the complexity $q_\text{max}$ as a function of $J$ for the confine model. At $J=J_c$, $q_\text{max}=q_\text{ms}$. (ii) Plots of $J_\text{max}$ (blue points) and $J_\text{sep}$ (red points) as a function of $g$. The former is the value of $J$ at which a local maximum of the complexity appears, the latter the value at which this local maximum becomes isolated (see third image in Fig.~\ref{app:fig:ferro_steps}). }
    \label{app:fig:q_max_vs_J}
\end{figure*}

\paragraph*{Ferromagnetic equilibria: discontinuous support of the complexity.}  
In this paragraph we investigate the behavior of the complexity for values of $J\in[J_-:=J_\Sigma,J_+]$, that is, when there are ferromagnetic unstable equilibria. In particular, we are interested in studying where the DMFT order parameter lies, and how the ferromagnetic branch of the complexity evolves with $J$ giving rise to the isolated ferromagnetic stable fixed point. In Fig.~\ref{app:fig:ferro_steps} we show how the complexity of ferromagnetic equilibria evolves in the chaotic phase as $J$ is increased towards $J_c$. 
The ferromagnetic complexity (blue curve) is monotonously decreasing for small $J$ (left panel); at larger $J$ it develops a local maximum (center panel) at a given value $q_\text{max}$; at larger $J$ (right panel) this local maximum belongs to  an  “island" of positive complexity, whose support is detached from the support of the other branch of positive complexity. Finally, as $J\to J_c$, this local maximum converges to $q_\text{max} \to q_{\rm ms}$, and the corresponding complexity vanishes: the system transitions to the Ferromagnetic Fixed Point phase. The value of $q_{\rm max}$ as a function of $J$ is presented in Fig.~\ref{app:fig:q_max_vs_J} (i), for the particular value of $g=1$. We see that $q_\text{max}$ increases monotonically up to $J=J_c$, where its value reaches $q_\text{ms}$. We can also compute, as a function of $g$, the value $J_\text{max}$ at which the local maximum appears, as well as the value $J_\text{sep}$ (“sep" for “separated") at which $q_\text{max}$ belongs to a region of positive complexity with support that is detached from the support of the remaining branch of ferromagnetic complexity. These are shown in Fig.~\ref{app:fig:q_max_vs_J}(ii), where the blue dots indicate $J_\text{max}$ and the red dots $J_\text{sep}$. 

\begin{figure*}
    \centering
    \includegraphics[width=0.5\textwidth, trim={1 2 2 2},clip]{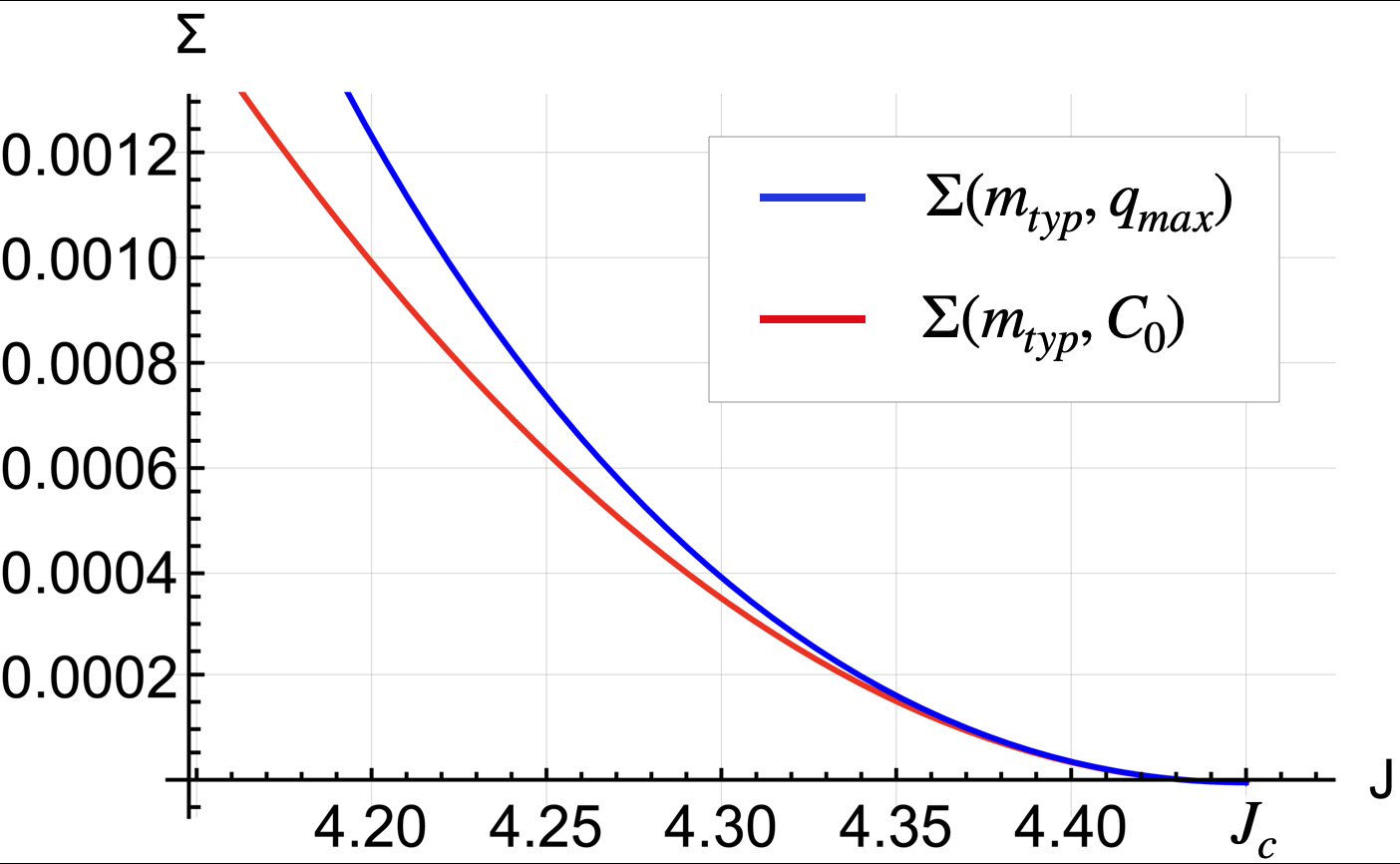}
    \caption{Complexity curves at $q_\text{max}$ (blue) and at the dynamical order parameter $C_0$ (red) in the ferromagnetic chaotic phase of the confined model, as a function of $J$. The complexity is optimized over $m$. At $J=J_c$ both curves collapse to zero, concomitantly with the transition to the FFP phase.}
    \label{app:fig:dmft_vs_qmax}
\end{figure*}

We remark that the order parameter $q=C_0$ selected asymptotically by the dynamics always lies within a zone of positive complexity, even for values of parameters for which the support of the ferromagnetic complexity is disconnected. This can be seen in  Fig.~\ref{app:fig:ferro_steps}. In Fig.~\ref{app:fig:dmft_vs_qmax} we plot the complexities $\Sigma(m_\text{typ}(C_0),C_0)$ and $\Sigma(m_\text{typ}(q_\text{max}),q_\text{max})$, that is, at the dynamical order parameter (red) and at the local maximum of the complexity (blue). We see that they differ, the blue one being slightly higher (obviously), than the red one. It is interesting to notice that at $C_0$ the complexity is always positive and close to the local maximum. Eventually, the two values diverge as the local maximum disappears when lowering $J$.

\subsubsection{Partially non-conservative case: $\alpha\neq 0$.}
The paramagnetic complexity of the confined model for non-zero $\alpha$ reads:
\begin{align}
   \Sigma(m=0,q)= -\frac{(q-\gamma)^2}{8 g^2 q(1 + 3 \alpha + 2 \alpha^2)} + \frac{\log2}{2},
\end{align}
and vanishes at 
\begin{align}
\label{app:eq:root_para_conf}
q=\gamma + g \left( g (1 + \alpha) (1 + 2\alpha) \log4 + 2 \sqrt{(\log2 + \alpha (\alpha \log4 + \log8)) 
\left(c + g^2 (\log2 + \alpha (\alpha \log4 + \log8))\right)} \right).
\end{align}
The value of $q$ at which equilibria become marginally stable, and above which are stable, is given by 
\begin{align}
\label{app:eq:marg_conf}
q_{\text{ms}}=\gamma + 2 g (1 + \alpha) \left(g + g \alpha + \sqrt{
    \gamma + g^2 (1 + \alpha)^2}\right).
\end{align}
By imposing that Eqs.~\eqref{app:eq:marg_conf} and \eqref{app:eq:root_para_conf} are equal, one finds that marginal equilibria appear at the same $\alpha=\alpha_c$ already determined for the spherical case. The stable (zero-complexity) ferromagnetic fixed point is characterized by
\begin{align}
    q_{\text{ffp}}=\frac{J (\gamma + J)}{J - 4 g^2 \alpha}, \quad \quad \quad   m_{\text{ffp}}=\frac{\sqrt{(\gamma + J) (-2 \gamma g^2 + J (J - 2 g^2 (1 + 2 \alpha)))}}{{J - 4 g^2 \alpha}},
\end{align}
and it becomes unstable at $J_c=2 g \left(g + g \alpha + \sqrt{c + g^2 (1 + \alpha)^2}\right)$. In this case as well one can show that for $q=q_{\text{ffp}}$ in the unstable ferromagnetic region, the quenched and annealed complexities coincide, upon optimization of $m$. \\

We conclude by generalizing our conjecture on the $\alpha$ dependence of the DMFT parameters to the confined model. In the paramagnetic chaotic phase, we know from the asymptotic DMFT solution that $C_0(\alpha=0)=\gamma + \frac{2}{3} g \left(g + \sqrt{3 \gamma + g^2}\right)$ and $C_0(\alpha=1)=\gamma + 4 g \left(2 g + \sqrt{\gamma + 4 g^2}\right)$. At these values, the complexities of equilibria read: $\Sigma(m=0,q=C_0(\alpha=0))=(\log8-1)/6$ and $\Sigma(m=0,q=C_0(\alpha=1))=(\log8-2)/6$. These functions $\gamma,g$ independent, but depend on $\alpha$. We conjecture that that for general $\alpha$, the dynamics selects values of $q=C_0$ that change with $\gamma,g$ in such a way that the corresponding complexity remains constant. Imposing that $\Sigma$ is $\gamma$-independent leads to 
\begin{align}
    -2 q(g,\gamma,\alpha) + [\gamma + q(g,\gamma,\alpha)]\partial_\gamma q(g,\gamma,\alpha)=0,
\end{align}
which is solved by 
\[
C_0(\gamma,g,\alpha)=\gamma+\frac{1}{2}\left(\nu(g,\alpha) + \sqrt{\nu(g,\alpha)} \sqrt{4 \gamma + \nu(g,\alpha)}\right)
\]
with $\nu$ a function of $g,\alpha$. Imposing also the invariance of the complexity with respect to $g$ we obtain $\nu(g,\alpha)=g^2 B(\alpha)$. Assuming a linear behavior $B(\alpha)=a+b\,\alpha$ as for the spherical model, we would have $a=\frac{4}{3}$ and $b=\frac{44}{3}$ using the boundary conditions in $\alpha=0$ and $\alpha=1$. 

\section{Detailed derivation of the complexity}\label{eq:Conto}

\subsection{The replicated Kac-Rice formalism}
We compute the quenched complexity of equilibria by means of the replicated Kac-Rice formalism, following the procedure outlined in \cite{ros2019complex, ros2019complexity} (see also \cite{ros2023high} for a review). The random variable $\mathcal{N}(m,\lambda,q)$ counting the number of fixed points of the dynamical equations for a fixed realization of the disorder admits the integral representation:
\begin{align}\label{eq:KR}
    \mathcal{N}(m,\lambda,q)=\int_{\mathbb{R}^N} d{\bf x}\,\Delta({\bf x})|\det \mathcal{H}({\bf x})|\delta({\bf f}({\bf x})-\lambda\,{\bf x})
\end{align}
where 
\begin{align*}
&\Delta({\bf x})=\delta\left(\sum_ix_i-Nm\right)\delta({\bf x}^2-Nq),\quad \quad 
[\mathcal{H}({\bf x})]_{ij}=\frac{\partial f_i({\bf x})}{\partial x_j}-\lambda\delta_{ij}.
\end{align*}
The $N \times N$ matrix $\mathcal{H}({\bf x})$ is a Jacobian that arises due to the non-linearity of the constraints $ {\bf f}({\bf x})-\lambda\,{\bf x}=0$. 
The integers powers $\mathcal{N}^n$ are obtained multiplying the integral representation $n$ times: they involve therefore $n$ integration variables, ${\bf x}^a$ with $a=1, \cdots, n$, which we refer to as replicas. Taking the average with respect to the random field ${\bf f}$, one gets the Kac-Rice formula for the moments of $\mathcal{N}$, which reads:
\begin{align}\label{eq:kRM}
\mathbb{E} [\mathcal{N}^n]=\int \prod d{\bf x}^a \prod_a\Delta({\bf x}^a)\mathbb{E}\left[\prod_a |\det\mathcal{H}({\bf x}^a)|\bigg| \begin{subarray}{l}
  {\bf f}^a({\bf x}^a)=\lambda{\bf x}^a,\\
  a=1,\ldots,n
   \end{subarray} \right] \, \mathbb{E}\left[\prod_a\delta({\bf f}({\bf x}^a)-\lambda{\bf x}^a)\right].
\end{align}
In this expression, the expectation value of the product of determinants of the matrices $\mathcal{H}({\bf x}^a)$ is conditioned to ${\bf f}({\bf x}^a)=\lambda {\bf x}^a$.
This formula arises taking the average over the randomness of the $n$-th power of the integral representation \eqref{eq:KR}, and making use of the law of total expectation, which implies:
\begin{equation}\label{eq:TotalLaw}
  \mathbb{E}\left[\prod_a\delta({\bf f}({\bf x}^a)-\lambda{\bf x}^a) \; \prod_a |\det\mathcal{H}({\bf x}^a)| \right]=  \mathbb{E}\left[\prod_a |\det\mathcal{H}({\bf x}^a)|\bigg| 
  \begin{subarray}{l}
  {\bf f}^a({\bf x}^a)=\lambda{\bf x}^a,\\
  a=1,\ldots,n
   \end{subarray}\right]  \mathbb{E}\left[\prod_a\delta({\bf f}({\bf x}^a)-\lambda{\bf x}^a)\right].
\end{equation}
This identity allows to replace the expectation of the product of determinants and delta functions with the product of the conditional expectation $\mathbb{E}\left[\prod_a |\det\mathcal{H}({\bf x}^a)|\bigg| 
  \begin{subarray}{l}
  {\bf f}^a({\bf x}^a)=\lambda{\bf x}^a,\\
  a=1,\ldots,n
   \end{subarray}\right] $ with the marginal probability $\mathbb{E}\left[\prod_a\delta({\bf f}({\bf x}^a)-\lambda{\bf x}^a)\right]$. This general identity follows from the law of total probability, which implies that the joint distribution $P(x,y)$ of two random variables $X, Y$ satisfies  $P(x,y)= P(x | Y=y) P(y)$, where $P(x | Y=y)$ is the  distribution of $X$ conditional to $Y=y$,  and  $P(y)$ the marginal \footnote{Indeed, this relation implies that $\mathbb{E}[X \cdot \delta(Y-y^*)]= \int dx dy P(x|y) P(y) \, x \cdot \delta(y-y^*) = \int dy [\int dx x P(x|y)] P(y) \delta(y-y^*)= \mathbb{E}[X |Y=y^*] P(y^*)$. Eq. \eqref{eq:TotalLaw} is a realization of these identities, where the forces ${\bf f}^a({\bf x}^a)$ play the role of the variable $Y$.}. The splitting of the expectation value is quite convenient both for technical reasons, and because each of the two terms in the right-hand side of \eqref{eq:TotalLaw} is interpretable: the second term is the probability that the chosen configurations ${\bf x}^a$ are all equilibria of the equations of motion; the first term is more interesting: to compute it, one has to determine the statistics of the Jacobian matrices $\mathcal{H}({\bf x}^a)$ at configurations that are not arbitrary, but are equilibria (this is enforced  by conditioning to the values of the random forces). This term therefore gives access to the information on the linear stability of these equilibria. 
We also remark that the annealed complexity can be derived setting $n=1$ in \eqref{eq:kRM}. \\
The expectation values in \eqref{eq:kRM} are functions of the configurations $ {\bf x}^a$; below, we show that for large $N$ the dependence on these configurations enters only through the overlaps $Q^{ab}=N^{-1}  {\bf x}^a \cdot {\bf x}^b$ between the different replicas $a \neq b$. This implies a huge dimensionality reduction in the integral \eqref{eq:kRM}, which can be represented in the form:
\begin{align}\label{eq:ActionQ}
\mathbb{E} [\mathcal{N}^n(m, \lambda, q)]=\int \prod_{a < b} dQ^{ab} e^{N n \tilde{\Sigma} \tonde{m,\lambda,q,Q^{ab}}+ o(Nn)}
\end{align}
for some function $\tilde{\Sigma} \tonde{m,\lambda,q,Q^{ab}}$. The leading order exponential behavior of this quantity can be determined via a saddle-point on the variables $Q^{ab}$. We solve the resulting problem within the RS (Replica Symmetric) ansatz, which corresponds to setting
\begin{equation}
    Q^{ab}= q \delta_{ab}+ \tilde Q (1-\delta_{ab}), \quad \quad \tilde{\Sigma} \tonde{m,\lambda,q,Q^{ab}} \to  \tilde{\Sigma}(m,\lambda,q,\tilde{Q}).
\end{equation}
 As a consequence of the saddle-point calculation and of the limiting formula \eqref{app.SQ}, the quenched complexity within the RS framework is given in terms of the variational problem \eqref{eq.compvar}.
 In the following subsections, we determine the form of the function $\tilde{\Sigma}(m,\lambda,q,\tilde{Q})$, by studying separately each of the three terms appearing in the integrand in \eqref{eq:kRM}.

\subsection{Joint probability that all replicas are equilibria}
Under the constraints imposed on the ${\bf x}^a$, the probability that all the configurations ${\bf x}^a$ are equilibria with prescribed values of $m, q$ and $\lambda$ reads:
\begin{align}
\label{app:proba_term_initial}
\begin{split}
\mathbb{P}:=\mathbb{E}\left[\prod_a\delta({\bf f}({\bf x}^a)-\lambda{\bf x}^a)\right]&=\frac{1}{\sqrt{(2\pi)^{Nn}\det\hat{C}}}\int \prod_a d{\bf f}^a\delta(-\lambda{\bf x}^a+{\bf f}^a) e^{-\frac{1}{2}\sum_{a,b}({\bf f}^a-Jm\mathbf{1})[\hat{C}^{-1}]^{ab}({\bf f}^b-Jm\mathbf{1})}\\
    &=\frac{1}{\sqrt{(2\pi)^{Nn}\det\hat{C}}}e^{-\frac{1}{2} \sum_{a,b}\tonde{\lambda {\bf x}^a - J m {\bm 1}} [\hat C^{-1}]^{ab}\tonde{\lambda {\bf x}^b - J m {\bm 1}}},
\end{split}
\end{align}
where we have used the fact that the field ${\bf f}({\bf x})$ is Gaussian with the statistics \eqref{app:eq:def_Covariance}, and where we defined the covariance matrix
\begin{align}\label{eq:CovMat}
    \hat{C}^{ab}_{ij}=\text{Cov}[f_i({\bf x}^a),f_j({\bf x}^b)]= \delta_{ij} \Phi_1\tonde{\frac{{\bf x}^a \cdot {\bf x}^b}{N}}+ \frac{x^a_j \, x^b_i}{N} \Phi_2\tonde{\frac{{\bf x}^a \cdot {\bf x}^b}{N}}.
\end{align}
{The joint probability \eqref{app:proba_term_initial} in principle depends on all the configurations ${\bf x}^a$; we now show that, due to the isotropic structure of the covariances of the random fields, the term \eqref{app:proba_term_initial} can actually be written solely as a functions of the overlaps $Q^{ab}= N^{-1} {\bf x}^a \cdot {\bf x}^b$ between the replicas. To show this, we follow the procedure introduced in \cite{ros2019complex}.
The key observation for computing the expression in the exponent of \eqref{app:proba_term_initial} is that there is no need to invert the full $N n \times N n$ covariance matrix $\hat{C}$; instead, it is sufficient to determine its inverse within an appropriately chosen subspace.} To proceed, we decompose the matrix $\hat{C}$ as the sum of a diagonal and non-diagonal part in replica space:
\begin{equation}
    \hat C= \hat D + \hat O, \quad \quad \hat C^{-1}= \hat D^{-1}[\mathbb{I}+ \hat O \hat D^{-1}]^{-1}
\end{equation}
where $\mathbb{I}$ is the $N n \times N n$ identity matrix, and
\begin{equation}
\begin{split}
     D^{ab}_{ij}&= \delta_{ab}\quadre{\delta_{ij} \Phi_1(q) + \frac{x_j^a x_i^a}{N} \Phi_2(q)}\\
     O^{ab}_{ij}&=(1-\delta_{ab})\quadre{\delta_{ij} \Phi_1(Q^{ab})+ \frac{x_j^a x_i^b}{N} \Phi_2(Q^{ab})}.
\end{split}
\end{equation}
We exploit the notation \eqref{eq:Notation}, and additionally define:
\begin{align}\label{eq:not2}
    \Phi_{i}(Q^{ab})\equiv \Phi_i^{ab},\quad\quad i\in\{1,2\},
\end{align}
with $\Phi_i^{aa}=\Phi_i^q$. 
By means of the Shermann-Morrison formula, we get:
\begin{equation}
\begin{split}
  &  [\hat D^{-1}]^{ab}_{ij}= \delta_{ab} \quadre{(\Phi_1^q)^{-1} \delta_{ij}- \eta \frac{x_j^a x_i^a}{N} }, \quad \quad \eta=\frac{\Phi_2^q/\Phi_1^q}{\Phi_1^q+ \Phi_2^q q}.
    \end{split}
\end{equation} 
We now introduce a set of vectors spanning a sub-space (of the $(Nn)$- dimensional space on which the matrix $\hat C$ acts) that is closed under the action of $\hat C$, and which is relevant to reconstruct the products appearing in the exponent of Eq.~\eqref{app:proba_term_initial}. In the following, we implement explicitly the RS assumption on the overlap matrix. We define the following three $Nn$-dimensional vectors:
\begin{equation}
\begin{split}
    &{\bm \xi}_1:= ({\bf x}^1, \cdots, {\bf x}^n),\\
    &{\bm \xi}_2:=({\bm 1}, \cdots, {\bm 1}),\\
    &{\bm \xi}_3:= \tonde{\sum_{b \neq 1} {\bf x}^b, \cdots, \sum_{b \neq n} {\bf x}^b},
    \end{split}
\end{equation}
and determine the action of the matrix $\hat O \hat D^{-1}$ on these vectors. From Eq.\eqref{app:proba_term_initial} it appears that the expression at the exponent can be expressed in terms of the action of $\hat C^{-1}$ on these vectors, as  $
-\frac{1}{2} \tonde{\lambda {\bm \xi}_1 - J m\, {\bm \xi}_2 } \hat C^{-1}\tonde{\lambda {\bm \xi}_1 - J m\, {\bm \xi}_2}$, 
from which it is evident that determining the action of $\hat{C}$ on this set of vectors suffices for our purposes.
The action on $\hat{D}^{-1}$  is given by the following expressions:
\begin{equation}
\begin{split}
    & [\hat D^{-1}{\bm \xi}_1]^a= (\Phi_1^q)^{-1}{\bm \xi}_1^a - \eta q {\bm \xi}_1^a,\\
    &[\hat D^{-1}{\bm \xi}_2]^a=(\Phi_1^q)^{-1}{\bm \xi}_2^a- \eta m{\bm \xi}_1^a,\\
    &[\hat D^{-1}{\bm \xi}_3]^a= (\Phi_1^q)^{-1}{\bm \xi}_3^a- \eta {\bm \xi}_1^a \sum_{b \neq a} Q^{ab}, 
    \end{split}
\end{equation}
while the action of $\hat{O}$ reads:
\begin{equation}
\begin{split}
    & [\hat O{\bm \xi}_1]^a_i=\sum_{b\neq a}\left[\Phi_1^{ab}+\Phi_2^{ab}Q^{ab}\right]x_i^b, \\
    &[\hat O{\bm \xi}_2]^a_i=\sum_{b\neq a}\left[\Phi_1^{ab}+m\Phi_2^{ab}x_i^b\right],\\
    &[\hat O{\bm \xi}_3]^a_i=\sum_{b\neq a}\left[\Phi_1^{ab}\sum_{c\neq a}x_i^c+x_i^b\left(\Phi_2^{ab}\sum_{c\neq a}Q^{ac}+q\Phi_2^{ab}-\Phi_1^{ab}-Q^{ab}\Phi_2^{ab}\right)+\Phi_1^{ab}x_i^a\right].
    \end{split}
\end{equation}
The RS ansatz assumes the form $Q^{ab}= q \delta_{ab} + \tilde{Q} (1-\delta_{ab})$ for the overlap matrix. Within this assumption, the previous expressions simplify, taking the form:
\begin{equation}
\label{app:action_B_xi}
\begin{split}
    &\hat D^{-1}{\bm \xi}_1= [(\Phi_1^q)^{-1}- \eta q]{\bm \xi}_1,\\
    &\hat D^{-1}{\bm \xi}_2=(\Phi_1^q)^{-1}{\bm \xi}_2- \eta m{\bm \xi}_1,\\
    &\hat D^{-1}{\bm \xi}_3= (\Phi_1^q)^{-1}{\bm \xi}_3- \eta(n-1)\tilde{Q}{\bm \xi}_1,
    \end{split}
\end{equation}
and:
\begin{equation}
\begin{split}
     \hat O{\bm \xi}_1&=\left[\tilde{\Phi}_1+\tilde{\Phi}_2\tilde{Q}  \right]{\bm\xi}_3, \\
    \hat O{\bm \xi}_2&=(n-1)\tilde{\Phi}_1{\bm\xi}_2+m\tilde{\Phi}_2{\bm\xi}_3,\\
    \hat O{\bm \xi}_3&=(n-1)\tilde{\Phi}_1{\bm\xi}_3+\left(\tilde{\Phi}_2\tilde{Q}(n-1)+q\tilde{\Phi}_2-\tilde{\Phi}_1-\tilde{Q}\tilde{\Phi}_2\right){\bm\xi}_3+(n-1)\tilde{\Phi}_1{\bm\xi}_1\\
    &=\left[(n-2)\tilde{\Phi}_1+\tilde{\Phi}_2\left(\tilde{Q}(n-2)+q\right)\right]{\bm\xi}_3+(n-1)\tilde{\Phi}_1{\bm\xi}_1.
    \end{split}
\end{equation}
Combining these formulas, we get:
\begin{align}
\label{app:action_xi_OD}
\begin{split}
\hat{O}\hat{D}^{-1}{\bm\xi}_1&=[(\Phi_1^q)^{-1}-\eta q][\tilde{\Phi}_1+\tilde{\Phi}_2\tilde{Q}]{\bm\xi}_3\\
\hat{O}\hat{D}^{-1}{\bm\xi}_2&=(\Phi_1^q)^{-1}\tilde{\Phi}_1(n-1){\bm\xi}_2+\left[(\Phi_1^q)^{-1}m\tilde{\Phi}_2
-\eta m[\tilde{\Phi}_1+\tilde{\Phi}_2\tilde{Q}]\right]{\bm\xi}_3\\
\hat{O}\hat{D}^{-1}{\bm\xi}_3&=(\Phi^q_1)^{-1}(n-1)\tilde{\Phi}_1{\bm\xi}_1 +\\&\Bigg\{-\eta(n-1)\tilde{Q}\left[\tilde{\Phi}_1+\tilde{\Phi}_2\tilde{Q}\right]+(\Phi^q_1)^{-1}\left[(n-2)\tilde{\Phi}_1+\left(\tilde{Q}(n-2)+q\right)\tilde{\Phi}_2\right]\Bigg\}{\bm\xi}_3.
\end{split}
\end{align}
In order to invert the matrix $\mathbb{I}+ \hat O \hat D^{-1}$, it is convenient to express it in a basis of the subspace spanned by ${\bm\xi}_1,{\bm\xi}_2,{\bm\xi}_3$, that is made of orthogonal vectors. We thus consider the following orthogonal basis vectors:
\begin{equation}
\label{app:def_v}
    \begin{split}
        {\bf v}_1&= \alpha_1{\bm \xi}_1,\\
       {\bf v}_2&= \alpha_2\tonde{{\bm  \xi}_2- \frac{m}{q} {\bm \xi}_1},\\
       {\bf v}_3&= {\bm \xi}_3 - \alpha_3{\bm \xi}_1 - \alpha_4 {\bm \xi}_2,
    \end{split}
\end{equation}
with
\begin{align}
    \begin{split}
        &\alpha_1=\frac{1}{\sqrt{N  n q}},\\
        &\alpha_2=\frac{\sqrt{q}}{\sqrt{N n (q-m^2)}}, \\
        &\alpha_3=\frac{(n-1) (\tilde{Q}-m^2 )}{q-m^2 },\\
        &\alpha_4=\frac{m (n-1) (q - \tilde{Q})}{q-m^2}.
    \end{split}
\end{align}
Notice that we did not normalize the last vector  ${\bf v}_3$, as the final expression we aim to compute depends only on the components of the matrix along the vectors  ${\bf v}_1$ and ${\bf v}_2$, making the normalization of ${\bf v}_3$ unnecessary. We denote by $\hat{M}^{{\bm\xi}}$ the $3 \times 3$ matrix expressing the action of  $\hat{M}:=\mathbb{I}+\hat{O}\hat{D}^{-1}$ on the subspace spanned by the vectors ${\bm\xi}$, such that $\hat{M} {\bm \xi}_k=\sum_{l=1}^3[\hat{M}^{{\bm\xi}}]_{lk}{\bm \xi}_l$ for $k=1, 2,3$. The components of this matrix can be easily read from Eq.\eqref{app:action_xi_OD}. We also denote by $P$ the $3 \times 3$ matrix encoding for the change of basis from the vectors ${\bm\xi}$ to ${\bf v}$, meaning that ${\bf v}_k=\sum_{l=1}^3 P_{lk} \, {\bm \xi}_l$ $k=1, 2,3$. We have:
\[
P=\begin{pmatrix}
\frac{1}{\sqrt{n \, N \, q}} & -\frac{m}{ \sqrt{n \, N \,q \left(q-m^2\right)}} & -\frac{\left(n-1\right) \, \left(m^2 - \tilde{Q}\right)}{m^2 - q} \\
0 & \frac{\sqrt{q}}{\sqrt{n \, N \, \left(q-m^2\right)}} & \frac{m \, \left(n-1\right) \, \left(q - \tilde{Q}\right)}{m^2 - q} \\
0 & 0 & 1
\end{pmatrix}
\]
which gives us an inverse:
\[
P^{-1}=\begin{pmatrix}
\sqrt{n  N  q} & m\sqrt{\frac{nN}{q}} & (n-1)\tilde{Q}\sqrt{\frac{nN}{q}} \\
0 & \sqrt{\frac{n  N  (q-m^2)}{q}} & m(n-1)(q-\tilde{Q})\sqrt{\frac{nN}{q(q-m^2)}} \\
0 & 0 & 1
\end{pmatrix},
\]
such that ${\bm \xi}_k=\sum_{l=1}^3 P^{-1}_{lk} \, {\bf v}_l$. Then the action of $\hat M$ on the subspace spanned by the vectors ${\bf v}$ is obtained as:
\begin{align}
    \hat{M}^{{\bf v}}=P^{-1}\hat{M}^{{\bm\xi}}P,
\end{align}
meaning that $\hat M {\bf v}_k= \sum_{l=1}^3 [\hat{M}^{{\bf v}}]_{lk} {\bf v_l}$ for $k=1, 2,3$.
Therefore
\begin{align}
    [\hat{M}^{{\bf v}}]^{-1}=P^{-1}[\hat{M}^{{\bm\xi}}]^{-1}P,
\end{align}
and by plugging the expressions for $\hat{M}^{\bm\xi}$ and $P$, we get:
\begin{align}\label{eq:Ys}
    \begin{split}
    Y_{11}:&=[\hat{M}^{{\bf v}}]^{-1}_{11}\Big|_{n=0}=
    \frac{(\Phi_1^q + q \Phi_2^q) (q \Phi_1^q - 
   2 q \tilde{\Phi}_1 + 
   \tilde{Q} \tilde{\Phi}_1 + (q - \tilde{Q})^2 \tilde{\Phi}_2)}{q\mathcal{A}}\\
   Y_{12}:&=[\hat{M}^{{\bf v}}]^{-1}_{12}\Big|_{n=0}=\frac{m (q - \tilde{Q}) ( \Phi_1^q + q \Phi_2^q) (\tilde{\Phi}_1 + 
   \tilde{Q} \tilde{\Phi}_2)}{q\mathcal{A}\sqrt{q-m^2}}\\
   Y_{21}:&=[\hat{M}^{{\bf v}}]^{-1}_{21}\Big|_{n=0}=\frac{m (q - \tilde{Q})\Phi_1^q (\tilde{\Phi}_1 + 
   \tilde{Q} \tilde{\Phi}_2)}{q\mathcal{A}\sqrt{q-m^2}}\\
   Y_{22}:&=[\hat{M}^{{\bf v}}]^{-1}_{22}\Big|_{n=0}=
   \frac{q\,\Phi_1^q}{(\Phi_1^q - \tilde{\Phi}_1)(q-m^2)}
   -\frac{\Phi_1^q\,m^2 (q (\Phi_1^q + q \Phi_2^q) - 
   \tilde{Q}(\tilde{\Phi}_1 + \tilde{Q} \tilde{\Phi}_2))}{q\mathcal{A}(q-m^2)}
    \end{split}
\end{align}
where
\begin{align}
\begin{split}
    \mathcal{A}=(\Phi_1^q)^2 + (\tilde{\Phi}_1)^2 - 2 q \tilde{\Phi}_1 \Phi_2^q + (q - \tilde{Q})^2 \Phi_2^q \tilde{\Phi}_2 + \tilde{Q} \tilde{\Phi}_1 (\Phi_2^q + \tilde{\Phi}_2) + \Phi_1^q (-2 \tilde{\Phi}_1 - 2 \tilde{Q} \tilde{\Phi}_2 + q (\Phi_2^q + \tilde{\Phi}_2)).
\end{split}
\end{align}
In these expressions, we have already taken the limit $n \to 0$ since we are interested in the linear term (in $n$) of the expansion of the exponent in \eqref{app:proba_term_initial}. As we shall show below, a multiplicative factor of order $n$ will be provided by the  normalization of ${\bf v}_1$, allowing us to set $n=0$ in the matrix elements. Now, recall that $\hat{C}^{-1}=\hat{D}^{-1}\hat{M}$. We have the action of $\hat{M}$ on the ${\bf v}$ basis, and so it remains to find the action of $\hat{D}^{-1}$. This is easily done from Eq.~\eqref{app:action_B_xi} and \eqref{app:def_v}. Using that  $\hat{D}$ is symmetric, we find:
\begin{align}
    \begin{split}
        &{\bf v}_1^\top\hat{D}^{-1}=[(\Phi_1^q)^{-1}- \eta q]{\bf v}_1^\top\\
        &{\bf v}_2^\top\hat{D}^{-1}=(\Phi_1^q)^{-1}{\bf v}_2^\top.
    \end{split}
\end{align}
The terms needed to compute the exponent in \eqref{app:proba_term_initial} are
\begin{align}
\begin{split}
U_{11}:= \lim_{n \to 0}\frac{{\bm\xi}_1^\top\hat{C}^{-1}{\bm\xi}_1}{N n}&=\lim_{n \to 0} \frac{1}{\alpha_1^2}\frac{{\bf v}_1^\top\hat{D}^{-1}\hat{M}^{-1}{\bf v}_1}{N n }= q[(\Phi_1^q)^{-1}-\eta q]Y_{11}\\
\end{split}
\end{align}
and 
\begin{align}
    \begin{split}
      U_{22}:&=\lim_{n \to 0} \frac{{\bm\xi}_2^\top\hat{C}^{-1}{\bm\xi}_2}{N n}=\lim_{n \to 0}  \frac{1}{N n}\left(\frac{{\bf v}_2}{\alpha_2}+\frac{m}{q}\frac{{\bf v}_1}{\alpha_1}\right)^\top\hat{C}^{-1}\left(\frac{{\bf v}_2}{\alpha_2}+\frac{m}{q}\frac{{\bf v}_1}{\alpha_1}\right)\\
        &=\lim_{n \to 0} \frac{1}{Nn} \quadre{ \frac{1}{\alpha_2^2}{\bf v}_2\hat{C}^{-1}{\bf v}_2+\frac{m}{q}\frac{1}{\alpha_1\alpha_2}\tonde{{\bf v}_2\hat{C}^{-1}{\bf v}_1+{\bf v}_1\hat{C}^{-1}{\bf v}_2}+\frac{m^2}{q^2}\frac{1}{\alpha_1^2}{\bf v}_1\hat{C}^{-1}{\bf v}_1}\\
        &= 
       \frac{q-m^2}{q}(\Phi_1^q)^{-1}Y_{22}+\frac{m \sqrt{q-m^2}}{q}\left((\Phi_1^q)^{-1}Y_{21}+[(\Phi_1^q)^{-1}- \eta q]Y_{12}\right)+\frac{m^2}{q}[(\Phi_1^q)^{-1}- \eta q]Y_{11}
    \end{split}
\end{align}
and 
\begin{align}
\begin{split}
U_{12}:&=\lim_{n \to 0}\frac{1}{Nn} \tonde{{\bm\xi}_2^\top\hat{C}^{-1}{\bm\xi}_1+{\bm\xi}_1^\top\hat{C}^{-1}{\bm\xi}_2}=\lim_{n \to 0}\frac{1}{Nn} \quadre{\frac{1}{\alpha_1}\left(\frac{{\bf v}_2}{\alpha_2}+\frac{m}{q}\frac{{\bf v}_1}{\alpha_1} \right)\hat{C}^{-1}{\bf v}_1+\frac{1}{\alpha_1}{\bf v}_1\hat{C}^{-1}\left(\frac{{\bf v}_2}{\alpha_2}+\frac{m}{q}\frac{{\bf v}_1}{\alpha_1} \right)}\\
&=\lim_{n \to 0}\frac{1}{Nn}\frac{1}{\alpha_1\alpha_2}\left[{\bf v}_2\hat{C}^{-1}{\bf v}_1+{\bf v}_1\hat{C}^{-1}{\bf v}_2\right]+\lim_{n \to 0}\frac{1}{Nn}\frac{2}{\alpha_1^2}\frac{m}{q}{\bf v}_1\hat{C}^{-1}{\bf v}_1\\
&=\sqrt{q-m^2}\left[(\Phi_1^q)^{-1}Y_{21}+ [(\Phi_1^q)^{-1}- \eta q]Y_{12}\right]+2m [(\Phi_1^q)^{-1}- \eta q]Y_{11}.
\end{split}
\end{align}
Substituting the expressions \eqref{eq:Ys} into these formulas, we finally get:
\begin{align}\label{eq:Ufin}
\begin{split}
U_{11}&=\frac{q \Phi_1^q - 2 q \tilde{\Phi}_1 + 
 \tilde{Q} \tilde{\Phi}_1 + (q - \tilde{Q})^2 \tilde{\Phi}_2}{\mathcal{A}}\\
 U_{22}&=\frac{1}{\Phi_1^q - \tilde{\Phi}_1}-\frac{m^2(\Phi_2^q-\tilde{\Phi}_2)}{\mathcal{A}}\\
U_{12}&=2m\frac{\tilde{\Phi}_2(q-\tilde{Q})+\Phi_1^q-\tilde{\Phi}_1}{\mathcal{A}}.
\end{split}
\end{align}

To complete the calculation of the joint probability \eqref{app:proba_term_initial}, it remains to determine the determinant of the matrix
\[
\hat{C}^{ab}_{ij}=\delta_{ij}\Phi_1(Q^{ab})+\frac{x_j^ax_i^b}{N}\Phi_2(Q^{ab}).
\]
It is useful to decompose $\mathbb{R}^N$ into a subspace $S=\text{span}\grafe{{\bf x}^1/\sqrt{N},\ldots,{\bf x}^n/\sqrt{N},{\bf 1}/\sqrt{N}}$ with ${\bf 1}=(1, \cdots, 1)^T$, and its orthogonal complement $S^\perp$. By doing so, the matrix $\hat{C}$ in this basis, and within the RS ansatz, reads:
\begin{align}
    \hat{C}^{ab} =(\Phi_1^q\delta_{ab}+(1-\delta_{ab})\tilde{\Phi}_1)\mathbb{I}+\frac{[{\bf x}^b]^\top}{\sqrt{N}}\frac{{\bf x}^a}{\sqrt{N}}\Phi_2^{ab}
\end{align}
Therefore, within the RS ansatz, by the matrix determinant lemma, to leading exponential order in $N$ the determinant reads:
\begin{align*}
    \det\hat{C}&=\quadre{\det\begin{pmatrix}
        \Phi_1^q & \ldots & \tilde{\Phi}_1\\
        \vdots & \ddots & \vdots\\
        \tilde{\Phi}_1 & \ldots &\Phi_1^q
    \end{pmatrix}}^N o(e^N)=\left[(\Phi_1^q-\tilde{\Phi}_1)^{n-1}(\Phi_1^q+(n-1)\tilde{\Phi}_1) \right]^N o(e^N) =e^{\grafe{nN\left[\text{log}(\Phi_1^q-\tilde{\Phi}_1)+\frac{\tilde{\Phi}_1}{\Phi_1^q-\tilde{\Phi}_1}\right]+ o(N n)}}.\\
\end{align*}

\paragraph*{Asymptotic behavior of the joint probability: quenched case}
Combining all the pieces derived above, the contribution of the joint probability to the quenched complexity reads:
\begin{align}
\label{app:P_quench}
\begin{split}
\mathcal{P}:=\lim_{n\to0,N\to\infty}\frac{\log\mathbb{P}}{nN}=-\frac{1}{2}\left\{\log(2\pi)+\log(\Phi_1^q-\tilde{\Phi}_1)+\frac{\tilde{\Phi}_1}{\Phi_1^q-\tilde{\Phi}_1}+\lambda^2U_{11}-\lambda Jm U_{12}+J^2m^2U_{22}\right\},
    \end{split}
\end{align}
where $U_{11}, U_{22}$ and $U_{12}$ are given in \eqref{eq:Ufin}. As we remarked, this contribution depends on the configurations ${\bf x}^a$ only through the overlaps $q$ and $\tilde Q$. \\

\paragraph*{Asymptotic behavior of the joint probability: annealed case}
o get the  annealed case we set $n=1$ and $\tilde{Q}=0$. This gives 
\begin{align}
\det\hat{C}=e^{N\log\Phi_1^q+ o(N)}
\end{align}
and 
\begin{align}
\begin{split}
U_{11}^{A}=\frac{q}{\Phi_1^q + q \Phi_2^q}, \quad \quad
U_{12}^A =\frac{2m}{\Phi_1^q + q \Phi_2^q}, \quad \quad
U_{22}^A=\frac{\Phi_1^q + (q-m^2) \Phi_2^q}{\Phi_1^q (\Phi_1^q + 
   q \Phi_2^q)}
\end{split}
\end{align}
which finally implies
\begin{align}
\label{app:P_ann}\mathcal{P}_A:=\lim_{N\to\infty}\frac{\log\mathbb{P}|_{n=1,\tilde{Q}=0}}{N}=-\frac{1}{2}\left[\log(2\pi)+\log(\Phi_1^q)+\lambda^2U_{11}^A-\lambda J m U_{12}^A+J^2m^2U_{22}^A \right].
\end{align}

\subsection{The conditional expectation of the Jacobians}
\label{app:determinant_calc}
Consider now the term:
\begin{align}\label{eq.det}
   \mathbb{D}:= \mathbb{E}\left[\prod_{a=1}^n|\det\mathcal{H}({\bf x}^a)|\Bigg| \substack{{\bf f}^b=\lambda{\bf x}^b \\ b=1, \cdots, n} \right]=\mathbb{E}\left[ e^{\sum_{a=1}^n\text{Tr} \log |\mathcal{H}({\bf x}^a)|}\Bigg| \substack{{\bf f}^b=\lambda{\bf x}^b \\ b=1, \cdots, n} \right], \quad \quad [\mathcal{H}({\bf x})]_{ij}=\frac{\partial f_i({\bf x})}{\partial x_j} -\lambda\delta_{ij}.
\end{align}
We define $\mathcal{H}^a:=\mathcal{H}({\bf x}^a)$  and  $G^a_{ij}:=\partial_jf_i({\bf x}^a)$
where $a$ is the replica index, and $\partial_j$ denotes the derivative with respect to $x_j^a$. Computing the expectation $  \mathbb{D}$ requires a priori to determine the joint distribution of the matrices $G^a$ for $a=1, \cdots, n$ conditioned to the forces ${\bf f}^b:={\bf f}({\bf x}^b)$ taking values $\lambda {\bf x}^b$. In fact, the second equality in \eqref{eq.det} shows that, since the trace can be expressed as a sum over eigenvalues, the quantity we are interested in depends on the random matrices $G^a$ only through their eigenvalue distribution: denoting with $\tilde \rho^a_N(z)$ the eigenvalue distribution of the \emph{conditioned} matrix $G^a$, it holds:
\begin{align}\label{eq.det2}
   \mathbb{D}= \mathbb{E}\left[ e^{N \sum_{a=1}^n \int d\tilde \rho_N^a(z) \log |z-\lambda|} \right].
\end{align}
We now argue that the problem simplifies drastically if one is interested only in the leading order behavior (in $N$) of this quantity.\\

\paragraph*{Statistics of the Jacobian matrices prior to conditioning. } We begin by discussing the statistics of the entries of the random Gaussian matrices $G^a$. It holds
\begin{align}\label{eq:mean}
    \mu_{G}^{aij}:=\mathbb{E}\left[ G_{ij}^a\right] = \frac{J}{N}
\end{align}
and (making use of he compact notation \eqref{eq:not2}):
\begin{align}
\label{app:eq:sigma_G}
\begin{split}
    \Sigma_G^{aij,bkl}=\text{Cov}\left[ G_{ij}^a,G_{kl}^b\right]&=\left(\delta_{ik}\delta_{jl}\frac{[\Phi_1']^{ab}}{N}+\delta_{il}\delta_{jk}\frac{\Phi_2^{ab}}{N} \right)+\delta_{ik}[\Phi_1'']^{ab}\frac{x_l^ax_j^b}{N^2}+\delta_{jl}[\Phi'_2]^{ab}\frac{x_i^b x_k^a}{N^2}\\
    &+\delta_{jk}[\Phi_2']^{ab}\frac{x_l^a x_i^b}{N^2}+\delta_{il}[\Phi'_2]^{ab}\frac{x_k^a x_j^b}{N^2}+[\Phi_2'']^{ab}\frac{x_l^a x_k^a x_i^b x_j^b}{N^3}.
\end{split}
\end{align}
To analyze the expression for the covariances, it is convenient to perform a change of the basis in which the matrices are expressed. We consider the same decomposition of $\mathbb{R}^N$ into the $(n+1)$-dimensional subspace $S$ and its $(N-n-1)$-dimensional orthogonal complement $S^\perp$ that we have made use of in the previous subsection. Consider a set of orthonormal basis vectors ${\bf e}_\alpha$ with $\alpha=1, \cdots, N-n-1$ spanning the subspace $S^\perp$: these vectors are orthogonal to ${\bf x}^a$ and to ${\bf 1}$. Let $G^a_{\alpha \beta}={\bf e}_\alpha \cdot  G^a \cdot  {\bf e}_\beta$ be the components of the matrix $G$ in this basis. From \eqref{eq:mean} and the fact that ${\bf e}_\alpha \perp {\bf 1}$ (for $\alpha=1, \cdots, N-n-1$) it follows that these components have zero average, while \eqref{app:eq:sigma_G} shows that they have covariances 
\begin{align}
\label{app:eq:sigma_G_rotate}
\begin{split}
    \Sigma_G^{a\alpha \beta,b \gamma \delta}=\text{Cov}\left[ G_{\alpha \beta}^a,G_{\gamma \delta}^b\right]&=\left(\delta_{\alpha \gamma}\delta_{\beta \delta}\frac{[\Phi_1']^{ab}}{N}+\delta_{\alpha \delta}\delta_{\beta \gamma}\frac{\Phi_2^{ab}}{N} \right) \quad \quad \alpha, \beta, \gamma, \delta \leq N-n-1.
\end{split}
\end{align}
These covariances between the components in $S^\perp$ thus depend on the configurations ${\bf x}^a$ only through the overlap $Q^{ab}$ between them. Moreover, in this subspace the statistics is isotropic: it is invariant with respect to changes of the basis spanning the subspace. This invariance will be crucial for the subsequent calculation. \\
The covariances of the components with respect to the basis vectors in $S$, as well as the covariances between mixed components, have a more complicated expression which depends explicitly on the choice of the basis vectors in $S$, and which therefore is not basis invariant. These covariances can be computed explicitly for specific choices of the basis vectors, see \cite{ros2019complex, ros2019complexity} for similar examples. Since we are however interested only to the leading order contribution (in $N$) of the expectation value, we can neglect computing such covariances explicitly. Indeed, in the subspace  decomposition of $\mathbb{R}^N$ that we have chosen the matrices $G^a$ have a block structure, 
\begin{equation}\label{eq:Block}
    G^a= 
\begin{pNiceArray}{ccc|c}
  \Block{3-3}<\Large>{E^a}  & && \\
  & & & b^a\\
  & & &\\
    \hline
&[b^a]^T & & c^a
\end{pNiceArray},
\end{equation}
where $E^a$ is a block of dimension $(N-n-1) \times (N-n-1)$, corresponding to the subspace $S^\perp$, where the entries have statistics \eqref{app:eq:sigma_G_rotate}, while $c^a$ and $b^a$ are $(n+1) \times (n+1)$ and $(N-n-1) \times (n+1)$ dimensional blocks with entries with covariances that we have not determined explicitly. To leading order in $N$, the determinant of the matrix $G^a$ equals to the determinant of the block $E^a$, which has dimension scaling with $N$: the continuous part of the eigenvalue distribution of $G^a$ is in fact determined solely by this block in the limit $N \to \infty$. The remaining components have a different statistics, that can be expressed in terms of finite-rank perturbations (both additive and multiplicative) to a $N \times N$ matrix with entries with the same statistics as $E^a$, namely, invariant with respect to changes of basis. These finite-rank perturbations can contribute to the eigenvalues distribution with sub-leading terms in $1/N$, corresponding to isolated eigenvalues (aka, outliers). Since these isolated eigenvalues, being sub-leading, do not contribute to the complexity, we do not perform their calculation in this work. Notice that such outliers must be tracked when discussing the stability of the equilibria counted by the complexity. Indeed, there may be cases where the isolated eigenvalues are the only ones with negative real parts, leading to a dynamical instability of the equilibrium that would be overlooked if these eigenvalues were ignored. However, as we discuss extensively below and in the main text, in the models we consider, most equilibria are already linearly unstable. Therefore, for now, we neglect the calculation of the Jacobian's outliers. \\

\paragraph*{The (negligible) effect of conditioning. }
We now discuss how the statistics of the matrices $G^a$ is affected by conditioning to ${\bf f}({\bf x}^b)= \lambda {\bf x}^b$ for all $b$. We use the shorthand notation $f^a_i := f_i({\bf x}^a)$. Prior to conditioning, it holds:
\begin{align}\label{eq:mixed1}
   \mu_f^{ai}:= \mathbb{E}[f^a_i]=J \,m,\quad\quad \Sigma_{ff}^{ai,bj}:=\text{Cov}[f^a_i,f^b_j]=\delta_{ij}\Phi_1^{ab}+\frac{x_j^ax_i^b}{N}\Phi_2^{ab}.
\end{align}
and
\begin{align}\label{eq:mixed2}
    \text{Cov}\left[f_i^a,G^b_{kl} \right]=\delta_{ik}[\Phi'_1]^{ab}\frac{x^a_l}{N}+[\Phi_2']^{ab}\frac{x_l^ax_j^ax_i^b}{N^2}+\delta_{il}\Phi_2^{ab}\frac{x_j^a}{N}.
\end{align}
The conditional law of $G$ can be determined with the standard laws for Gaussian conditioning, following the procedure discussed in detail in \cite{ros2019complex}. However, Eqs.~\eqref{eq:mixed1} and \eqref{eq:mixed2} show that the only components $G_{\alpha \beta}^a= {\bf e}_\alpha \cdot G \cdot {\bf e}_\beta$ whose statistics is affected by the conditioning are those such that either ${\bf e}_\alpha$ or ${\bf e}_\beta$ belong to the subspace $S$. These are precisely the components whose statistics we are neglecting, since it gives a sub-leading contribution to the eigenvalue density.  Such subleading contributions are irrelevant for the calculation of the complexity of equilibria, as motivated in \cite{ros2019complex}, and in the previous paragraph. Again, these sub-leading contributions can contribute to eventual outliers: if one is interested in such outliers, the effect of the conditioning has to be worked out explicitly.\\

\paragraph*{Large-$N$ factorization and concentration. } The final simplifying ingredient to proceed with the calculation consists in the observation that the correlations between the Jacobian matrices evaluated at different ${\bf x}^a$, which are non-zero according to \eqref{app:eq:sigma_G}, are not relevant when computing the expectation value \eqref{eq.det2} to leading exponential order in $N$. In fact, to leading order in $N$ the expectation value factorizes,
\begin{equation}
         \mathbb{D}= \mathbb{E}\left[ e^{N \sum_{a=1}^n \int d\tilde \rho_N^a(z) \log |z-\lambda|} \right]= \prod_{a=1}^n\mathbb{E}\left[ e^{N  \int d\tilde \rho_N^a(z) \log |z-\lambda|} \right]o(e^N).
\end{equation}
An argument for such factorization can be found in \cite{ros2019complex}; it relies on the fact that the eigenvalues distribution of the Gaussian matrices $G^a$ has a large-deviation law with speed higher than $N$. For a rigorous proof of this factorization in the case of symmetric matrices and $n=2$, see \cite{subag2017complexity}.   \\

\paragraph*{The elliptic ensemble determinant. } Combining all the arguments reported above, we conclude that: (a) the joint expectation value of the conditioned Jacobian is, to leading order in $N$, determined only in terms of the single-matrix eigenvalue distributions $\tilde\rho_N^a(x)$: the correlations between the different Jacobians do not enter in the calculation; (b) the matrices $G^a$, both prior and after conditioning to the forces, have the block structure \eqref{eq:Block}; (c) to leading order in $N$, the eigenvalue distribution of matrices of this form coincides with the eigenvalue density of the block $E^a$, which has extensive size in $N$; (d) the block $E^a$ has Gaussian components with the statistics \eqref{app:eq:sigma_G_rotate}: this block is therefore a matrix belonging to the real elliptic ensemble \cite{girko1986elliptic, sommers1988spectrum, nguyen2015elliptic}. Hence, to compute the contribution of $\mathbb{D}$ to the complexity, it suffices to exploit results on the asymptotic eigenvalue density of matrices belonging to the real elliptic ensemble. This eigenvalue density is well known: it is uniform, with a support with elliptic shape in the complex plane \cite{sommers1988spectrum}. Consider the rescaled matrices $\tilde G^a= G^a/ \sqrt{\Phi_1'(q)}$ having covariances $\mathbb{E}\left[\tilde G^a_{ij}\tilde G^a_{kl}\right]=N^{-1}\left(\delta_{ik}\delta_{jl}+\alpha_q\delta_{il}\delta_{jk}\right)$, where we recall that $ \alpha_q= {\Phi_2^q}/{\dot \Phi_1^q}$. The eigenvalue density of the matrices $\tilde G^a$ is equal for each $a$. It is supported on a domain with elliptic shape in the complex plane $z=x+i y$, 
\begin{align}
\frac{x^2}{(1+\alpha_q)^2}+\frac{y^2}{(1-\alpha_q)^2}\leq 1.
\end{align}
The density is uniform, equal to
\begin{equation}
    \rho_{ \tilde G}(x,y)=\frac{1}{\pi(1-\alpha_q^2)}
\end{equation}
. We therefore find that
\begin{equation}
         \mathbb{D}= [\Phi_1'(q)]^{\frac{N n}{2}}  \tonde{ e^{N  \int dx dy\, \rho_{\tilde G}(x,y) \log |x+i y-\kappa|} }^n o(e^N), \quad \quad \kappa=\frac{\lambda}{\sqrt{\Phi_1'(q)}}.
\end{equation}
Consider now
\begin{equation}
\begin{split}
    I(\kappa)&:= \int dx dy\, \rho_{\tilde G}(x,y) \log |x+i y-\kappa|=\frac{1}{2\pi}\frac{1}{1-\alpha_q^2}\int dx dy \,\log\left[(x-\kappa)^2 + y^2\right]\\
    &= \frac{1}{\pi}\int_{-1}^1 dx\int_0^{\sqrt{1-x^2}} dy\log\left[\left(x(1+\alpha_q)-\kappa
    \right)^2+y^2(1-\alpha_q)^2\right].
    \end{split}
\end{equation}
This integral can be computed explicitly, see Appendix A.3 in \cite{RosEcoQuenched2023}. One finds that for arbitrary real $\kappa$ and $\alpha_q> -1$:
\begin{align}
   I(\kappa)=\begin{cases}
    \frac{1}{2}\left(\frac{\kappa^2}{1+\alpha_q}-1\right)\quad\quad\text{if } |\kappa|\leq 1+\alpha_q\\
    \frac{1}{8\alpha_q}\left(\kappa-\text{sign}(\kappa)\sqrt{\kappa^2-4\alpha_q}\right)^2+\log\left|\frac{\kappa+\text{sign}(\kappa)\sqrt{\kappa^2-4\alpha_q}}{2}\right|\quad\quad\text{if }|\kappa|>1+\alpha_q
    \end{cases},
\end{align}
and therefore:
\begin{align}
\label{app:det_eqn2}
    \Theta:= \lim_{N \to \infty} \lim_{n \to 0} \frac{\log \mathbb{D}}{Nn}= \begin{cases}
    \frac{\log\Phi'_1(q)}{2}+\frac{1}{2}\left(\frac{\kappa^2}{1+\alpha_q}-1\right)\quad\quad\text{if } |\kappa|\leq 1+\alpha_q\\
    \frac{\log\Phi'_1(q)}{2}+\frac{1}{8\alpha_q}\left(\kappa-\text{sign}(\kappa)\sqrt{\kappa^2-4\alpha_q}\right)^2+\log\left|\frac{\kappa+\text{sign}(\kappa)\sqrt{\kappa^2-4\alpha_q}}{2}\right|\quad\quad\text{if }|\kappa|>1+\alpha_q.
    \end{cases}
\end{align}
Notice that one gets the same contribution for both the annealed and quenched case. Whenever $\alpha_q=0$, this reduces to 
\begin{align}
\label{app:eq:det_alpha_q=0}
\Theta|_{\alpha_q=0}=\begin{cases}
    \frac{\log\Phi'_1(q)}{2}+\frac{1}{2}\left(\kappa^2-1\right)\quad\quad\text{if } |\kappa|\leq 1\\
    \frac{\log\Phi'_1(q)}{2}+\log |\kappa|\quad\quad\text{if }|\kappa|>1.\\
    \end{cases}
\end{align}
We remark that the results in this section are obtained in explicit form thanks to the simplicity of the statistics of the Jacobian matrices. In the family of models we consider, these statistics are directly and simply related to those of the Ginibre ensemble. This simplicity motivates our choice of model and underlies our analysis of the distribution of equilibria for arbitrary values of the parameter  $\alpha$. In contrast, for other models with randomness —such as the one considered in \cite{ChaosSompo88}— the random matrix structure of the Jacobians is less straightforward, particularly for arbitrary values of $\alpha$.

\paragraph*{A note on the spherical model. } As we have remarked in Sec. \ref{Sec:IntroKR}, the topological complexity of equilibria in the spherical model can be obtained from $\Sigma(m,\lambda, q)$ by setting $q \to 1$ and optimizing over $\lambda$. To be more precise, the Kac-Rice formula in the spherical case differs from \eqref{eq:KR} in the fact that the Jacobian $\mathcal{H}({\bf x})$ has to be replaced by its projection on the tangent plane to the sphere $\mathcal{S}_N(\sqrt{N})$ at the point ${\bf x}$. In other words, when expressing the matrix $\mathcal{H}({\bf x})$ in a basis such that ${\bf e}_1={\bf x}/\sqrt{N}$, the correct Jacobian is obtained neglecting the first line and column of the matrix $\mathcal{H}({\bf x})$. To leading order in $N$, however, this removal in irrelevant. The contribution of the Jacobians for the spherical model is identical to that obtained in this section, provided $q \to 1$.

\subsection{The constrained integral over the replicas}
As illustrated in the sections above, the isotropy of the correlations of the force field implies that the asymptotic behavior of both $\mathbb{P}$ and $\mathbb{D}$ depends on the vectors ${\bf x}^a$ only through the overlap parameters $Q^{ab}$. Therefore, the expression \eqref{eq:kRM} for the moments simplifies to \eqref{eq:ActionQ}. We define the volume factor:
\begin{align*}
V(Q^{ab})&:=\int \prod_a d{\bf x}^a\Delta({\bf x}^a) \prod_{a\leq b}\delta({\bf x}^a\cdot {\bf x}^b-Q^{ab})=\int \prod_a d{\bf x}^a \delta\left(\sum_i x^a_i-N m\right)\delta({\bf x}^a\cdot{\bf x}^a-N q) \prod_{a\leq b}\delta({\bf x}^a\cdot {\bf x}^b-Q^{ab}).
\end{align*}
Using the integral representation of the delta distributions and performing a rotation in the complex plane, we obtain
\begin{align*}
V(Q^{ab})
&=\int \prod_a d{\bf x}^a \int \prod_{a\leq b}\frac{d\lambda_{ab}}{\sqrt{2\pi}}\prod_a \frac{dw_a}{\sqrt{2\pi}}e^{-i\sum_{a\leq b}\lambda_{ab}({\bf x}^a\cdot {\bf x}^b-Q^{ab}N)-i\sum_aw_a(\sum_i x_i-Nm)}\\
&=\int \prod_a d{\bf x}^a \int \prod_{a\leq b}\frac{d\lambda_{ab}}{\sqrt{2\pi}}\prod_a \frac{dw_a}{\sqrt{2\pi}}e^{\frac{N}{2}\text{Tr}(\hat{\Lambda}\hat{Q})+Nm\sum_a\hat{w}_a}\left[\int \prod_a d x^a e^{-\sum_{a\leq b}x^a\lambda_{ab}x^b-\sum_aw_ax^a}  \right]^N,
\end{align*}
where $\Lambda_{ab}=2\lambda_{aa}\delta_{ab}+(1-\delta_{ab})\lambda_{ab}$. We consider the RS ansatz $w_a=\hat{w}$ and 
$\Lambda_{ab}=2\hat{\lambda}_1\delta_{ab}+\hat{\lambda}_0(1-\delta_{ab}).$
Then integral in square brackets then reads 
\begin{align}
    \int d{\bf x}\, e^{-\frac{1}{2}{\bf x}\hat{\Lambda}{\bf x}-\hat{w}{\bf 1}^\top {\bf x}}=(2\pi)^{\frac{n}{2}}(\det \hat{\Lambda})^{-\frac{1}{2}}e^{\frac{1}{2}\hat{w}^2{\bf 1}^\top\hat{\Lambda}^{-1}{\bf 1}},
\end{align}
where now in bold we denote $n$-dimensional vectors.
The inverse of $\hat{\Lambda}$:
\begin{align}
\hat{\Lambda}=
    \begin{pmatrix}
        2\hat{\lambda}_1 & \ldots & \hat{\lambda}_0\\
        \vdots & \ddots & \vdots\\
        \hat{\lambda}_0 & \ldots & 2\hat{\lambda}_1
    \end{pmatrix},
    \quad\quad\quad\quad
    \hat{\Lambda}^{-1}=
    \begin{pmatrix}
        \tilde{h} & \ldots & \tilde{z}\\
        \vdots & \ddots & \vdots\\
        \tilde{z} & \ldots & \tilde{h}
    \end{pmatrix}
\end{align}
has entries
\begin{align*}
\tilde{h}=\frac{2\hat{\lambda}_1+(n-2)\hat{\lambda}_0}{(2\hat{\lambda}_1-\hat{\lambda}_0)(2\hat{\lambda}_1+(n-1)\hat{\lambda}_0)}, \quad \quad
&\tilde{z}=-\frac{\hat{\lambda}_0}{(2\hat{\lambda}_1-\hat{\lambda}_0)(2\hat{\lambda}_1+(n-1)\hat{\lambda}_0)}.
\end{align*}
It follows that 
\begin{align*}
\text{Tr}(\hat{\Lambda}\hat{Q})=n\left[2\hat{\lambda}_1q+(n-1)\tilde{Q}\hat{\lambda}_0\right], \quad \quad
&{\bf 1}\hat{\Lambda}^{-1}{\bf 1}=n\tilde{h}+(n^2-n)\tilde{z}, \quad \quad \log\det\hat\Lambda\approx n\left[\frac{\hat{\lambda}_0}{2\hat{\lambda}_1-\hat{\lambda}_0}+\log(2\hat{\lambda}_1-\hat{\lambda}_0)\right].
\end{align*}
Summing up, we obtain for arbitrary $n$:
\begin{align*}
\frac{\log V}{N}&=\text{extr}_{\hat{\lambda}_0,\hat{\lambda}_1, \hat{w}}\Bigg\{\frac{1}{2}n\left[2\hat{\lambda}_1q+(n-1)\tilde{Q}\hat{\lambda}_0\right]
+n\hat{w}m+\frac{n}{2}\log(2\pi)-\frac{1}{2}n\left[\frac{\hat{\lambda}_0}{2\hat{\lambda}_1-\hat{\lambda}_0}+\log(2\hat{\lambda}_1-\hat{\lambda}_0)\right]\\
&+\frac{1}{2}\hat{w}^2(n\tilde{h}+(n^2-n)\tilde{z})
\Bigg\}.
\end{align*}
The contribution to the annealed complexity is obtained setting $n \to 1$, with $\hat{\lambda}_0,\tilde{Q}\to 0$. It reads:
\begin{align}
\label{app:vol_ann}
    \mathcal{V}_A(m,q):=\frac{1}{2}+\frac{1}{2}\log(2\pi (q-m^2)).
\end{align}

The contribution to the quenched complexity is instead obtained taking $n \to 0$,
\begin{align*}
    \mathcal{V}:=\lim_{N\to\infty,n\to0}\frac{\log V}{nN}=\frac{1}{2}\text{extr}_{\hat w,\hat\lambda_0,\hat\lambda_1}\Bigg\{
    2 m \hat w - \tilde{Q} \hat\lambda_0 - 
   \frac{\hat w^2}{\hat\lambda_0 - 2 \hat\lambda_1} + \frac{\hat\lambda_0}{\hat\lambda_0 - 
    2\hat\lambda_1} + 2 q \hat\lambda_1 + \log(2 \pi) - 
   \log(2\hat\lambda_1-\hat\lambda_0)
    \Bigg\}.
\end{align*}
Optimizing over the parameters results in:
\begin{align}
\label{app:vol_quench}
    \mathcal{V}(m, q , \tilde Q)=
\frac{ q-m^2 + (q - \tilde{Q}) \log(2 \pi) + (q - \tilde{Q}) \log\left(q - \tilde{Q}\right)}{2 (q - \tilde{Q})}.
\end{align}
Notice that quenched and annealed contributions coincide only for $m=0,\tilde{Q}=0$. Finally, we see that we recover all the expressions reported in  Sec.~\ref{app:sec:general_family} of the Supplemental Material.

\section{The maximal Lyapunov exponent and the complexity}\label{sec:touboul}
In this final section of the Supplemental Material, we discuss a characterization of the chaotic phase in terms of the maximal Lyapunov exponent associated to the dynamics.
A natural diagnostic of chaos is the sensitivity of the dynamical trajectories to small changes in the initial conditions. In a chaotic system, two copies of the system initialized at two points that are very close in configuration space exhibit rapidly diverging trajectories. A measure of this sensitivity to initial conditions is given by the maximal Lyapunov exponent $\Lambda_0$, which controls the exponential divergence of the distance between the two trajectories. 
The purpose of this section is to investigate to which extent the maximal Lyapunov exponent can be related to the complexity of equilibria in the chaotic phases of the models that we have studied.

For the family of dynamical systems considered in this work, the maximal Lyapunov exponent  $\Lambda_0$ can be computed explicitly when $\alpha=0$ by extending the methods of  \cite{schuecker_functional_2016, crisanti2018path}.
It has been conjectured in \cite{wainrib2013topological} that $\Lambda_0$ may be related in a quantitative way with the growth of the complexity of equilibria at the  transition to chaos.
This conjecture is rather appealing from a theoretical point of view, as it represents another attempt to formulate a dynamics-static correspondence and it is part of the questions we consider in this work. In \cite{wainrib2013topological} some affirmative indications have been shown in a standard model of recurrent neural networks \cite{ChaosSompo88}.

The fact that both the Lyapunov exponent and the topological complexity can be explicitly computed for the class of models that we are considering puts us in a position to test this conjecture. Specifically, we focus on two different models: the first is the ``pure'' model presented in the main text and in Sect.~\ref{sect: stationnary solution of the DMFT equations for alpha 0} of this Supplemental Material. The study of this model provides relevant indications; however, the model does not exhibit a transition from a Paramagnetic Fixed Point (PFP) phase to a Paramagnetic Chaotic (PC) phase, which is the one that was investigated in the work \cite{wainrib2013topological} where the conjecture was proposed. Motivated by this discrepancy, we consider as a second model the ``mixed'' one that was studied in \cite{fournier2023statistical}, which corresponds to choosing $\Phi_1(u) = 2g^2u^2 + g^2u$ in \eqref{app:eq:def_Covariance} and to considering a confining potential of the form $\lambda({\bf x})=1+||{\bf x}||^2/N$. This model exhibits a PFP-to-PC transition as a function of the strength of interactions between the degrees of freedom. Below, we show that in neither of these models do we find a quantitative relation between the behavior of the complexity and the maximal Lyapunov exponent.

\subsection{The maximal Lyapunov exponent}
\label{sectsupp: maximal lyapunov exponent}
 The maximal Lyapunov exponent is defined as \cite{RegStochMotion,paladin1987anomalous}
\begin{equation}
\begin{split}
    \Lambda_0 = \lim_{t_0\to \infty} \; \lim_{t-t_0\to\infty} \frac1{2(t-t_0)}\ln \left[\frac1N\sum_{i,j=1}^N \chi^2_{ij}(t,t_0)\right], \quad \quad \quad
    \chi_{ij}(t,t_0)= \left(\frac{\delta x_i(t)}{\delta h_j(t_0)}\right)^2,
\end{split}
    \label{Lyap_def}
\end{equation}
where $\delta h_i(t_0)$ is a small external field added at a time $t_0<t$ and acting as an infinitesimal perturbation to the dynamics.
Note that we added the $t_0\to \infty$ limit to the definition in Eq.~\eqref{Lyap_def} to ensure that we compute the Lyapunov exponent in the stationary state of the chaotic dynamics.
Eq.~\eqref{Lyap_def} defines a random quantity, which depends on the initial condition of the dynamics as well as on the realization of the random interactions between the degrees of freedom. However one can observe that
$\sum_{ij}\chi^2_{ij}(t,s)/N$ is self-averaging for $N\to \infty$, implying that:
\begin{equation}
   \Lambda_0 = \lim_{t_0\to \infty} \; \lim_{t-t_0\to\infty} \frac1{2(t-t_0)}\ln \left[\frac1N\sum_{i,j=1}^N \mathbb{E}\quadre{\chi^2_{ij}(t,t_0)}\right],
\end{equation}
where $\mathbb{E}[\cdot]$ denotes the average with respect to the quenched randomness.

\begin{figure}
  \centering
  \includegraphics[width=1\textwidth]{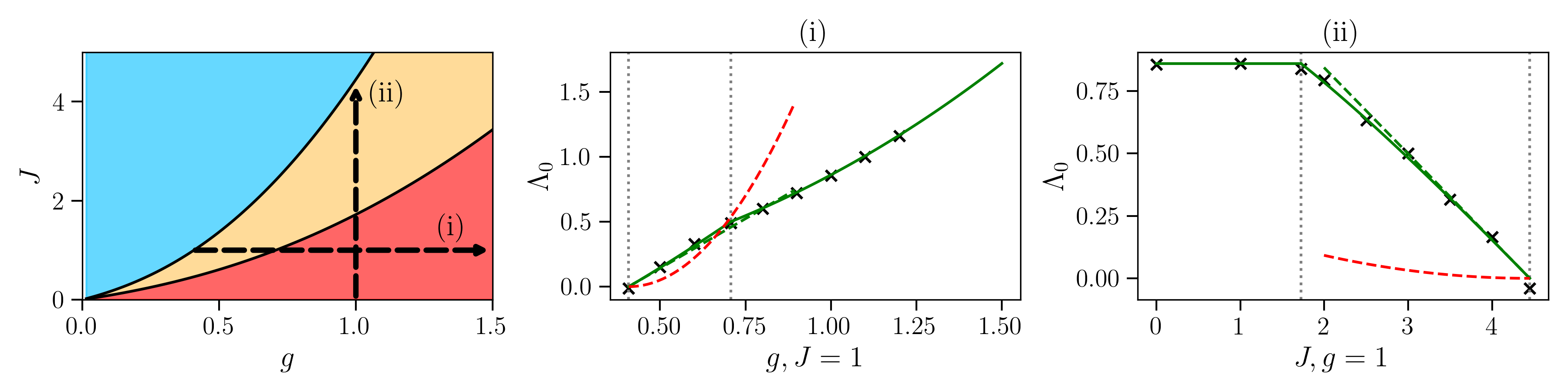}
  \caption{Phase diagram of the pure model in the spherical case. The Ferromagnetic Fixed Point (FFP) phase is in light blue, the Paramagnetic Chaotic (PC) phase in red, and the Ferromagnetic Chaotic (FC) phase in orange. Maximal Lyapunov exponent $\Lambda_0$ as a function of $g$ for fixed $J=1$ (\emph{Middle}) and as a function of $J$ for fixed $g=1$ (\emph{Right}), corresponding to the two dashed arrows in the Left panel. The analytic expression for $\Lambda_0$ is shown in the green full lines; the black marks are the results of simulations. The green dashed lines are the first order expansions of $\Lambda_0$ at the transition to the Fixed Point phase, while the red dashed lines are the expansions of the complexity at that same transition. The  vertical gray dotted lines mark the locations of the phase transitions.
 }
  \label{fig supp: phase diagram and lyapunov of pure model confined}
\end{figure}

\subsubsection{Analytical expressions}
We present here the explicit expressions of $\Lambda_0$ for the pure and mixed models, in their chaotic steady-state. These results are obtained using standard methods developed to compute Lyapunov exponents in the context of recurrent neural networks \cite{schuecker_functional_2016, crisanti2018path}. The analytical results are checked against numerical simulations in Sect.~\ref{sect: supp simulations for lyapunov}.\\
\paragraph*{\textbf{Pure model --}}
We consider first the pure model defined by $\Phi_1(u) = 2g^2u^2$, in both the spherical and confined setting.
For $\alpha=0$, the maximum Lyapunov exponent reads
\begin{equation}
    \Lambda_0 = -\lambda_\infty + g\sqrt{C_\infty + 3C_0},
\end{equation}
where $\lambda_\infty$, $C_\infty$ and $C_0$ are the stationary order parameters defined in Sect.~\ref{sect: stationnary solution of the DMFT equations for alpha 0}. In the confined case with $\lambda_\infty=C_0-\gamma$, their expression is given in Eq.~\eqref{ParaChaPhaseConfPar}
in the PC phase and by Eq.~\eqref{FChaPhaseConfPar} in the FC phase. In the spherical case instead, we have that $C_0=1$ while $C_\infty, \lambda_\infty$ are given by Eq.~\eqref{eq supp: stationary order parameters in PC phase for spherical model} in the PC phase and by Eq.~\eqref{dmftsphfc} in the FC phase.\\

\begin{figure}
  \centering
  \includegraphics[width=1\textwidth]{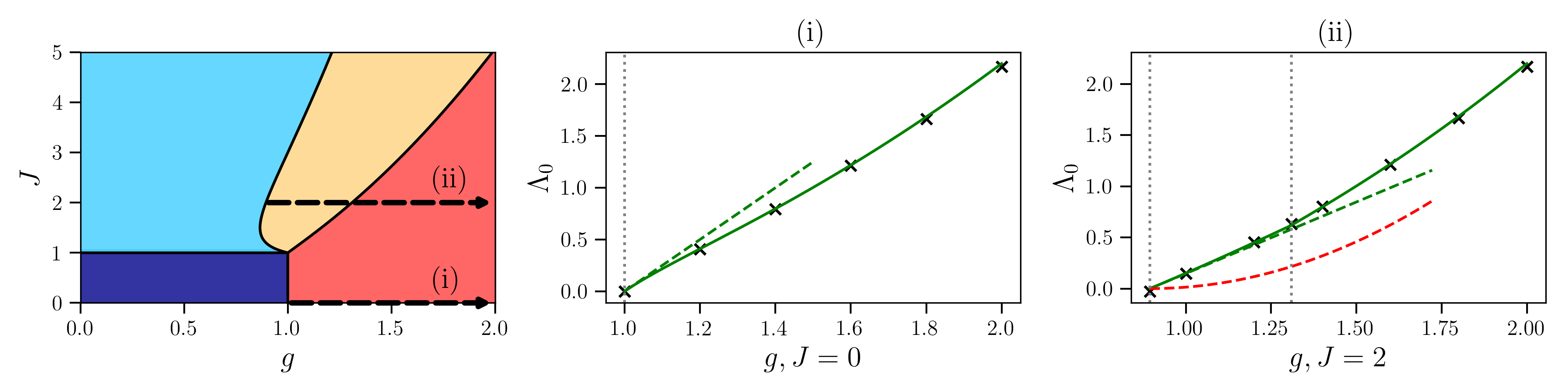}
  \caption{(\emph{Left.}) Phase diagram of the mixed model in the confined case. The Paramagnetic Fixed Point (PFP) phase is in dark blue, the Ferromagnetic Fixed Point (FFP) phase in light blue, the Paramagnetic Chaotic (PC) phase in red, and the Ferromagnetic Chaotic (FC) phase in orange. Maximal Lyapunov exponent $\Lambda_0$ as a function of $g$ for fixed $J=0$ (\emph{Middle}) and  $J=2$ (\emph{Right}), corresponding to the two dashed arrows in the Left panel. The analytic expression for $\Lambda_0$ is shown in the green full lines; the black marks are the results of simulations. The green dashed lines are the first order expansions of $\Lambda_0$ at the transition to the Fixed Point phase, while the red dashed lines are the expansions of the complexity at that same transition. The  vertical gray dotted lines mark the locations of the phase transitions.}
  \label{fig supp: phase diagrame and lyapunov of mixed model}
\end{figure}

\paragraph*{\textbf{Mixed model --}}
The conjecture about the link between the complexity and maximal Lyapunov exponent was formulated in \cite{wainrib2013topological} in connection to a model that exhibits a phase transition from a PFP phase to a PC phase. The  models that we have discussed in details in this work do not show such a transition, since the transition to chaos happens in the ferromagnetic regime. Motivated by this, we  extend our analysis to the model studied in \cite{fournier2023statistical}, which  corresponds to $\Phi_1(u) = 2g^2u^2 + g^2u$ and $\lambda({\bf x})=1+||{\bf x}||^2/N$. The dynamical phase diagram of this model is given in Fig.\ref{fig supp: phase diagrame and lyapunov of mixed model} (\emph{Left.}). The dark blue phase corresponds to the PFP phase, absent in the pure models. 
 The maximal Lyapunov exponent for $\alpha=0$ reads
\begin{equation}
    \Lambda_0 = -\lambda_\infty + g\sqrt{1+C_\infty + 3C_0}\:.
\end{equation}
The order parameters $C_\infty$ and $C_0$ can be obtained by solving the the DMFT equations for the stationary state of the dynamics, which in turn can be derived following the same recipe developed in Sec.~\ref{sect: stationnary solution of the DMFT equations for alpha 0}. The explicit expressions of the stationary order parameters are given by
\begin{gather}
    C_0 = \frac23 g^2 -1 + \frac{g}{\sqrt{3}} \sqrt{\frac43g^2-1}, \;\;\;\;\; C_\infty = 0
\end{gather}
in the PC phase, and by
\begin{gather}
    C_0 = J-1, \;\;\;\;\; C_\infty = \frac12 \left(\frac{J_0^2}{\frac43g^2} - J_0 + \frac14\right)
\end{gather}
in the FC phase. For the mixed case, we do not consider the spherical version of the model due to the fact that the latter lacks the PFP-to-PC phase transition.

\begin{figure}
  \centering
  \includegraphics[width=1\textwidth]{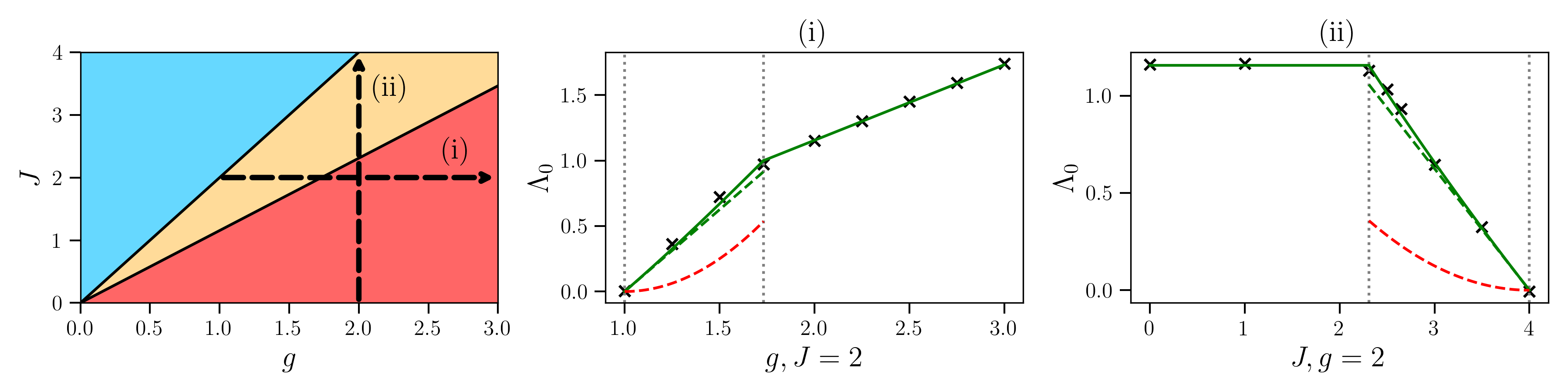}
  \caption{(\emph{Left.}) Phase diagram of the pure model in the spherical case. The Ferromagnetic Fixed Point (FFP) phase is in light blue, the Paramagnetic Chaotic (PC) phase in red, and the Ferromagnetic Chaotic (FC) phase in orange. Maximal Lyapunov exponent $\Lambda_0$ as a function of $g$ for fixed $J=2$ (\emph{Middle}) and as a function of $J$ for fixed $g=2$ (\emph{Right}), corresponding to the two dashed arrows in the Left panel. The analytic expression for $\Lambda_0$ is shown in the green full lines; the black marks are the results of simulations. The green dashed lines are the first order expansions of $\Lambda_0$ at the transition to the Fixed Point phase, while the red dashed lines are the expansions of the complexity at that same transition. The  vertical gray dotted lines mark the locations of the phase transitions. }
  \label{fig supp: phase diagram and lyapunov of pure model spherical}
\end{figure}

\subsubsection{Simulations}
\label{sect: supp simulations for lyapunov}
\begin{figure}
  \centering
  \includegraphics[width=0.4\textwidth]{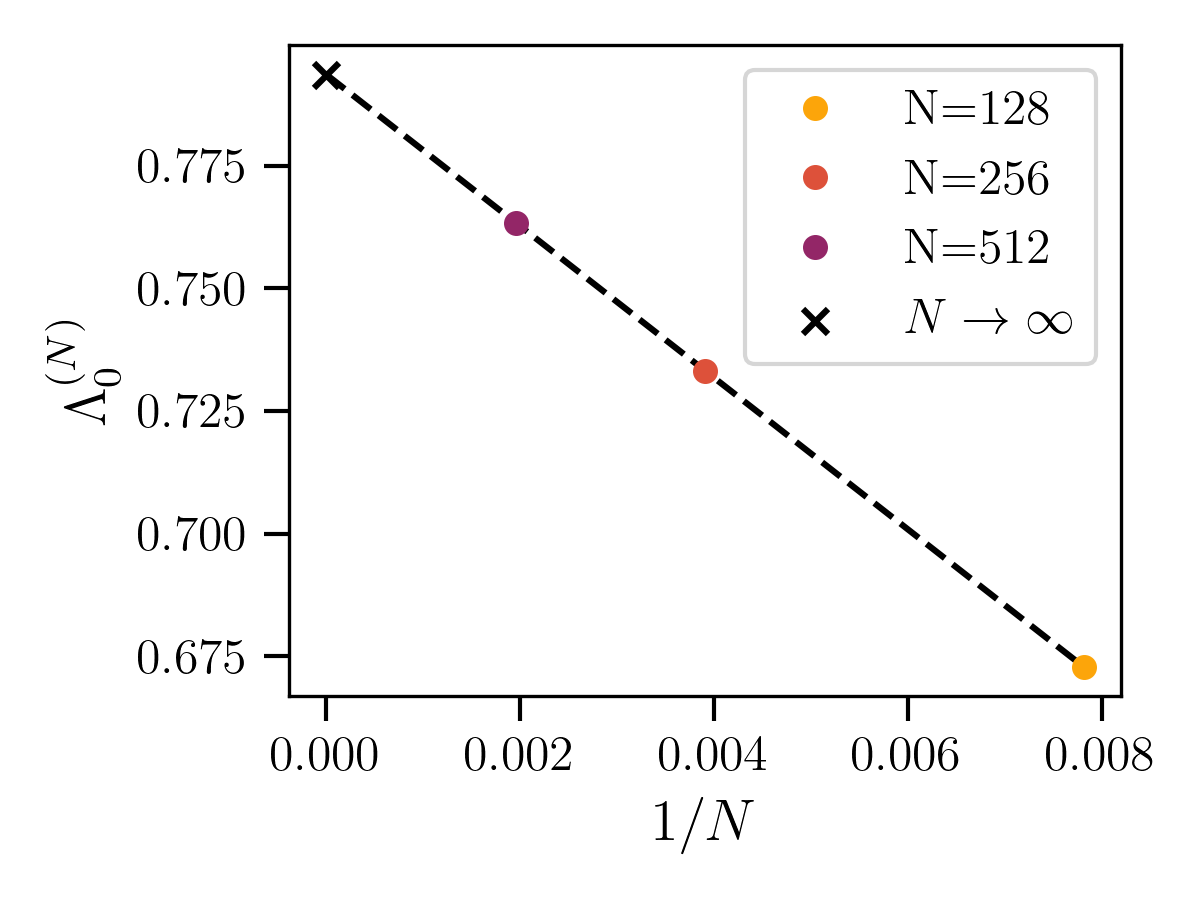}
  \caption{Numerical extrapolation (cross) of $\Lambda_0$ from the finite-$N$ values   $\Lambda_0^{(N)}$, for the mixed model at $J=0, g=1.4$. The dashed line is the fit done with the scaling ansatz Eq.~\eqref{scalin_Lya_N}.}
  \label{fig supp: extrapolation lyapunov}
\end{figure}

The maximal Lyapunov exponent can be estimated numerically by simulating the dynamics of the corresponding dynamical systems with a large system size $N$ when they are in their stationary states. We implemented a numerical estimation of the maximal Lyapunov exponent $\Lambda_0^{(N)}$ for a system of size $N$ using the Bennetin-Wolf algorithm \cite{BennetinWolf1, BennetinWolf2}. The pseudocode of the algorithm is given below. This algorithm computes a quantity $\gamma(t)$ that is related to the maximal Lyapunov exponent by
\begin{gather}
    \Lambda_0^{(N)} = \left\langle \lim_{t \to \infty}  \frac{\gamma(t)}{t} \right\rangle_S,
    \label{finit_N_L}
\end{gather} 
where $\langle \cdot \rangle_S$ is an average over $S$ samples of the dynamical system taken with different initializations of the dynamics and different realisations of the randomness.
We then define the maximal Lyapunov exponent  as
\begin{gather}
    \Lambda_0 = \lim_{N\to\infty}  \Lambda_0^{(N)}\:.
    \label{extrap_L}
\end{gather}
In order to extract this limit from the finite-$N$ simulations, we use the scaling
\begin{gather}
   \Lambda_0^{(N)}\simeq \Lambda_0 + \frac{\ell_1}{N} + \frac{\ell_2}{N^2} + \mathcal{O}\tonde{\frac 1{N^3}}\:.
    \label{scalin_Lya_N}
\end{gather}
Practically, we compute numerically $\Lambda_0^{(N)}$ for different values of $N=[128,\, 256,\, 512]$ using $S=100$ samples, and  $N_\mathrm{iter}=300/dt$, $N_\mathrm{transient}=100/dt$ and $dt=0.01$, see the pseudocode of the algorithm. We find that when the control parameters $g, J$ are not too large, a linear fit is sufficient, that is we set $\ell_2=0$ and fit $\Lambda_0, \ell_1$. Instead for large values of the control parameters, specifically for $J\geq2$ in the pure spherical model and $J\geq4$ in the pure confined model, a quadratic fit is needed.
In Fig.\ref{fig supp: extrapolation lyapunov} we show the result of the numerical extrapolation of the Lyapunov exponent for increasingly large values of $N$.

\begin{algorithm}[H]
    \caption{Bennetin-Wolf algorithm to compute the maximal Lyapunov exponent $\Lambda_0^N$}
    \label{algosupp: simulations for lyapunov}
    \begin{algorithmic}[1]
        \For{samples $s=1$ to $S$}
            \State Initialize $t=0,\, \mathbf{u}=\frac1{\sqrt{N}}(1,\ldots,1)$,\, draw random initialization $\textbf{x}_0,\,$ draw random disorder $J$
            \While{step number $<N_\mathrm{iter}$}
                \State $\textbf{x} \gets \textbf{x} + dt\left[-\lambda({\textbf{x}}) \, \textbf{x}+\textbf{f}({\textbf{x}}) \right]$
                \State $\tilde{\textbf{u}} \gets T(\textbf{x}) \textbf{u}$ \Comment{$T_{ij}(\textbf{x})=\delta_{ij} + dt \left[-\frac{\partial \lambda({\bf x})}{\partial x_j}x_i -\lambda({\bf x}) \delta_{ij} + \frac{\partial f_i({\bf x})}{\partial x_j} \right]$ for $i,j\in\{1,\ldots,N\}$}
                \State $\textbf{u} \gets \tilde{\textbf{u}} /  \|\tilde{\textbf{u}}\|$
                \If{step number $>N_\mathrm{transient}$}
                    \State $t \gets t+dt$
                    \State $\gamma \gets \gamma + \log(\|\tilde{\textbf{u}}\|)$
                \EndIf
            \EndWhile
            \State \Return $\gamma(t)/t$
        \EndFor
    \end{algorithmic}
\end{algorithm}

The data presented in Figs.~\ref{fig supp: phase diagram and lyapunov of pure model confined}--\ref{fig supp: phase diagrame and lyapunov of mixed model} show the final results of this analysis, used to validate our analytical predictions for the maximal Lyapunov exponent.
The numerical simulations are in excellent agreement with the analytical predictions.

\subsection{Comparison between the maximal Lyapunov exponent and the complexity}
\begin{figure*}[t!]
    \centering
    \includegraphics[width=0.7\textwidth]{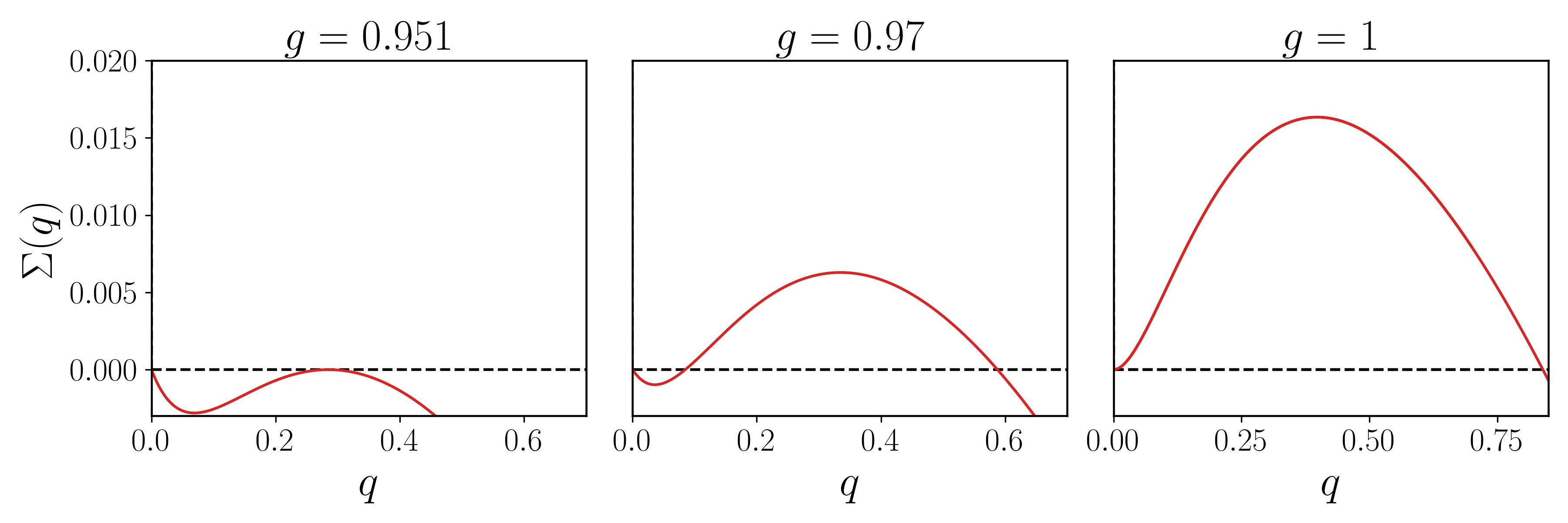}\\
    \includegraphics[width=0.7\textwidth]{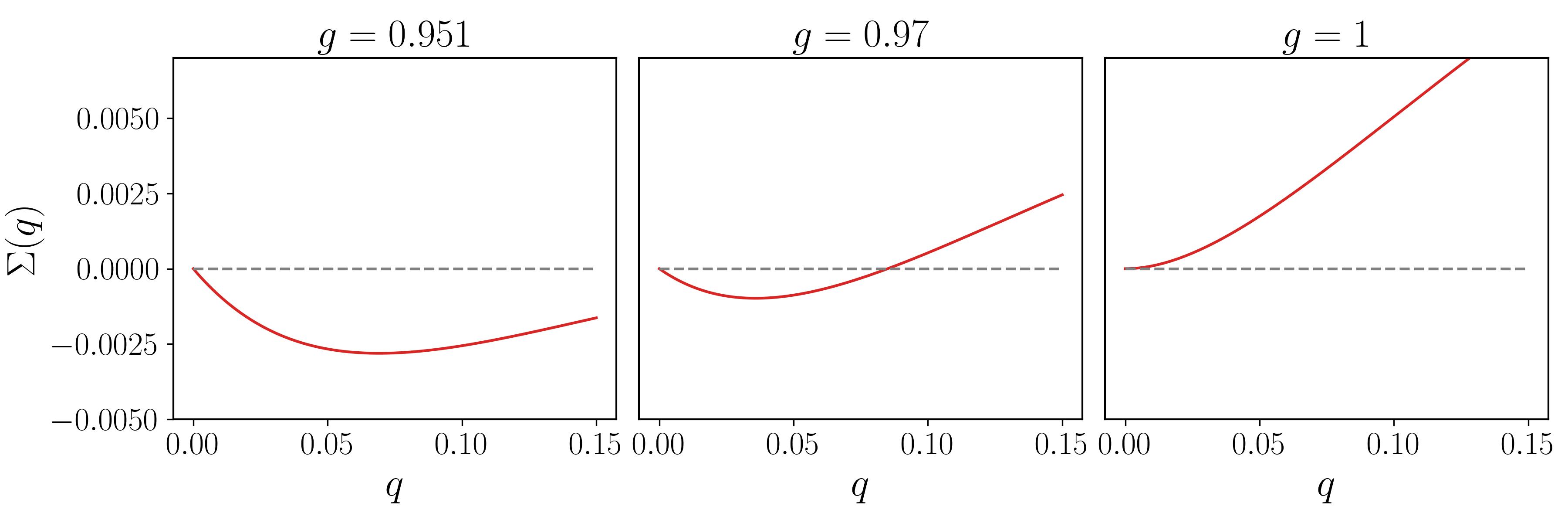}
    \caption{Plots of the complexity of unstable equilibria in the mixed confined model, at the transition PFP-PC. \textit{Top}. Plot of the paramagnetic complexity for various $g$, to observe that the complexity vanishes at a pair $(g_\text{trv},q_\text{trv})\neq (1,0)$. \textit{Bottom} Zoom of the complexity near the origin to show how the convexity changes at $q=0$ as we cross the transition $g=1$, where an exponential number of fixed points close to $q=0$ appears. The middle picture (both bottom and top) with $g=0.97$ shows the resilience gap between $q=0$ and the point where the complexity becomes positive.}
    \label{fig:mixed_kac_3_gs}
\end{figure*}

In the work \cite{wainrib2013topological} it is suggested that at the transition between a fixed point and a chaotic phase, the 
complexity and the maximal Lyapunov exponent vanish as a power law controlled by the same critical exponent. 
We find that this is violated in the models that we have analyzed, in the following two ways:
\begin{itemize}
    \item In the context of the FFP-to-FC phase transitions, the complexity of ferromagnetic equilibria (namely equilibria for which the magnetization is positive) grows with a different power law  with respect to the maximal Lyapunov exponent: while the complexity grows quadratically with the distance from the phase transition, the Lyapunov exponent grows linearly. The comparison between the Lyapunov exponent and the complexity close to the transition point is presented in Figs.\ref{fig supp: phase diagram and lyapunov of pure model confined} and \ref{fig supp: phase diagram and lyapunov of pure model spherical}. The scalings of the complexity at the FFP-to-FC transition are reported below for the various models.
    
\item In the mixed model we find that at the PFP-to-PC phase transition the complexity of equilibria does not vanish as a power law, but it is strictly positive.
Indeed, in this model the onset of chaos and the explosion of the number of equilibria happen at two different points: an exponentially large number of equilibria emerges in the PFP phase, before reaching the transition to chaos. These equilibria are all dynamically unstable, and coexist with the unique stable paramagnetic fixed point that attracts the dynamics. The complexity for $J=0$ can be derived on the same lines as the previous sections. 
For $g>1$, all fixed points are unstable, and the complexity is easily found from Eq.~\eqref{app:eq:unstable_compl}:
\begin{align}
    \Sigma(q)=\frac{1}{2}\left(-\frac{2 q (1 + q)^2}{g^2 (1 + 6 q + 8 q^2)} - \log\left(\frac{1+4q}{1 + 2 q}\right) \right).
\end{align}
For $g>1$ the point ${\bf x}={\bf 0}$ (corresponding to $q=0$) is an unstable fixed point of the dynamics. As we cross $g=1$, the discussion around Eq.~\eqref{app:eq:det_alpha_q=0} applied to this case implies that this point becomes stable. However, as we see in Fig.~\ref{fig:mixed_kac_3_gs}, the complexity of unstable equilibria remains positive (and well separated from the point ${\bf x}={\bf 0}$). The topology trivialization (that is, the vanishing of the complexity) only happens as we cross a point $g_\text{trv}$ which is solution to $\Sigma(q)=\Sigma'(q)=0$, and is given by $g_\text{trv}\approx 0.95,q_\text{trv}\approx 0.28$. Hence, the complexity is still positive for $1>g>g_\text{trv}$ even if the maximal Lyapunov exponent vanishes in this region and the system's dynamics is not chaotic, but converges to the PFP. This \textit{resilience gap} shown in Fig.~\ref{fig:mixed_kac_3_gs}-Top between the stable fixed point and the region where the complexity is positive was already observed in \cite{Fyodorov_resilient_2021}. Our analysis shows that, despite the exponential abundance of fixed points, the system (initialized at a random initial condition) is not chaotic for $g<1$. Moreover, the transition to chaos is associated with a change in convexity of the complexity at the point $q=0$. As soon as the PFP becomes unstable, there is an exponential number of unstable fixed points arbitrarily close to it (see Fig.~\ref{fig:mixed_kac_3_gs}-Bottom).

\end{itemize}

\subsection{Scalings of the Kac-Rice complexity at the FFP-to-FC transition}
Here we report the scaling of the quenched complexity at the FFP-to-FC transition, by either varying $g$ or $J$. When we get close to the transition line the complexity develops a small "island" (as we explained around Fig.~\ref{app:fig:ferro_steps}) which vanishes exactly at the transition, where the maximum of this island touches zero, thus indicating that we entered the phase with a stable fixed point. Hence, the most natural thing to study is the scaling of this local maximum of complexity, as we get close to the transition line. Below we denote by $\Sigma_{lm}$ this local maximum.\\

\noindent\textbf{Confined model.}\\
First we fix $g$ and we scale $J$ as $J=2 g (g + \sqrt{\gamma + g^2})(1-\delta)$, the scaling of the complexity is then:
\begin{equation}
    \Sigma_{lm}(\delta)|_g=\frac{(g^2 + \gamma) \left(\gamma + 
   2 g \left[g - \sqrt{g^2 + \gamma}\right]\right)}{\gamma^2}\delta^2+\mathcal{O}(\delta^3).
\end{equation}
For $\gamma=0$ it reduces to $\Sigma_{lm}(\delta)|_g=\frac{\delta^2}{4}+\mathcal{O}(\delta^3)$.\\
If instead we fix $J$ and we scale $g$ as $g=\frac{J}{2\sqrt{J_0 + \gamma}}(1+\delta)$, the scaling of the complexity is then:
\begin{equation}
    \Sigma_{lm}(\delta)|_{J}=\delta^2+\mathcal{O}(\delta^3)
\end{equation}

\noindent\textbf{Spherical model.}\\
First let us fix $g$ and scale $J$ as $J=2g(1-\delta)$, the scaling of the complexity is then:
\begin{equation}
    \Sigma_{lm}(\delta)|_g=\delta^2+\mathcal{O}(\delta^3).
\end{equation}
If instead we fix $J$ and scale $g$ as $g=\frac{J}{2}(1+\delta)$ we also get the following scaling of the complexity:
\begin{align}
    \Sigma_{lm}(\delta)|_g=\delta^2+\mathcal{O}(\delta^3).
\end{align}

\noindent\textbf{Mixed model.}\\
If we fix $g$ and scale $J$ as $J=[2 g^2 + g \sqrt{-3 + 4 g^2}](1-\delta)$, the scaling of the complexity is then:
\begin{align}
\Sigma_{lm}(\delta)|_{g}=-\frac{1}{9} (4 g^2-3) (3 + 4 g (-2 g + \sqrt{-3 + 4 g^2})) \delta^2+\mathcal{O}(\delta^3)
\end{align}
If instead we fix $J$ and scale $g$ as $g=\frac{J_0}{\sqrt{-3 + 4 J}}(1+\delta)$ at fixed $J$, the scaling of the complexity is then:
\begin{align}
\Sigma_{lm}(\delta)|_{J}=\delta^2+\mathcal{O}(\delta^3)
\end{align}

\end{document}